\documentclass[longauth]{aa} 

\newcommand{\closeappendix}{
  \setcounter{section}{0}
  \renewcommand{\thesection}{\arabic{section}}}

\usepackage{booktabs}
\usepackage{placeins}
\usepackage{lipsum} 
\usepackage{graphicx}
\usepackage{amsmath}
\usepackage{txfonts}
\usepackage{verbatim}
\usepackage{appendix}
\usepackage{hyperref}
\usepackage{booktabs}
\usepackage{tabularx}
\usepackage{ulem}
\usepackage{marvosym}
\usepackage{caption}
\usepackage[switch]{lineno}
\usepackage[dvipsnames]{xcolor}
\usepackage{orcidlink}
\hypersetup{
    colorlinks=true,
    citecolor=blue,
    linkcolor=blue,
    urlcolor=red,
}

\usepackage{graphicx}
\usepackage{txfonts}
\begin{document}

   \title{Hydride ion continuum hides absorption signatures in the NIRPS near-infrared transmission spectrum of the ultra-hot gas giant WASP-189b\thanks{Based on Guaranteed Time Observations collected at the European Southern Observatory under ESO programme 111.2506 by the NIRPS consortium.}}


\author{V. Vaulato}
\authorrunning{Vaulato et al.} 
\titlerunning{Hydride hides metals in the near-infrared spectrum of WASP-189b}


\author{
Valentina Vaulato\inst{1}\thanks{Corresponding author: \textbf{\texttt{valentina.vaulato@unige.ch}}}\orcidlink{0000-0001-7329-3471},
Stefan Pelletier\inst{1,2},
David Ehrenreich\inst{1,3},
Romain Allart\inst{2},
Eduardo Cristo\inst{4,5},
Michal Steiner\inst{1},
Xavier Dumusque\inst{1},
Hritam Chakraborty\inst{1},
Monika Lendl\inst{1},
Avidaan Srivastava\inst{1,2},
\'Etienne Artigau\inst{2,6},
Fr\'ed\'erique Baron\inst{2,6},
Susana C. C. Barros\inst{4,5},
Bj\"orn Benneke\inst{2},
Xavier Bonfils\inst{7},
Fran\c{c}ois Bouchy\inst{1},
Marta Bryan\inst{8},
Bruno L. Canto Martins\inst{9},
Ryan Cloutier\inst{10},
Neil J. Cook\inst{2},
Nicolas B. Cowan\inst{11,12},
Jose Renan De Medeiros\inst{9},
Xavier Delfosse\inst{7},
Ren\'e Doyon\inst{2,6},
Jonay I. Gonz\'alez Hern\'andez\inst{13,14},
David Lafreni\`ere\inst{2},
Izan de Castro Le\~ao\inst{9},
Christophe Lovis\inst{1},
Lison Malo\inst{2,6},
Claudio Melo\inst{15},
Lucile Mignon\inst{1,7},
Christoph Mordasini\inst{16},
Francesco Pepe\inst{1},
Rafael Rebolo\inst{13,14,17},
Jason Rowe\inst{18},
Nuno C. Santos\inst{4,5},
Damien S\'egransan\inst{1},
Alejandro Su\'arez Mascare\~no\inst{13,14},
St\'ephane Udry\inst{1},
Diana Valencia\inst{8},
Gregg Wade\inst{19},
Khaled Al Moulla\inst{1},
Jose Manuel Almenara\inst{7},
Babatunde Akinsanmi\inst{1},
Luc Bazinet\inst{2},
Vincent Bourrier\inst{1},
Charles Cadieux\inst{2},
Andres Carmona\inst{7},
Yann Carteret\inst{1},
Ana Rita Costa Silva\inst{4,5,1},
Antoine Darveau-Bernier\inst{2},
Laurie Dauplaise\inst{2},
Roseane de Lima Gomes\inst{2,9},
Jean-Baptiste Delisle\inst{1},
Thierry Forveille\inst{7},
Yolanda Frensch\inst{1,20},
Jonathan Gagn\'e\inst{21,2},
Fr\'ed\'eric Genest\inst{2},
Jo\~ao Gomes da Silva\inst{4},
Nolan Grieves\inst{1},
Melissa J. Hobson\inst{1},
Vigneshwaran Krishnamurthy\inst{11},
Alexandrine L'Heureux\inst{2},
Pierrot Lamontagne\inst{2},
Pierre Larue\inst{7},
Olivia Lim\inst{2},
Gaspare Lo Curto\inst{20},
Yuri S. Messias\inst{2,9},
Leslie Moranta\inst{2},
Dany Mounzer\inst{1},
Nicola Nari\inst{22,13,14},
Ares Osborn\inst{10},
L\'ena Parc\inst{1},
Caroline Piaulet\inst{2},
Mykhaylo Plotnykov\inst{8},
Angelica Psaridi\inst{1,23,24},
Atanas K. Stefanov\inst{13,14},
M\'arcio A. Teixeira\inst{9},
Thomas Vandal\inst{2},
Joost P. Wardenier\inst{2},
Drew Weisserman\inst{10},
Vincent Yariv\inst{7}
}

\institute{
\inst{1}Observatoire de Gen\`eve, D\'epartement d’Astronomie, Universit\'e de Gen\`eve, Chemin Pegasi 51, 1290 Versoix, Switzerland\\
\inst{2}Institut Trottier de recherche sur les exoplan\`etes, D\'epartement de Physique, Universit\'e de Montr\'eal, Montr\'eal, Qu\'ebec, Canada\\
\inst{3}Centre Vie dans l’Univers, Facult\'e des sciences de l’Universit\'e de Gen\`eve, Quai Ernest-Ansermet 30, 1205 Geneva, Switzerland\\
\inst{4}Instituto de Astrof\'isica e Ci\^encias do Espa\c{c}o, Universidade do Porto, CAUP, Rua das Estrelas, 4150-762 Porto, Portugal\\
\inst{5}Departamento de F\'isica e Astronomia, Faculdade de Ci\^encias, Universidade do Porto, Rua do Campo Alegre, 4169-007 Porto, Portugal\\
\inst{6}Observatoire du Mont-M\'egantic, Qu\'ebec, Canada\\
\inst{7}Univ. Grenoble Alpes, CNRS, IPAG, F-38000 Grenoble, France\\
\inst{8}Department of Physics, University of Toronto, Toronto, ON M5S 3H4, Canada\\
\inst{9}Departamento de F\'isica Te\'orica e Experimental, Universidade Federal do Rio Grande do Norte, Campus Universit\'ario, Natal, RN, 59072-970, Brazil\\
\inst{10}Department of Physics \& Astronomy, McMaster University, 1280 Main St W, Hamilton, ON, L8S 4L8, Canada\\
\inst{11}Department of Physics, McGill University, 3600 rue University, Montr\'eal, QC, H3A 2T8, Canada\\
\inst{12}Department of Earth \& Planetary Sciences, McGill University, 3450 rue University, Montr\'eal, QC, H3A 0E8, Canada\\
\inst{13}Instituto de Astrof\'isica de Canarias (IAC), Calle V\'ia L\'actea s/n, 38205 La Laguna, Tenerife, Spain\\
\inst{14}Departamento de Astrof\'isica, Universidad de La Laguna (ULL), 38206 La Laguna, Tenerife, Spain\\
\inst{15}European Southern Observatory (ESO), Karl-Schwarzschild-Str. 2, 85748 Garching bei M\"unchen, Germany\\
\inst{16}Space Research and Planetary Sciences, Physics Institute, University of Bern, Gesellschaftsstrasse 6, 3012 Bern, Switzerland\\
\inst{17}Consejo Superior de Investigaciones Cient\'ificas (CSIC), E-28006 Madrid, Spain\\
\inst{18}Bishop's University, Dept of Physics and Astronomy, Johnson-104E, 2600 College Street, Sherbrooke, QC, Canada, J1M 1Z7\\
\inst{19}Department of Physics and Space Science, Royal Military College of Canada, PO Box 17000, Station Forces, Kingston, ON, Canada\\
\inst{20}European Southern Observatory (ESO), Av. Alonso de Cordova 3107, Casilla 19001, Santiago de Chile, Chile\\
\inst{21}Plan\'etarium de Montr\'eal, Espace pour la Vie, 4801 av. Pierre-de Coubertin, Montr\'eal, Qu\'ebec, Canada\\
\inst{22}Light Bridges S.L., Observatorio del Teide, Carretera del Observatorio, s/n Guimar, 38500, Tenerife, Canarias, Spain\\
\inst{23}Institute of Space Sciences (ICE, CSIC), Carrer de Can Magrans S/N, Campus UAB, Cerdanyola del Valles, E-08193, Spain\\
\inst{24}Institut d’Estudis Espacials de Catalunya (IEEC), 08860 Castelldefels (Barcelona), Spain\\
\inst{*}\email{valentina.vaulato@unige.ch}
}

   \date{Received 12 November 2024; Accepted 3 March 2025}

 
  \abstract
    {Ultra-hot Jupiters showcase one-of-a-kind extreme atmospheric conditions, including the dissociation of molecules into atomic species, ionisation, and significant day-to-night temperature contrasts. 
    The proximity to their host stars exposes ultra-hot Jupiters to intense stellar irradiation, enabling high temperatures that drive noteworthy contributions to the overall opacity by hydride ions (H$^-$), potentially obscuring features of metals in the near-infrared transmission spectrum.}
    {This work aims to detect atomic, ionic, and molecular species in the atmosphere of WASP-189b (H, He, Fe, Ti, V, Mn, Na, Mg, Ca, Cr,
Ni, Y, Ba, Sc, Fe$^+$, Ti$^+$, TiO, H$_2$O, CO, and OH). A focus is placed on (i) understanding the role of H$^-$ as a source of absorption continuum opacity, and (ii) retrieving the relative hydride-to-Fe abundance using combined optical and near-infrared data.}
    {We present two transits of WASP-189b gathered simultaneously in the optical with HARPS and in the near-infrared with NIRPS, supported by photometric light curves from EulerCam and ExTrA. Transmission spectra were analysed via cross-correlation to detect a planet's absorption features and to increase the signal-to-noise ratio of potential detections. Additionally, atmospheric retrievals quantified relative abundances by fitting the overall metallicity, and abundance proxies for TiO, H$^-$, and e$^-$.}
    {Only atomic iron is detected in HARPS data (S/N$\sim$5.5). However, no Fe is detected at near-infrared wavelengths, likely due to the H$^{-}$ continuum dampening. 
    Atmospheric retrievals on HARPS only and HARPS+NIRPS combined suggest that the hydride-to-Fe ratio exceeds equilibrium model predictions by $\sim$0.5 dex, hinting at a strong ionisation rate for hydrogen atoms.
    Including NIRPS data helps to constrain the H$^{-}$ abundance, as well as set an upper limit on the free electron density, which is unconstrained from the HARPS-only retrieval. These results emphasise the impact of H$^{-}$ as a non-negligible source of continuum absorption opacity impeding the detection of planetary absorption features in the near-infrared transmission spectrum of WASP-189b.}
{}
   \keywords{instrumentation: spectrographs – methods: observational – techniques: spectroscopic – planets and satellites: atmospheres – planets and satellites: composition – planets and satellites: gaseous planets
               }

   \maketitle
%
\section{Introduction}
\label{sec:Introduction}
Ultra-hot Jupiters are a class of gas giant exoplanets with short orbital periods around hot, early-type (typically A or F) host stars 
\citep[e.g.][]{Hellier+2009,A.CollierCameron+2010,West+2016,Delrez+2016,Gaudi+2017,Lund+2017,Talens+2018,Anderson+2018}. 
These planets are exposed to intense stellar irradiation, leading to equilibrium temperatures above 2000~K and giving rise to extreme physico-chemical and climatic conditions. Ultra-hot Jupiters experience 
cloud-free daysides, the dissociation of molecules into atomic and ionic species 
\citep{Kitzmann+2018, Parmentier+2018, Hoeijmakers+2018, Hoeijmakers+2019}, and strong day-to-night temperature contrasts and winds \citep{Arcangeli+2019, Seidel+2019, Seidel+2021, Seidel+2023} resulting in the dissociation or recombination and ionisation of molecular hydrogen \citep[H$_2$;][]{BellAndCowan2018}. Condensation of metals can happen at the terminator as material from the dayside is transported to the nightside by the day-to-night flow \citep{Ehrenreich+2020Nature}. On the other hand, evaporation of metals happens on the dayside, where the overall temperature of the atmosphere of the planet increases. Metallic species such as iron (Fe) or titanium (Ti) and vanadium (V) and their oxides are thought to play an important role in driving inversions in the temperature profile \citep{Evans+2017, Lothringer+2018}, boosting the atmospheric scale height $H(z) = k_{B} T(z) / \mu(z) g(z)$ (where $z$ is the altitude; $k_{B}$ is the Boltzmann constant; $T(z)$, $\mu(z)$, and $g(z)$ are the temperature, mean-molecular weight, and acceleration of gravity, respectively) at high altitudes. Indeed, the scale height is impacted by the local mean-molecular weight and temperature of the atmosphere (gravity can be assumed to be constant).
As a result of the dependence of the scale height on the temperature and mean-molecular weight, the dayside is “puffier” than the nightside. The effect is twofold. Firstly, the dayside is much hotter than the nightside. Secondly, atomic hydrogen on the dayside is subject to thermal dissociation, resulting in a factor~2 decrease in mean-molecular weight. This dichotomy gives ultra-hot gas giants a peculiar terminator shape as seen during transit, characterised by two halves of a different thickness \citep{Pluriel+2020, Wardenier+2021, Wardenier+2023}, making them particularly well-suited for transit spectroscopy. 
Transmission spectroscopy benefits from the stellar light being filtered by the atmosphere of the planet. Indeed, during the transit event, the stellar photons are absorbed and scattered depending on the wavelength, the chemical composition, and the structure of the planetary atmosphere \citep{SeagerAndSasselov2000}.
The transmission spectrum of a planetary atmosphere can exhibit the following: (i) absorption features caused by electronic transitions in atoms and ions or vibrational, rotational, and rovibrational transitions in molecules, and (ii) scattering continuum due to atoms and molecules (Rayleigh regime) or aerosols (Mie regime). Ultra-hot gas giants are predicted to bear atomic and ionic species on their dayside atmospheres, with a composition close to chemical equilibrium \citep{Kitzmann+2018} and patchy nightside clouds due to the difficulty to form aerosols at such high temperatures \citep{Komacek+2022}.

Under clear atmospheres (i.e. in the absence of Mie scattering and significant continuum absorption caused by clouds, hazes, and aerosols), one would expect the contrast of absorption features to be maximised in the transmission spectrum of exoplanetary atmospheres. However, \cite{Arcangeli+2018} argue that in the atmospheres of ultra-hot gas giants, the contrast of potential spectroscopic features must be dampened by the absorption continuum caused by the negative hydrogen ion (hereafter, hydride or H$^-$). This anion has long been known to constitute `a major factor ruling the structure of late-type stellar atmospheres' \citep[][`late-type' meaning here cooler than F-type ones]{Wildt1939}, which share effective temperatures akin to those of ultra-hot gas giant daysides. 
In atmospheric layers where the electron density is high (due to the photoionisation of alkali metals), the hydrogen atom can hold a second electron in a bound state due to the highly polarised structure of the neutral H atom \citep{Gray_2021}. This extra electron can be set free by any photon impacting with an energy larger than 0.7542~eV, or a wavelength shorter than 1.644~$\mu$m. This photoelectric ionisation of the hydride ion thus results in a so-called bound-free absorption continuum covering all spectral domains bluer than the \textit{H} band (from 1.1 to 1.4 $\mu$m). In the optical, bound-free transitions are dominant, but their contribution decreases beyond 1.4 $\mu$m, where free-free interactions to form hydride become more significant in the infrared domain. 

Furthermore, the H$^-$ opacity overlaps with important absorbing species such as titanium and vanadium oxide \citep{Parmentier+2018}. Similarly, the near-infrared domain typically dominated by strong molecular bands in cooler planets becomes muted in ultra-hot Jupiters due to the combination of molecular dissociation and the increased presence of the hydride continuum.  Constraining the abundance of hydride relative to other species in the near-infrared could thus help disentangling the different sources of opacity at shorter wavelengths, or vice versa.

In this work, we report on new transit observations of the ultra-hot Jupiter WASP-189b \citep{Anderson+2018} obtained in the framework of the Guaranteed Time Observations (GTO) of the NIRPS consortium. The Near-InfraRed Planet Searcher \citep[NIRPS\footnote{\url{https://www.eso.org/public/teles-instr/lasilla/36/nirps/}\\ \url{https://www.eso.org/sci/facilities/lasilla/instruments/nirps.html}}, ][]{ Bouchy+2017, Wildi+2022, Bouchy+2025} is a new ultra-stable, near-infrared ($YJH$, from 0.95 to 1.8~$\mu$m), high-resolution fibre-fed spectrograph installed at the ESO 3.6~m telescope in La Silla, Chile. NIRPS is designed to observe simultaneously with HARPS \citep{Mayor+2003}, which covers the optical range from 378 up to 691~nm. 
Thanks to HARPS and NIRPS observations, we can obtain high-resolution spectra ($R = \lambda / \Delta \lambda$ ranging from $\sim$80\,000 in the near-infrared to 120\,000 in the optical) from the edge of the near-ultraviolet to the \textit{H} band, with a $\sim$250~nm gap between $\sim$700 and 950~nm. The NIRPS GTO consortium exploited these capabilities to design a large survey of exoplanetary atmospheres \citep{Allart+2025}, composed of several sub-programmes: (i) a transit survey, (ii) a dayside emission survey, as well as programmes dedicated to study (iii) young exoplanets, (iv) multi-planetary systems, (v) planetary system architectures and (vi) a TRAPPIST-1 dedicated initiative. The transit survey aims at observing 75 giant planets over 5 years for a wide range of irradiation levels and masses to derive population-level trends in the context of planet formation and evolution.

WASP-189b is the most irradiated target of the NIRPS GTO transit survey and one of the first transiting planets observed with NIRPS. 
WASP-189b \citep{Anderson+2018} is one of the most highly irradiated Jovian planets known to date: it is a massive ($M_p \approx 2~\mathrm{M_J}$) gas giant orbiting the hot ($T_\mathrm{eff} = 8\,000$~K) A-type star on a circular and strongly misaligned, almost polar orbit \citep[3D obliquity $\Psi = 85.4~\pm~4.3$\degr;][]{Lendl+2020}. A summary of the WASP-189 planetary system is provided in Table~\ref{Table1:StarPlanetParameters}. Observations of occultations of the planet by its star with the CHEOPS space telescope~\citep{Benz+2021} yielded an estimation of the dayside brightness temperature of $3\,435~\pm~27$~K \citep{Lendl+2020} and a precise planet radius of $1.619~\pm~0.021~\mathrm{R_J}$. Analysing two archival optical transits obtained with HARPS, \citet{Stangret+2022} reported on the detections of absorption features by neutral and singly ionised iron (\ion{Fe}{i} and \ion{Fe}{ii}, respectively) and neutral titanium (\ion{Ti}{i}). These chemical species were identified from the cross-correlation function (CCF) of the transmission spectrum with template models, as well as tentative, single-line detections of the hydrogen Balmer lines (H$_\alpha$, H$_\beta$), and the calcium \ion{Ca}{ii} H and K lines.  
Furthermore, \citet{Prinoth+2022,Prinoth+2023, Prinoth+2024} performed a broad optical survey of atomic and molecular species in the atmosphere of WASP-189b by jointly analysing transit observations obtained with the ESPRESSO \citep{Pepe+2021}, MAROON-X \citep{Seifahrt+2020, Seifahrt+2018}, HARPS, and HARPS-N \citep{Cosentino+2012} spectrographs. \citet{Prinoth+2023} reported significant detections of a variety of species (H, Na, Mg, Ca, Ca$^{+}$, Ti, Ti$^{+}$, TiO, V, Cr, Mn, Fe, Fe$^{+}$, Ni, Sr, Sr$^{+}$, and Ba$^{+}$). 
\cite{Gandhi+2023} retrieved abundances of a subset of these metals using four archival HARPS-South and HARPS-North transits combined, finding most to be in agreement with chemistry predictions of a solar-composition atmospheric model; only Fe is found to be depleted (Fe constraint to be below the stellar Fe/H value by $\sim$0.7 dex) likely because a large amount of iron is thought to be ionised. 
Additionally, \cite{Yan+2020} detected atomic Fe emission lines in dayside spectra of WASP-189b with HARPS-N. At longer wavelengths, \cite{Yan+2022} detected carbon monoxide (CO) in the thermal emission spectra observed with the GIANO-B near-infrared spectrograph \citep{Claudi+2017}. At shorter wavelengths, \cite{Sreejith+2023} observed the magnesium \ion{Mg}{ii} h and k lines with the CUTE \citep{France+2023} ultraviolet space telescope. Recently, \cite{Lesjak+2025} reported on blueshifted cross-correlation signals of CO and Fe likely due to day-to-night winds from CRIRES+ $K$-band dayside observations of WASP-189b. Together, these studies provide a comprehensive view of the atmospheric dynamics and chemical complexity of WASP-189b, highlighting the interplay between neutral and ionised species and the mechanisms that maintain the thermal structure of this ultra-hot Jupiter.

In this work, we report on the first near-infrared high-resolution spectroscopic NIRPS observations of the transit of WASP-189b. Section \ref{sec:Observations} describes the spectroscopic and photometric observations. Section \ref{sec:Stellar properties of WASP-189} describes the stellar properties, the telluric and preparatory corrections, and details on the cross-correlation transmission spectroscopy method used. In Section \ref{sec:Results and discussion}, we present the results of both the cross-correlation and the atmospheric retrieval analyses, and discuss their implications.

    \begin{table*}[htbp]
    \caption{Planetary and stellar relevant parameters}
    \begin{tabularx}{\textwidth}{XXXX}
    \toprule
    \midrule
                                           & \text{Unit} & \text{Value} & \text{Reference}\\
    \midrule
    \multicolumn{4}{l}{Planet parameters} \\
    \midrule
    Planet radius, $R_{p}$                 & R$_\mathrm{J}$        & $1.619 \pm 0.021$         & \cite{Lendl+2020}  \\ 
    Planet mass, $M_{p}$                   & M$_\mathrm{J}$        & $1.99^{+0.16}_{-0.14}$    &  \cite{Lendl+2020}\\
    Dayside Eq. temperature, $T_{eq}$              & K            & $3\,353^{+27}_{-34}$      &  \cite{Lendl+2020} \\
    Eccentricity, $e$                      & -            & 0 (fixed)                 & \cite{Lendl+2020} \\
    Orbital inclination, $i$                  & deg          & $84.03\pm0.14$            & \cite{Lendl+2020}\\
    True orb. obliquity, $\Psi$            & deg          & $85.4\pm4.3$              & \cite{Lendl+2020} \\
    $a/R_{\star}$                          & -            & $4.600^{+0.031}_{-0.025}$ & \cite{Lendl+2020} \\
    Transit duration, $T_{14}$ & h & 4.04 & This work\\
    Transit midpoint, $T_c$ & days & $2458926.5416960^{+0.0000650}_{-0.0000640}$ & \cite{Lendl+2020} \\
    Orbital period, $P$ & days & $2.7240308\pm0.0000028$ & \cite{Lendl+2020} \\
    \midrule
    \multicolumn{4}{l}{Stellar parameters} \\
    \midrule
    Spectral type                          & -            & A4/5/6(m?)\,\sc{iv/v}     &  $\dagger$  \\
    Stellar radius, $R_{\star}$            & R$_{\odot}$  & $2.36 \pm 0.03$           &  \cite{Lendl+2020}  \\
    Stellar mass, $M_{\star}$              & M$_{\odot}$  &   $2.030 \pm 0.066$       & \cite{Lendl+2020} \\
    Stellar magnitude, $J$                 & -            & 6.166                     & \citet{Skrutskie+2006} \\
    Proj. rot. velocity, $v \sin{i_\star}$ & km~s$^{-1}$  & $93.1\pm1.7$              & \cite{Lendl+2020}\\
    Reflex vel. semi-amp., $K_\star$ & m~s$^{-1}$ & 182 $\pm$ 13 & Adopted solution by \cite{Anderson+2018} \\
    \midrule
    \bottomrule                             
    \end{tabularx}
    \label{Table1:StarPlanetParameters}
    \begin{minipage}{\linewidth}
    \vspace{0.1cm}
    \small  Notes: $^\dagger$See Sect.~\ref{sec:Stellar properties of WASP-189} about the stellar classification.
    \end{minipage}
    \end{table*}

\section{Observations}
\label{sec:Observations}

\subsection{Spectroscopic transit observations with NIRPS and HARPS}
\label{subsec:Spectroscopic transit observations with NIRPS and HARPS}
\begin{figure}
   \centering
   \includegraphics[width=\columnwidth]{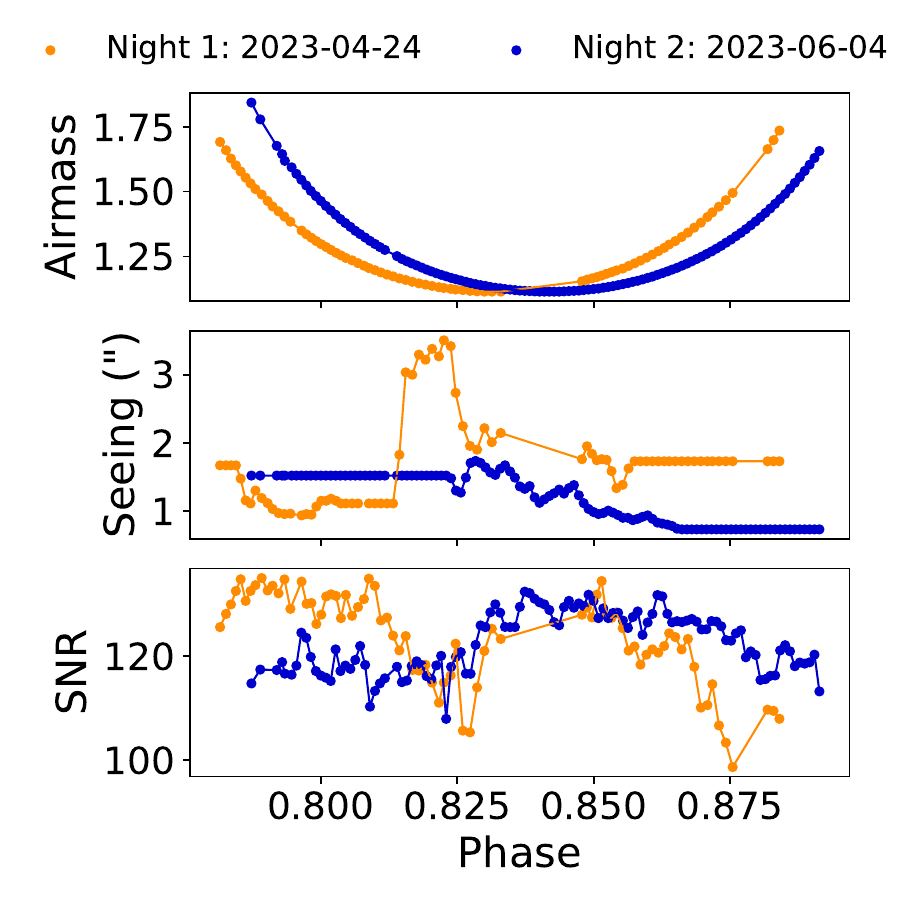}
   \caption{Observing conditions during Night 1 and Night 2 as a function of phase \citep[phase = $(BJD - T_\mathrm{c})/ \mathrm{P}$, where $BJD$ is the barycentric Julian date, $T_\mathrm{c} = 2458926.5$ days is the mid-transit time, and $\mathrm{P}=2.724033$ days is the period, which are all from][]{Lendl+2020}. The top panel and mid panel illustrate the airmass and the average atmospheric Dimm seeing evolving during the nights, respectively. The Dimm seeing is averaged for each observations (i.e. average between the start and the end of each observation). The bottom panel shows the S/N (extracted from the fits header) changing during the nights.}
   \label{Figure:NightConditions}
\end{figure}

We observed two transits (Figure \ref{Figure:NightConditions}) of WASP-189b across the disc of its bright host star \object{HD~133112} ($V=6.6$, $J = 6.2$) on 2023-04-24 (Night 1) and on 2023-04-06 (Night 2), as part of the ESO GTO program 111.2506 (PI: F. Bouchy). The transit duration (from contact point 1 and contact point 4) is $\sim$4 hours (see~Table~\ref{Table1:StarPlanetParameters}), and the total observing time (transit duration + baseline before ingress and after egress) was $\sim$6 hours for both nights. We observed the target simultaneously with HARPS in High Accuracy Mode (HAM) and NIRPS in High Efficiency (HE) mode. The adaptive optics system provides NIRPS with a main High Accuracy (HA) fibre covering a diameter of 0.4 arcsec on the sky for a greater spectral resolution (maximum of 96\,300 for the science fibre), and a larger HE fibre with a field of view of 0.9 arcsec \citep[see][for more details]{Bouchy+2025}. The NIRPS HE mode allows to reduce the impact of modal noise at the reddest wavelengths while keeping a good spectral resolution (maximum of 80\,300 for the science fibre) with respect to the HA mode. Both instruments have two fibres A and B, located on the target and the sky, respectively. 
Between 1.3765 and 1.3973~$\mu$m two echelle spectral orders (104 and 105) are missing due to OH doped absorption in the optical fibre train. These two orders also correspond to the unusable domain of the deep telluric water band between \textit{J} and \textit{H} photometric bands. The transit time series acquired on 2023-04-24 exhibits an approximately one hour gap which occurred during the transit window (from 05:38 UT until 06:35 UT) due to a communication issue between NIRPS and the telescope. This explains why we gathered a different total number of spectra between NIRPS and HARPS during the first night (difference of 20 spectra).
The observation log can be found in Table~\ref{Table2:Summary_HARPSNIRPS_transit_obs} and in Figure~\ref{Figure:NightConditions}.

%

    \begin{table}[]
    \caption{Log of NIRPS and HARPS transit observations.}
    \begin{tabularx}{\linewidth}{XlXX}
    \toprule
    \midrule
         & \text{Night 1} & \text{Night 2} \\
    Date &  2023-04-24 & 2023-06-04 \\
    \midrule
    NIRPS & & \\
    \midrule
    Number of spectra & 78 & 112 \\
    Exp. time [s] & 200 & 200 \\
    Avg. seeing ["] & 1.61 & 0.6 \\
    Avg. S/N order 57 $^{\star}$& 162 & 160 \\
    S/N \textit{Y} band & 147 & 148 \\
    \midrule
    HARPS & & \\
    \midrule
    Number of spectra & 98 & 101 \\
    Exp. time [s] & 200 & 200 \\
    Avg. seeing ["] & 1.61 & 0.6 \\
    S/N at 550 nm & 93 & 111\\
    \midrule
    \bottomrule  
    \end{tabularx}
    \label{Table2:Summary_HARPSNIRPS_transit_obs}
    \begin{minipage}{\linewidth}
    \vspace{0.1cm}
    \small Notes: $^{\star}$In NIRPS spectra, we choose as reference the spectral order 57 centred at 16\,285 $\AA$ (\textit{H} band) because it is little affected by telluric absorption lines.
    \end{minipage}
    \end{table}

\subsection{Simultaneous photometry with EulerCam and ExTrA}
\label{subsec:Simultaneous photometry with EulerCam}
Photometric observations contemporaneous to spectroscopic ones may refine the ephemeris of the system, improving orbital parameters such as the planet's orbital period, the transit duration, mid-transit time, and to assess the stellar variability (e.g. absence of flares and spots) if any. This allows for more accurate predictions of future transit events.

\begin{figure}
   \centering
   \includegraphics[width=\columnwidth]{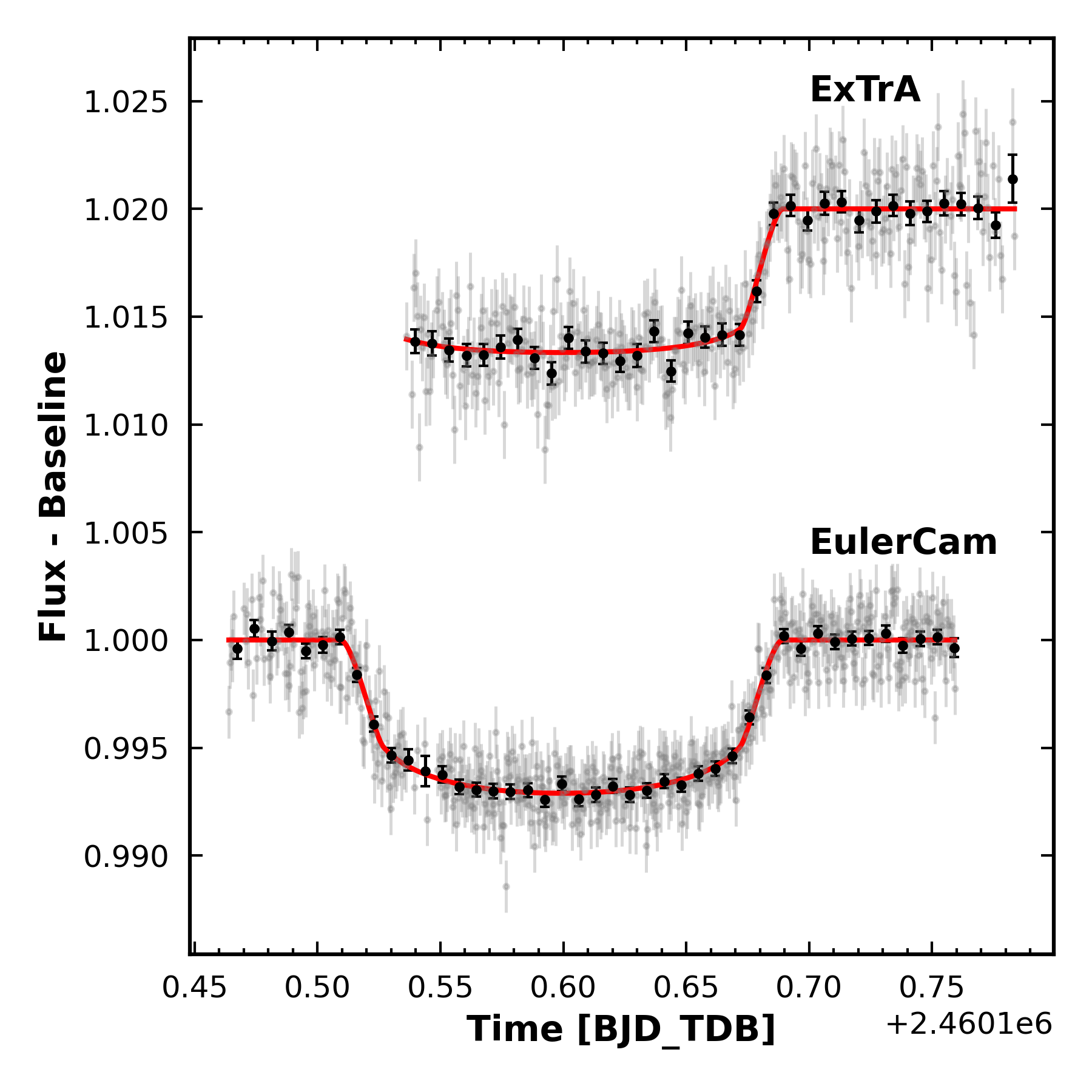}
   \caption{De-trended light curve obtained with ExTrA and EulerCam, respectively. The transit model is over-plotted in red. The root-mean-squares of EulerCam and ExTrA light curves are 1206 and 1620 ppm, respectively.}
   \label{Figure:WASP189b_EulerPhotometry}
\end{figure}

In parallel to our NIRPS-HARPS observation on 2023-04-06, we observed the transit with EulerCam \citep{Lendl+2012} and ExTrA \citep{ExTrA_Bonfils}. The lightcurves are reported in Figure \ref{Figure:WASP189b_EulerPhotometry}. EulerCam is a 4k $\times$ 4k CCD camera installed at the 1.2 metre Swiss Euler telescope in La Silla, Chile. The observations were taken in Johnson-$V$ filter and the exposure time was set to 30 seconds. The ExTrA facility consists of three 0.6-meter telescopes feeding a near-infrared, multi-object spectrograph in La Silla, Chile. The observations were taken on the three telescopes mode and the exposure time was set to 60 seconds. We lost ingress of the transit observation as the dome was vignetting in the eastern horizon up to 42 degrees. Additionally, we had to discard light curves from telescopes one and two, due to high photometric scatter.

To obtain the system parameters, we fit the observed light curves with transit models computed using \texttt{CONAN} \citep{Lendl+2017}.  The free parameters include the mid-transit time, duration, transit depth, impact parameter, and quadratic limb-darkening coefficients. We choose wide Gaussian priors centred on the values from \cite{Lendl+2020}. The quadratic limb-darkening coefficients were centred around estimates from the \texttt{LDCU\footnote{\url{https://github.com/delinea/LDCU}}} package \citep{Deline+2022}. \texttt{LDCU} is a modified version of the Python routine implemented by \cite{Espinoza2015} that computes the limb-darkening coefficients and their corresponding uncertainties using a set of stellar intensity profiles accounting for the uncertainties on the stellar parameters. The stellar intensity profiles are generated based on two libraries of synthetic stellar spectra: ATLAS \citep{Kurucz1979} and PHOENIX \citep{Husser2013}. In addition, we fix the period and eccentricity to literature values of 2.724033 days and 0, respectively. To account for the correlated noise in our observations, we fit a photometric baseline model along with the pure transit model. The baseline model is determined by iteratively fitting different models involving air mass, exposure time, sky
background, shifts (both coordinate and abscissa), and the Full-With at Half-Maximum (FWHM) of the stellar Point Spread Function (PSF), minimising the Bayesian Information Criterion (BIC). The optimal baseline for EulerCam observations comprises of a second-order polynomial on the x-shifts of stellar PSF and a first order polynomial on time and air mass. In addition, we simultaneously fit the light curves with a Gaussian Process (GP), using a Matérn 3/2 kernel, to account for the correlated noise at short timescales. The full set of priors and posteriors from our analysis are listed in Table \ref{Table:CONAN prior and posterior distribution parameters}. From our analysis, we do not refine the system parameters reported by \cite{Lendl+2020}. Even though simultaneous photometry does not always improve the system parameters, it still strengthens the reliability of the overall analysis by offering additional context and cross-validation. In addition, the derived system parameters are consistent with \cite{Lendl+2020}.

\section{Stellar properties} 
\label{sec:Stellar properties of WASP-189}
The most irradiated giant exoplanets are found around early-type stars, typically A- and F-type stars \citep[e.g. KELT-9b, TOI-2109b, WASP-189b, and WASP-33b,][]{Gaudi+2017, Wong+2021, Anderson+2018, A.CollierCameron+2010}. These stars are hotter than the Sun and their photospheres exhibit fewer spectroscopic features broadened by their fast rotation speed. Indeed, WASP-189b orbits a catalogued fast rotating A-type star, HD~133112 (HR~5599) (see Table \ref{Table1:StarPlanetParameters} for the main system parameters).  
Generally, A-type stars are not favourable targets for the detection of exoplanets in radial velocities because they exhibit few, rotationally broadened spectral lines \citep[requiring specific approaches or tools, e.g.][]{Galland+2005}. Despite the expected small signal induced by a close-in, planetary origin companion orbiting a line-poor and fast rotating star, \cite{A.CollierCameron+2010} confirmed for the first time a gas giant planet transiting the rapidly rotating, A-type, main-sequence star HD 15082 (also known as WASP-33). 
HD~133112 was first classified as an A5m star by \citet{Cowley1968} and listed with a `doubtful' A4m classification in the catalogue of chemically peculiar stars \citep{RensonAndManfroid2009}. In the discovery paper for WASP-189b \citep{Anderson+2018}, the Am classification was ruled out based on the strength of the \ion{Ca}{ii} K line in their HARPS spectra. 
\citet{Lendl+2020} re-analysed the stellar properties using archival HARPS spectra as well as synthetic models. They derived stellar abundances typical of Am stars and measured a stellar rotational velocity of $v \sin{i_\star} = 93.1~\pm~1.7$~km~s$^{-1}$. 
They concluded from this that the stellar inclination should be close to 90\degr. These authors also found clear evidence for gravity darkening of the stellar photosphere. \citet{Saffe+2021,Saffe+2022} performed a new analysis of the chemical abundances of the star and challenged its Am classification. 
Both \cite{Prinoth+2022} and \cite{Prinoth+2023} classify WASP-189 as A-type star, without further investigation about the peculiar subclass.

\section{Methods}
\label{sec:Methods}
\label{subsec:Data analysis}

All spectra are reduced by the automated HARPS and NIRPS Data Reduction Softwares \citep[DRS version 3.5 and 3.2.0, respectively;][]{Pepe+2021}, which yield the one- and two-dimensional stellar spectra (\texttt{S1D} and \texttt{S2D}) corrected for the Barycentric Earth Radial Velocity (BERV). We describe below the additional data correction and analysis steps that are still needed to obtain the transit transmission spectra of the planetary atmosphere. The starting point of this analysis are the HARPS \texttt{S1D\char`_\char`A} and NIRPS \texttt{S1D\char`_TELL\char`_CORR\char`_\char`A}\footnote{Telluric corrected.} FITS files (Subsection \ref{subsubsec:Cleaning steps})
, which are the one-dimensional spectra produced by the stitching of the deblazed 71 and 72 echelle spectral orders of NIRPS and HARPS, respectively. After the flux extraction, the spectra have $\mathrm{N_{order}}$ $\times$ 4\,096. Indeed, the detector is provided with 4\,096 $\times$ 4\,096 pixels containing all the spectral orders. For each night of observations, we thus work on a temporal series of spectra $f(\lambda,t) = \left. f(\lambda,t) \right|_\mathrm{BERV}$ (i.e. provided in the barycentric frame of reference).

\subsection{Correction for telluric features}
\label{subsubsec:Cleaning steps}
To mitigate the influence of Earth's atmosphere on HARPS stellar spectra, we use \texttt{molecfit} \citep{Smette+2015, Kausch+2015}, as the 3.5 version of the data reduction pipeline does not include corrections for telluric features. \texttt{molecfit} corrects for telluric lines (H$_2$O, O$_2$) impacting the optical band covered by HARPS. For this purpose, we follow the methodology outlined in \cite{Allart+2017}. The \texttt{molecfit} workflow takes our one-dimensional spectra $f(\lambda,t)$ as input and fits them using a synthetic transmission spectrum generated by a line-by-line radiative transfer code, which includes the list of telluric absorption lines within the relevant wavelength ranges. 

Infrared spectra are significantly contaminated by hydroxyl (OH) telluric emission lines originating in Earth's mesosphere from 0.61 $\mu$m to 2.62 $\mu$m \citep{Rousselot+2000}, as well as telluric absorption lines (H$_2$O, O$_2$, CH$_4$, CO$_2$) heavily polluting \textit{J} and \textit{H} photometric bands. Moreover, deep water bands make spectral orders 104 and 105 unusable (between 1376.5 and 1397.3 nm). The NIRPS data reduction pipeline adapted from the ESPRESSO pipeline \citep{Pepe+2021} takes care of correcting the telluric OH airglow in the sky-fibre channel, and the telluric absorption lines to produce \texttt{S1D\char`_TELL\char`_CORR\char`_\char`A} and \texttt{S2D\char`_TELL\char`_CORR\char`_\char`A} FITS files\footnote{It is worth mentioning a second reduction pipeline, APERO~\citep[A PipelinE to Reduce Observations,][]{cook_apero_2022} to process near-infrared NIRPS data. The spectra reduced by APERO are fully consistent with those produced by the ESPRESSO pipeline.}. For near-infrared NIRPS spectra, we use the automated correction of telluric features developed by \cite{allart_automatic_2022} and implemented in the data reduction pipeline (version 3.2.0). We choose this over using \texttt{molecfit} as a telluric correction approach as we found that, while both perform similarly (see Appendix~\ref{Appendix-Sec:TelluricCorrectionsComparison} for further details), the DRS correction can correct for both absorbing and emitting telluric features up to three times the continuum noise according to cross-correlation computations performed on stellar data using a built telluric mask. Indeed, the derived root-mean-square of the correlation peak is in agreement with what was obtained and shown for ESPRESSO in the visible wavelength range by \cite{allart_automatic_2022}.
To address the presence of telluric residuals, especially prevalent in the infrared domain around 1.4 $\mu$m and 1.8 $\mu$m where H$_{2}$O deep telluric bands dominate, we perform an iterative sigma-clipping process over the telluric profile used to fit the scientific stellar spectra.
The threshold optimisation has also been done by visual inspection of the post-correction spectra and cross-correlation maps (obtained as described in Section \ref{subsubsec:Cross-correlation functions of other atoms and molecules}). We adopt a threshold of 50\% to filter out persistent telluric absorption lines in the optical range of wavelengths. HARPS optical spectra are little impacted by deep telluric lines and, consequently, less affected by telluric residuals needy to be removed. For near-infrared wavelengths, where telluric lines can saturate (especially those belonging to water bands), and residuals are more likely to remain after the telluric correction, we increase the threshold to 90\%, resulting in 36.7\% of the NIRPS data being masked. 
After correcting for the Earth's atmosphere, we identify and mask any spurious spikes remaining in the data. 

\subsection{Normalisation of the spectra and outlier correction}

We normalise the telluric-corrected spectra by dividing them by the median flux of each spectrum, $\tilde{f}(\lambda,t) = f(\lambda,t)/f(\left<\lambda\right>,t)$. 
We also correct for cosmic rays which may randomly hit the detector and produce signals unrelated to the star or planet, and any other effect resulting in an outlier value that escaped the automated reduction pipeline. To do this, we calculate a median spectrum $\tilde{f}(\lambda,\left< t\right>)$ for each night and we subtract it from the normalised, telluric-corrected spectra. We apply a 5$\sigma$-clipping filter along the spectral dimension 
to identify outlying pixels and discard them from our spectra \citep{Allart+2017}.

\subsection{Transmission spectroscopy}
\label{subsubsec:Transmission spectroscopy}
Once the spectra have been cleaned and normalised, we Doppler-shift all the in-transit and out-of-transit spectra into the stellar rest frame,
\begin{equation}
\left. \tilde{f}(\lambda,t)\right|_\star = \tilde{f}\left[\lambda\left(1 + \frac{v_\mathrm{sys} + \Delta rv_\star(t)}{c}\right), t\right],
\end{equation}
where $v_\mathrm{sys}$ is the systemic velocity \citep[$v_\mathrm{sys}$ = $-$24.5 km~s$^{-1}$, ][]{Anderson+2018}, and $\Delta rv_\star(t)$ is the radial velocity variation of the star caused by the reflex motion of WASP-189b measured from the Sun barycentre rest frame,
to create the `master-in' $\mathcal{F}_\mathrm{in}(\lambda)$ and `master-out' spectra $\mathcal{F}_\mathrm{out}(\lambda)$, respectively,
\begin{eqnarray}
\left.\mathcal{F}_\mathrm{out}(\lambda)\right|_\star = \frac{1}{n_\mathrm{out}} \sum_{t \in \mathrm{out}} \tilde{f}(\lambda,t)|_\star,
\end{eqnarray}
where $n_\mathrm{out}$ is the number of spectra taken out-of-transit for nights~1 and 2 (see Table~\ref{Table2:Summary_HARPSNIRPS_transit_obs}). 

To obtain the spectral ratio $\mathfrak{R}(\lambda)$ \citep{Brown2001}, we divide the corrected, normalised in-transit spectra $\tilde{f}(\lambda, t\in\mathrm{in})$ by the master-out spectrum. Since the planet radial velocity is substantially changing during the transit, 
we then realign the calculated spectral ratios into the planetary rest frame before co-adding them \citep[e.g.][]{Wyttenbach+2015,Allart+2017,Seidel+2019,Mounzer+2022},
\begin{equation}
\mathfrak{R}(\lambda) = \sum_{t\in\mathrm{in}}\left[ {\left.\frac{\tilde{f}(\lambda,t)|_\star}{\mathcal{F}_\mathrm{out}(\lambda)|_\star}\right|_p} \right] = \sum_{t\in\mathrm{in}} \left[\mathfrak{r}(\lambda,t)|_p\right],
\end{equation}
with
\begin{equation}
\mathfrak{r}(\lambda,t)|_p = \left.\mathfrak{r}\left[\lambda \left(1+\frac{rv_p(t)}{c}\right),t\right]\right|_\star,
\end{equation}
where $\mathfrak{r}(\lambda,t)$, the `temporal spectral ratio', represents our time series of transit transmission spectra \citep[e.g.][]{Ehrenreich+2020Nature}, and $rv_p$ is the radial velocity \textit{of the planet} in the stellar rest frame (see Equation~\ref{eq:rv(t,Kp)_equation}).

\subsubsection{Removal of the Doppler shadow}
\label{subsubsec:Removal of the Doppler shadow}
\begin{figure}[!h]
   \centering
   \includegraphics[width=\columnwidth]{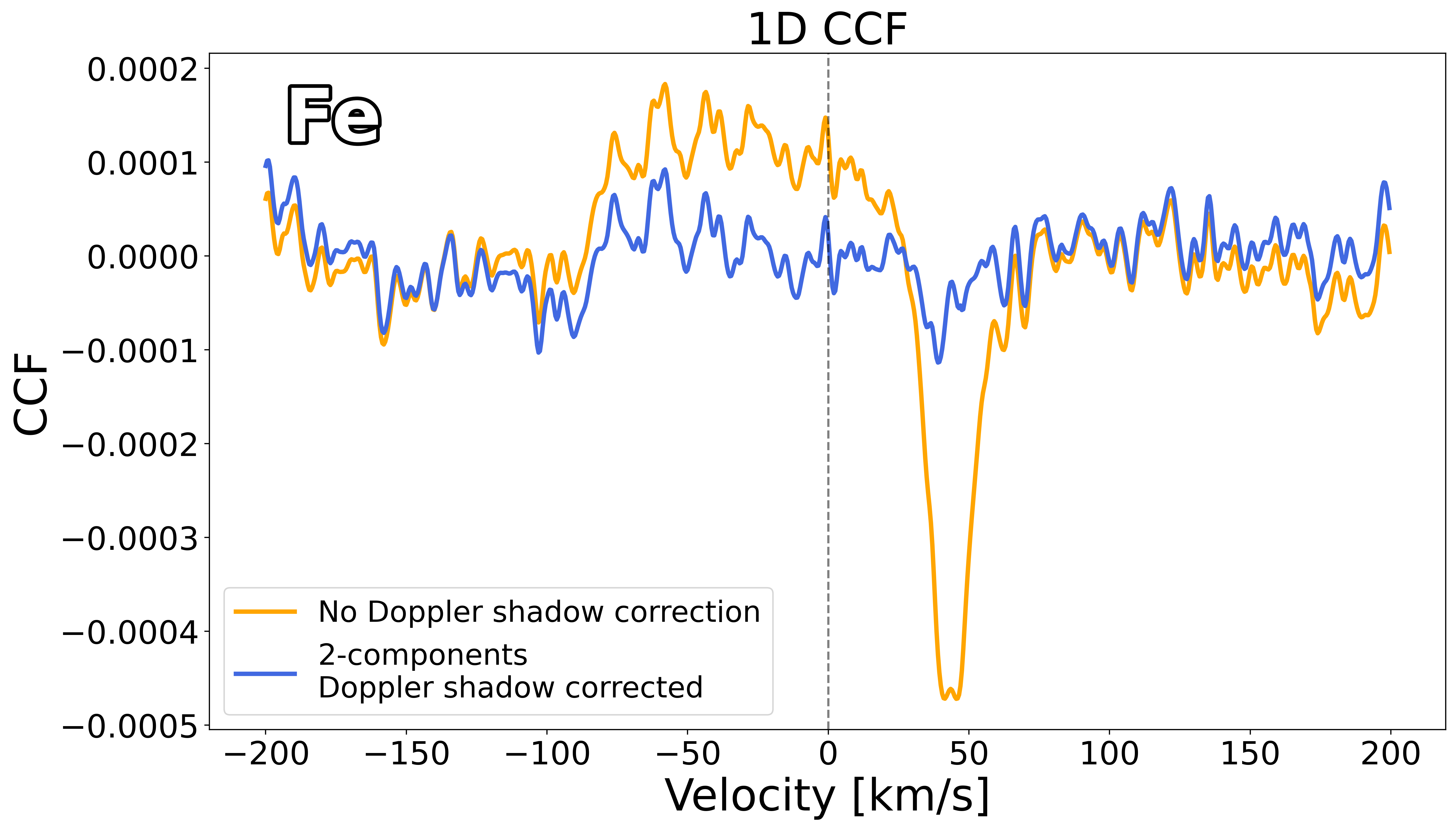}
   \caption{One-dimensional cross-correlation function based on HARPS data calculated for neutral iron before (orange) and after (blue) correcting for the stellar contamination in the form of the two-component  Rossiter-McLaughlin effect.}
   \label{Figure:DopplerShadowCorrection}
\end{figure}

\begin{figure}[!h]
   \centering
   \includegraphics[width=\columnwidth]{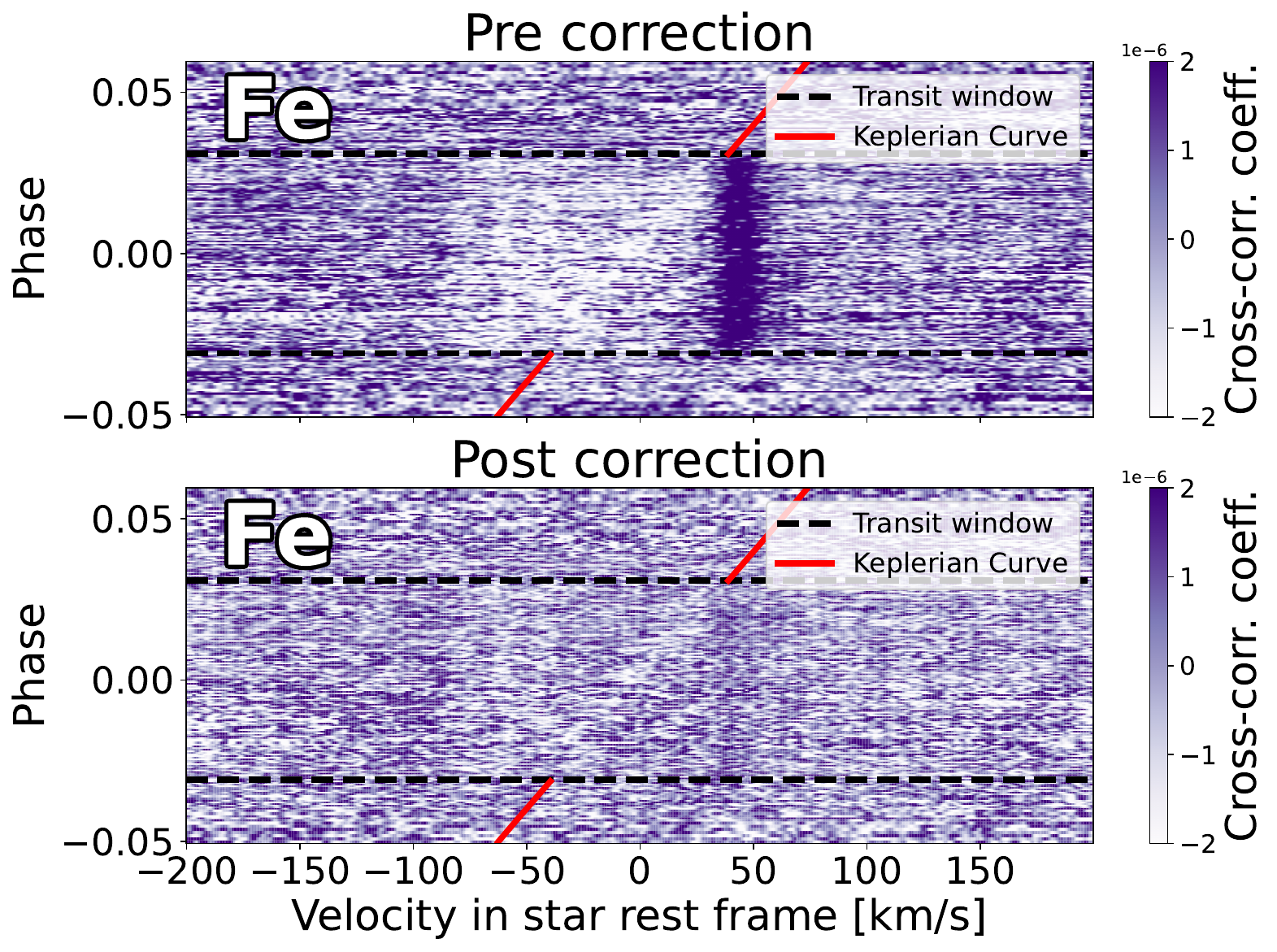}
   \caption{Cross-correlation map based on HARPS data with no correction of the Rossiter-McLaughlin effect. The top panel shows the pre-correction scenario. The thick emitting dark vertical strip around 50 km~s$^{-1}$ is the first component of the Rossiter-McLaughlin contamination, that is called Doppler shadow. The wide absorbing feature from $\sim$ $-$80 km~s$^{-1}$ up to $\sim$ 25 km~s$^{-1}$ is the second component of the stellar contamination. Bottom panel: the post-correction scenario once the best Gaussian fits are subtracted from the cross-correlation.}
   \label{Figure:PrePostDopplerShadowCorrection}
\end{figure}

While transiting the star, the planet partially obscures the rotating stellar surface, producing the well-known Rossiter-McLaughlin effect \citep{Rossiter1924,McLaughlin1924}, usually presented as a redshift followed by a blueshift in the stellar radial velocities integrated over the full stellar disc (for an aligned, prograde system). The Rossiter-McLaughlin effect distorts all spectral lines in the integrated stellar spectrum and typically creates residuals in transit 
cross-correlation time series
when it is not accounted for. 
Specifically, the distortions appear in the cross-correlation functions during the transit phases as a combination of a dark trace in emission (the `Doppler shadow') and a second negative broad component in the velocity versus\ time space (cf.\ Figure~\ref{Figure:DopplerShadowCorrection} and~\ref{Figure:PrePostDopplerShadowCorrection}).
The Doppler shadow can create absorption-like features that can be mistaken from atmospheric features \citep{Casasayas-Barris+2022}. The slope and width of the Doppler shadow depend on the projected spin-orbit angle ($\lambda$), and the projected stellar rotation velocity ($v \sin{i_\star}$), respectively. In simple cases (low $\lambda$, low $v \sin{i_\star}$), the Doppler shadow can be reasonably well fitted with a Gaussian, then subtracted to reveal the absorption signature of the transiting planetary atmosphere \citep[e.g.][]{Ehrenreich+2020Nature}. WASP-189b, however, is transiting its rapidly rotating host star ($v\sin{i_\star} = 93.1$~km~s$^{-1}$) on a strongly misaligned, almost polar orbit (true obliquity $\Psi=85.4$\degr\footnote{For WASP-189b, $\Psi$ is known thanks to the measurement of $i_\star$; it is related to $\lambda$ through the formula $\cos{\Psi}=\cos{i_\star}\cos{i} + \sin{i_\star}\sin{i}\cos{\lambda}$.}), leaving a noticeable residual structure on the maps (see Figure~\ref{Figure:PrePostDopplerShadowCorrection}) obtained from cross-correlating our transit spectra with a template model (see Sections~\ref{subsubsec:Cross-correlation functions of other atoms and molecules} and~\ref{subsubsec:Velocity-velocity maps}). 
To correct for the Rossiter-McLaughlin effect (both components) in the cross-correlation (Subsection \ref{subsubsec:Cross-correlation functions of other atoms and molecules}) and velocity-velocity maps (Subsection \ref{subsubsec:Velocity-velocity maps}), we use the Gaussian fitting approach. Step (i) is to fit the Doppler shadow with a Gaussian function for each time series as varying the phase \citep{Ehrenreich+2020Nature, Prinoth+2022}, and after to subtract the best-fit model from the cross-correlation. The residuals are the cross-correlation values which cannot be modelled and explained by the first Gaussian function, most likely linked to the second component of the Rossiter-McLaughlin effect. Therefore, step (ii) consists in fitting and subtracting a second Gaussian model to the residual data.
We set initial guesses for the parameters of the Gaussian profiles, namely for the amplitude ($\mathrm{A}$), the peak velocity ($\mu$), the width of the Gaussian ($\sigma$), and a constant value for the continuum baseline ($\mathrm{c}$). The initial guesses for both Gaussian components are reported in Table \ref{Table:GaussianPriors}. Correcting for the two-component  Rossiter-McLaughlin effect smooths the continuum level and increases the signal-to-noise ratio of the cross-correlation signal. Figure~\ref{Figure:DopplerShadowCorrection} illustrates the 1D cross-correlation function (cross-correlation function versus velocity) based on HARPS data calculated for neutral iron before and after correcting for the stellar contamination in form of the two-component  Rossiter-McLaughlin effect.

\section{Results and discussion}
\label{sec:Results and discussion}

\subsection{Signatures in the transit transmission spectrum of WASP-189b}

\subsubsection{Hydrogen and helium lines}
\begin{figure}
   \centering
   \includegraphics[trim={0.7cm 0 1.8cm 0}, clip, width=\columnwidth]{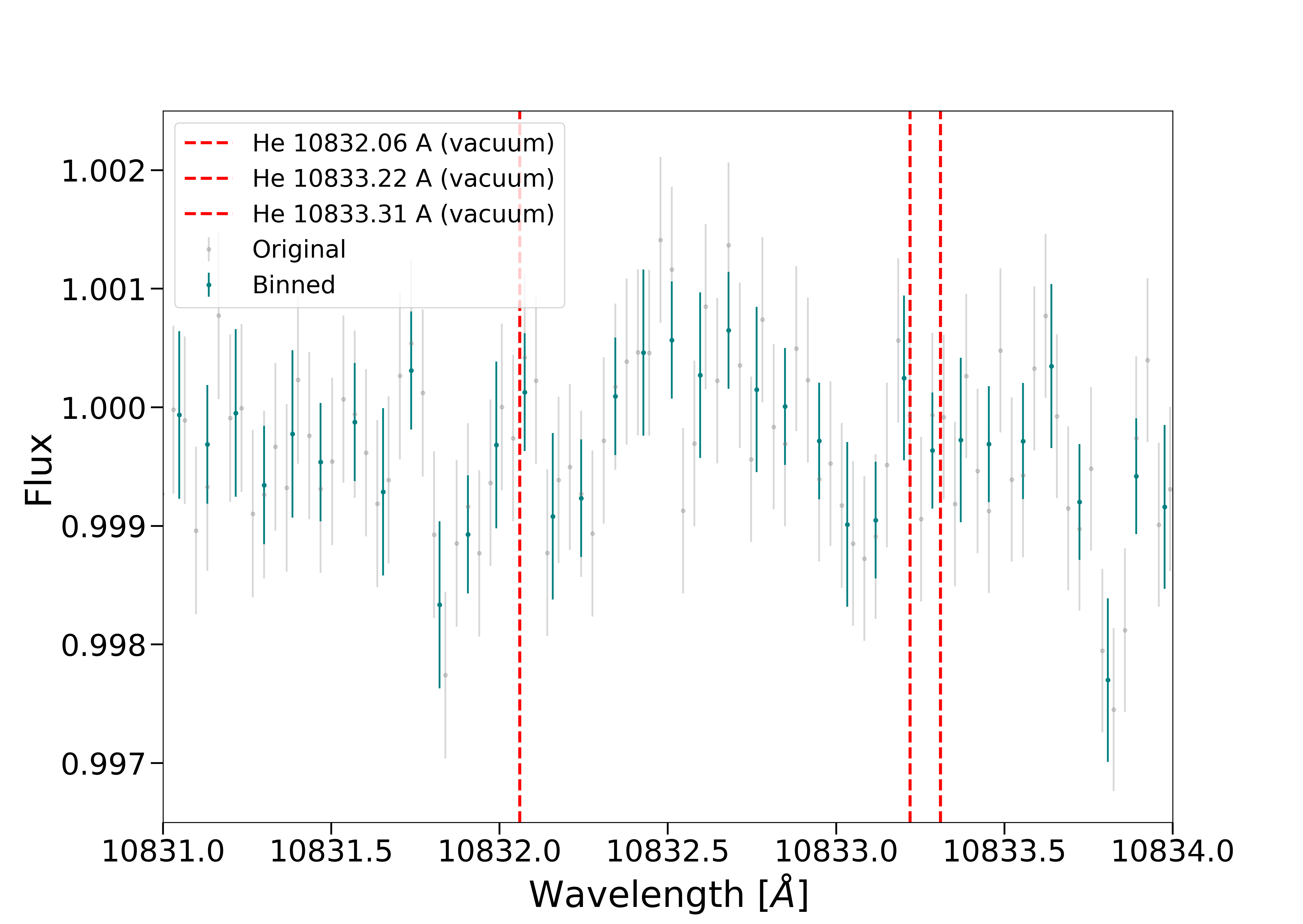}
  \caption{Transmission spectrum of WASP-189b around the metastable helium triplet (vertical dashed red lines). Grey dots with error bars represent the entire transmission spectrum at native resolution, and cyan dots with uncertainties are the binned spectrum. No trace of metastable helium triplet is found in WASP-189b from atomic line inspection of its transmission spectrum, as expected given the theory of \cite{Oklopcic&Hirata2018}.}
   \label{Figure:WASP189b_TS@HeTriplet}
\end{figure}
We perform a visual inspection of the transit transmission spectrum $\mathfrak{R}(\lambda)$ to search for signatures of the excited atomic hydrogen in the HARPS and NIRPS range, and metastable helium in the NIRPS range. Several Balmer lines of \ion{H}{i} fall within the HARPS range (H$_\alpha$ to H$_\zeta$), with H$_\alpha$ ($\lambda = 656.3$~nm), H$_\beta$ ($\lambda = 486.1$~nm), H$_\gamma$ ($\lambda = 434.0$~nm), and H$_\delta$ ($\lambda = 410.2$~nm) having been previously detected in the atmosphere of the ultra-hot gas giant KELT-9b \citep{Yan+2018,Wyttenbach+2020}. H$_\alpha$ (and possibly H$_\beta$) absorption signal has been detected in several other ultra-hot or bloated gas giants \citep[e.g.][]{Borsa+2021,Seidel+2023}. The NIRPS range covers lines from the Paschen series (Pa$_\beta$ $\lambda = 1\,282$~nm to Pa$_\epsilon$ $\lambda = 954.6$~nm) as well as the Brackett break (Br$_\infty$ $\lambda = 1\,458$~nm). To date, one detection of the Pa$_\beta$ line was reported in the literature \citep{SanchezLopez+2022}, also for KELT-9b.
We do not detect absorption signatures in any of the above-mentioned hydrogen lines (see Appendix~\ref{Appendix-Sec:Hydrogen lines in the transit spectrum of WASP-189b}). While \citet{Prinoth+2022,Prinoth+2023, Prinoth+2024} report on a detection of hydrogen lines from their CCF analysis of WASP-189b (H$_\alpha$ at 656.46 nm), these authors made use of eight observation epochs using not only HARPS and HARPS-N but also MAROON-X and ESPRESSO at 8-meter-class telescopes. As a result, they obtained a much higher signal-to-noise ratio in the optical than this study, which is based on two observation epochs with HARPS. 
Given that here we only utilise two HARPS transits, it is not  surprising that we do not have the sensitivity to recover the same detections as \citet{Prinoth+2022,Prinoth+2023, Prinoth+2024}.

The triplet lines of metastable helium around 1.083~$\mu$m represent another potential probe of the upper atmosphere of a gaseous planet \citep{Spake+2018,Allart+2018}. Indeed, helium atoms can accumulate in the upper atmosphere and experience strong irradiation from the star. The intense stellar radiation causes atmospheric heating and expansion. The heating can push the helium atoms into an excited state, specifically into a metastable state. Thus, the helium atoms in metastable state can strongly absorb near-infrared radiation at 1.083~$\mu$m in form of a triplet \citep{Spake+2018}. The signature of escaping helium atoms has been detected in several exoplanets and near-infrared high-resolution spectroscopy is especially well suited to achieve this \citep[e.g.][]{Allart+2018,Nortmann+2018,Salz+2018}. However, the transit signature of helium arises from the absorption of light by excited helium atoms \citep{Oklopcic+2018}.~\citet{Allart+2023} also argued that a specific irradiation environment is needed to allow for a large population of excited atoms to build. These environments include the vicinity of K dwarfs, while A-type dwarfs are thought not to be suited due to a lack of chromospherically driven extreme ultraviolet flux. It is thus not a surprise that none of the triplet lines are visible in the transit transmission spectrum of WASP-189b (see Figure~\ref{Figure:WASP189b_TS@HeTriplet}). A non-detection of the helium triplet is either a lack of helium in the atmosphere (unlikely for such a Jupiter planet), or it is not populated. In the latter, either the planet does not receive the correct amount of energy to excite the escaping helium atoms, or the helium atoms are shielded or deep enough in the atmosphere of the planet to not escape.
Meanwhile,~\cite{Allart+2025} report on a clear detection of helium around the warm Saturn WASP-69b --in orbit around a K star-- from three observation epochs, showing that NIRPS is indeed sensitive to this signature and thus, given the similar sensitivity of these data sets, our non-detection can be indeed attributed to the lack of a sufficient population of metastable helium in the exosphere of WASP-189b to produce significant excess absorption.

\subsubsection{Cross-correlation functions of other atoms and molecules}
\label{subsubsec:Cross-correlation functions of other atoms and molecules}

With the inspection of \ion{H}{i} Balmer, \ion{He}{i} and \ion{Na}{i} lines (the latter being not detected from the transmission spectrum of WASP-189b), most chemical species have spectral features too weak to allow to detect the effect of absorption features on individual lines. Instead, it is possible to search for transit signals in their cross-correlation functions $\mathcal{C}$ \citep{Snellen+2010}. This approach has proven extremely successful on ultra-hot gas giants, allowing to unveil a large amount of atoms and ions in several such objects \citep{Hoeijmakers+2018,Hoeijmakers+2019,Borsa+2021, Kesseli+2022}. \citet{Prinoth+2022,Prinoth+2023} recently adopted this approach to probe the atmosphere of WASP-189b in the optical. These authors reported a vast amount of atomic, ionic, and molecular signatures, for example TiO being present in the gaseous state in ultra-hot planets (thanks to their high temperatures), and known to be a relevant source of opacity due to its strong absorption bands in both the optical and near-infrared.

While most previous works studying the atmosphere of WASP-189b have been in the optical, the HARPS+NIRPS combination presented here covers
a broader spectral domain than previous studies \citep[including][]{Prinoth+2022,Prinoth+2023}, albeit with a lower sensitivity, since it is only based on two transits. Thus, we focus our attention on retrieving the species with the most ubiquitous spectral signatures, namely neutral iron, while still attempting to detect other species: Ti, V, Mn, Na, Mg, Ca, Cr, Ni, Y, Ba, Sc, Fe$^+$, Ti$^+$, TiO, H$_2$O, CO, and OH. 

We cross-correlate our transit spectra $\mathfrak{r}(\lambda,t)$ with a template model of a transmission spectrum ($T(v)$) computed assuming a solar composition in chemical equilibrium at a uniform temperature $T = 3\,000$~K \citep[adapted from][]{Anderson+2018}. The atmospheric model used is further described in Section \ref{subsec:Atmospheric modelling}.  
We calculate $\mathcal{C}$ at time $t$, for velocity shifts, $v$, between $\pm$ 200 km~s$^{-1}$ in steps of 0.5  km~s$^{-1}$,
   \begin{equation}
   \label{eq:CCF_function}
      \mathcal{C}(v, t) = \sum_{i=0}^{N} \mathfrak{r}_{i}(\lambda, t) T_{i}(v),
   \end{equation}
where $\mathfrak{r}_{i}(\lambda, t)$ is the time series of $N$ transmission spectra as function of time $t$ and wavelength $\lambda$, namely each pixel in the transmission spectrum at a given time $t$, and $T_{i}(v)$ are the computed model spectra Doppler shifted to a velocity $v$.
The cross-correlation thus computed generates two-dimensional cross-correlation maps for each chemical element tested.

Following the cross-correlation method, we recover a neutral iron signal in the HARPS optical data of WASP-189b combining the two-transit observations
(Figure~\ref{Figure:PrePostDopplerShadowCorrection}, ~\ref{Figure:WASP189b_HARPS_FeDetectionDopplerShadowCorrected}, and~\ref{Figure:WASP189b_HARPS_FeDetectionDopplerShadowCorrected_Blueshift_DELTAkp}). 
The excess absorption falls at a velocity (between $-50$ and +50~km~s$^{-1}$ in the cross-correlation trail map) consistent with numerous previous works \citep{Stangret+2022, Prinoth+2022, Prinoth+2023} and matches the expected amplitude based on injection-recovery tests (Section \ref{subsubsec:Injection-recovery results}).  
In contrast, we do not detect a neutral iron signature in the near-infrared NIRPS data\footnote{The absence of neutral iron in the near-infrared NIRPS transmission spectrum has been confirmed through analysis with both ESPRESSO and APERO-reduced transmission spectra.} (also when combining the two transits; see Figure \ref{Figure:WASP189b_NIRPS_FeNonDetection}). 

\subsubsection{Velocity-velocity maps}
\label{subsubsec:Velocity-velocity maps}
\begin{figure*}[ht!]
    \centering

    \begin{minipage}{0.48\hsize}  
        \centering
        \textsf{\textbf{\large HARPS - 2 transits}}  
        \vspace{0.01em}  
        \includegraphics[width=\linewidth]{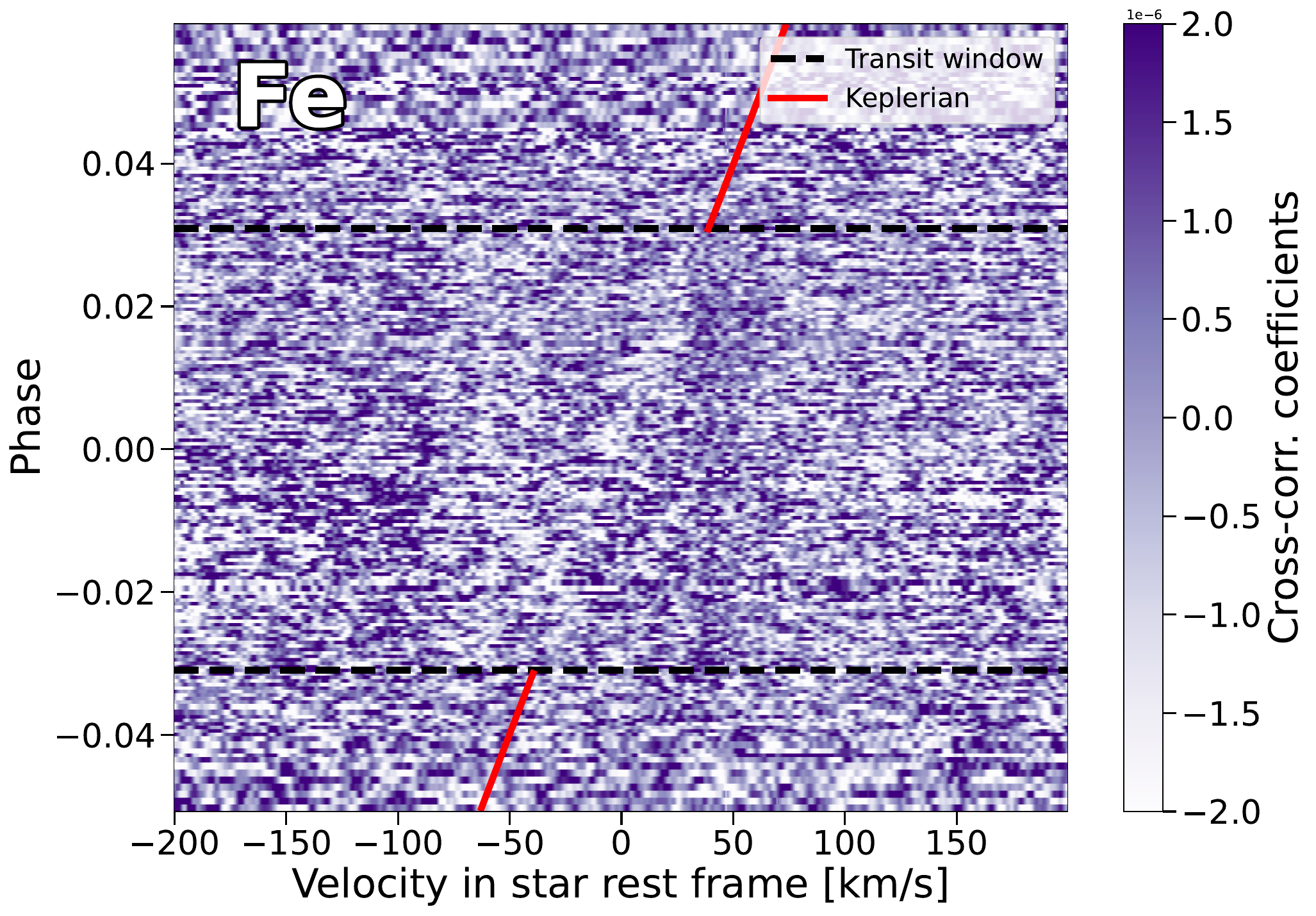}
    \end{minipage}
    \hspace{0.1em}  
    \begin{minipage}{0.48\hsize}  
        \centering
        \textsf{\textbf{\large HARPS - 2 transits}}  
        \vspace{0.01em}  
        \includegraphics[width=\linewidth]{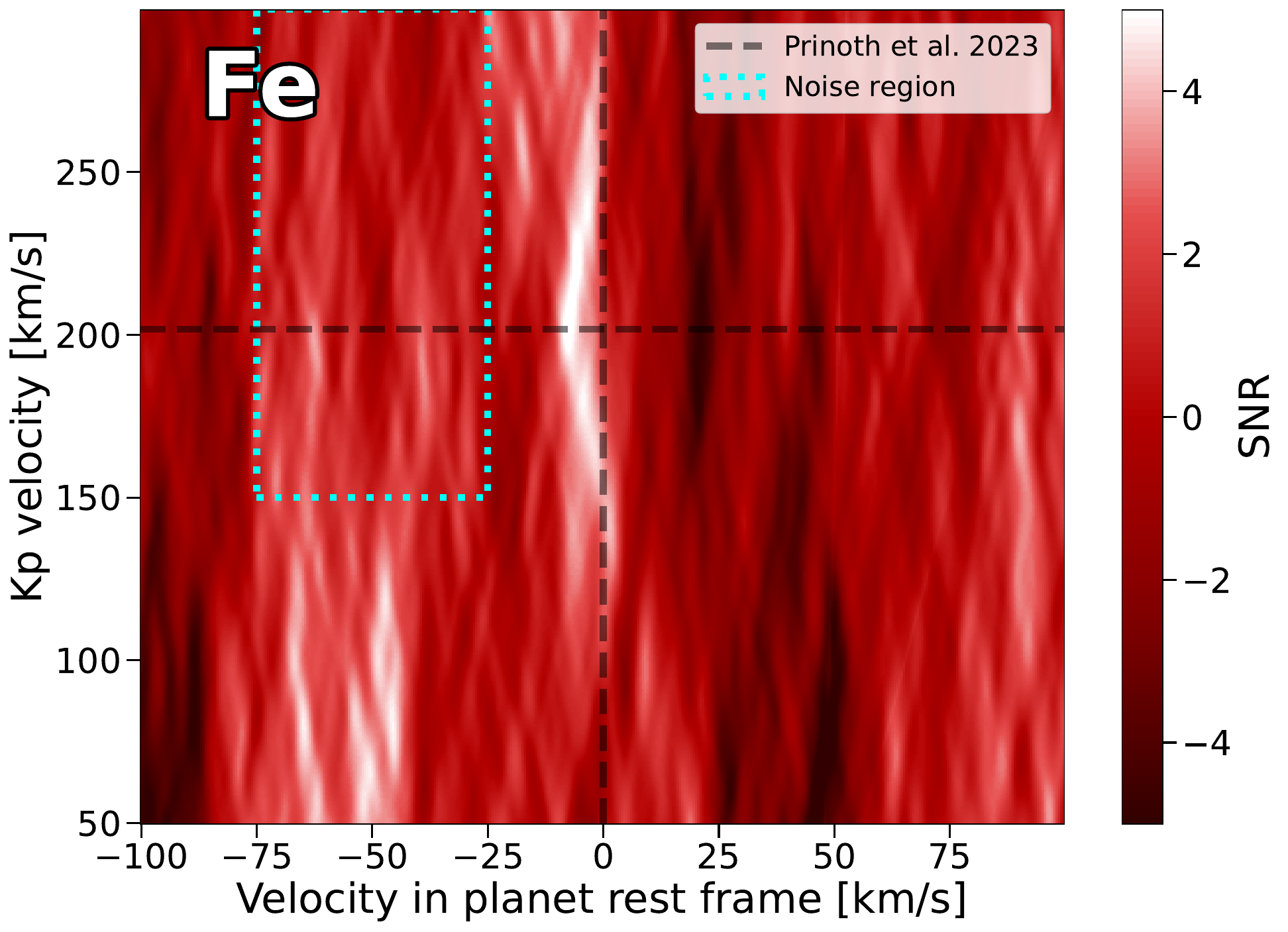}
    \end{minipage}

    \caption{Neutral Fe cross-correlation from two transits of WASP-189b observed with the HARPS spectrograph. Left panel: Cross-correlation time series corrected for the Rossiter-McLauglin contribution. The planetary trail follows the Keplerian motion in red, and it is bound within the transit window (first and fourth contact transit phases indicated by horizontal black dashed line). Right panel: $K_\mathrm{p} - v_\mathrm{sys}$ diagram. The black dashed lines denote the expected $v_\mathrm{sys}$ and $K_\mathrm{p}$ from \cite{Prinoth+2023}. The region enclosed in the cyan dotted box is used for the calculation of the S/N used throughout the entire map. Features around +50 and $-$50 km/s are residuals of the Rossiter-McLaughlin correction fitting and not due statistical fluctuations (Section \ref{subsubsec:Removal of the Doppler shadow}).}
    \label{Figure:WASP189b_HARPS_FeDetectionDopplerShadowCorrected}
\end{figure*}

\begin{figure*}[ht!]
\centering
\begin{minipage}{0.44\hsize}  
        \centering
        
        \textsf{\textbf{\large HARPS - 2 transits}}  
        
    \includegraphics[width=\columnwidth]{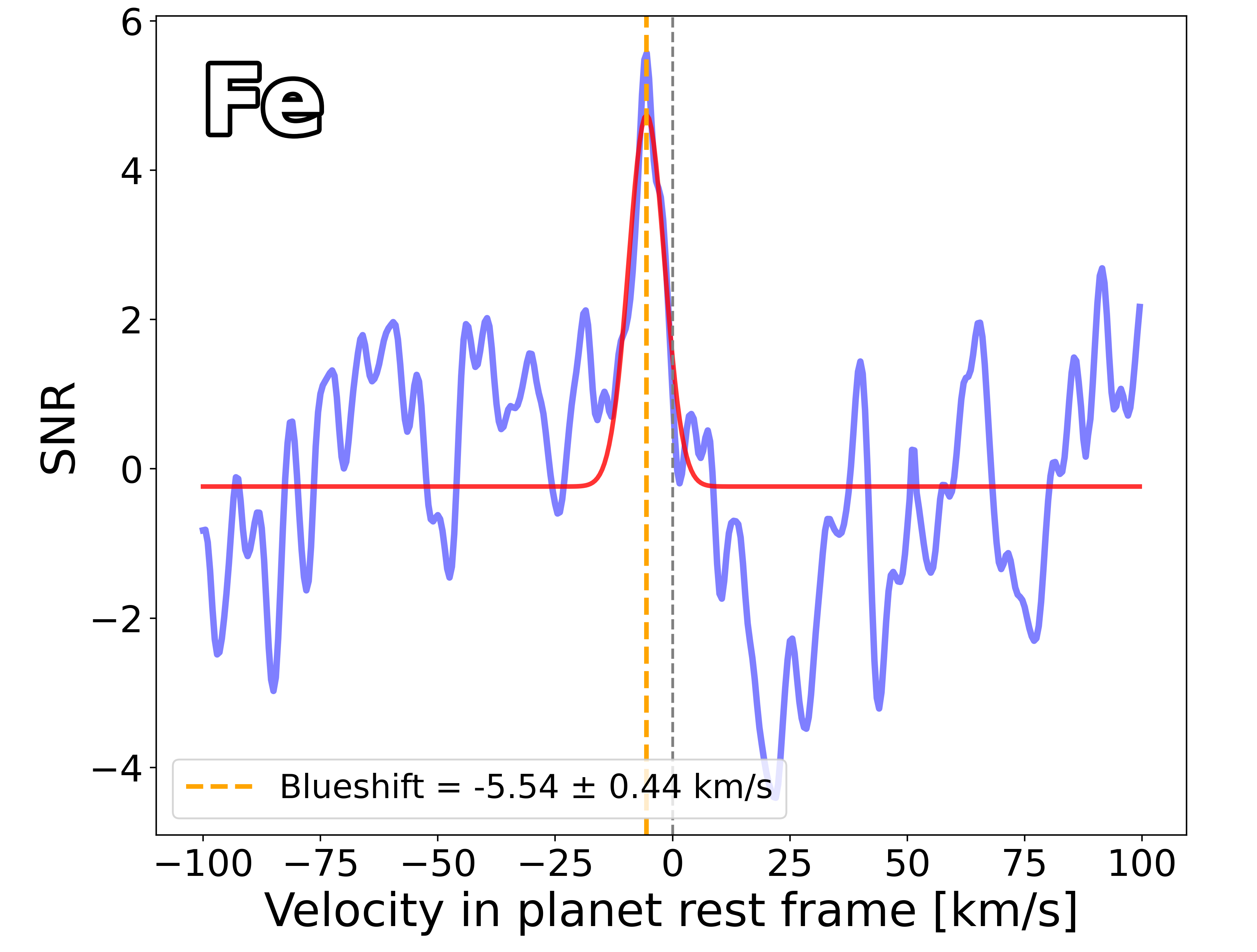}
    \label{Figure:WASP189b_HARPS_FeDetectionDopplerShadowCorrected_DELTAkp}
    \end{minipage}
    \hspace{0.1em}
\begin{minipage}{0.48\hsize}  
        \centering
        \textsf{\textbf{\large HARPS - 2 transits}}  
        \vspace{0.5em}
    \includegraphics[width=\linewidth]{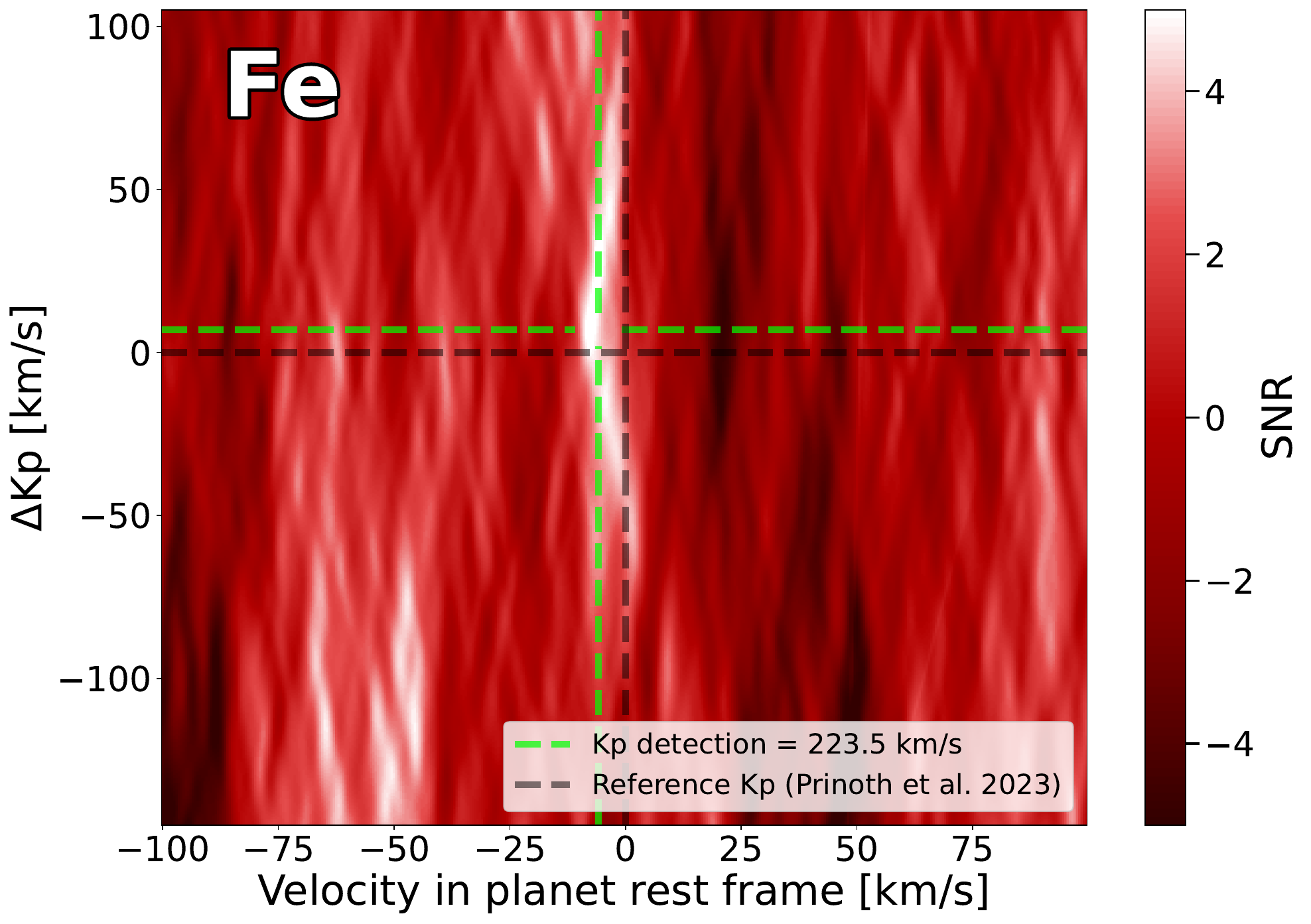}
    \label{Figure:WASP189b_HARPS_FeDetectionDopplerShadowCorrected_Blueshift}
    \end{minipage}
        \caption{Left panel: Maximum S/N versus velocities at the measured $K_\mathrm{p}$ (blue curve). In red, the best-fit Gaussian to the Fe cross-correlation peak, whose centroid results blueshifted by $-$5.54$~\pm~$0.44 km/s with respect to zero km/s. The positive peaks around $\sim$40 km~s$^{-1}$, $\sim -$60 km~s$^{-1}$, and the negative peak at $\sim -$20 km~s$^{-1}$ are artefacts of the Rossiter-McLaughlin correction and not statistical fluctuations. Right panel: $\Delta K_\mathrm{p} - v_\mathrm{sys}$ diagram. The $\Delta K_\mathrm{p}$ is calculated as $\Delta K_\mathrm{p}$ = $K_\mathrm{p} - K_\mathrm{p, measured}$ \citep{Wardenier+2023}. The black dashed lines denote the expected $v_\mathrm{sys}$ and $K_\mathrm{p}$ from \cite{Prinoth+2023} (i.e. $K_\mathrm{p}$ = 200.7 km~s$^{-1}$). Features around +50 and $-$50 km/s are residuals of the Rossiter-McLaughlin correction fitting and not due statistical fluctuations (Section \ref{subsubsec:Removal of the Doppler shadow}).}
     \label{Figure:WASP189b_HARPS_FeDetectionDopplerShadowCorrected_Blueshift_DELTAkp}   
\end{figure*}

\begin{figure*}[h!]
    \centering
    \begin{minipage}{0.48\hsize}  
        \centering
        \textsf{\textbf{\large NIRPS - 2 transits}}  
        \vspace{0.01em}  
        \includegraphics[width=\linewidth]{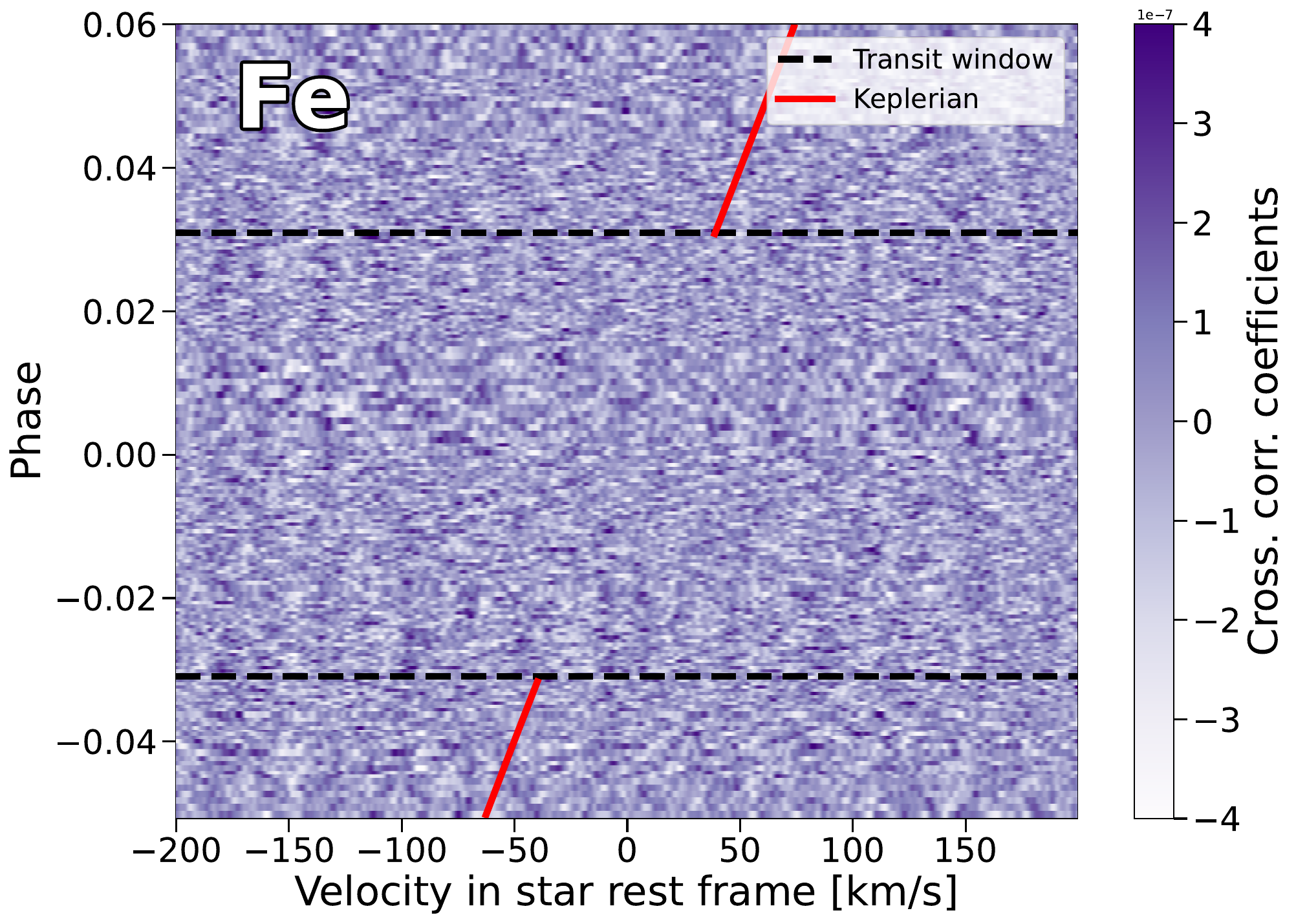}
    \end{minipage}
    \hspace{0.1em}  
    \begin{minipage}{0.48\hsize}  
        \centering
        \textsf{\textbf{\large NIRPS - 2 transits}}  
        \vspace{0.01em}  
        \includegraphics[width=\linewidth]{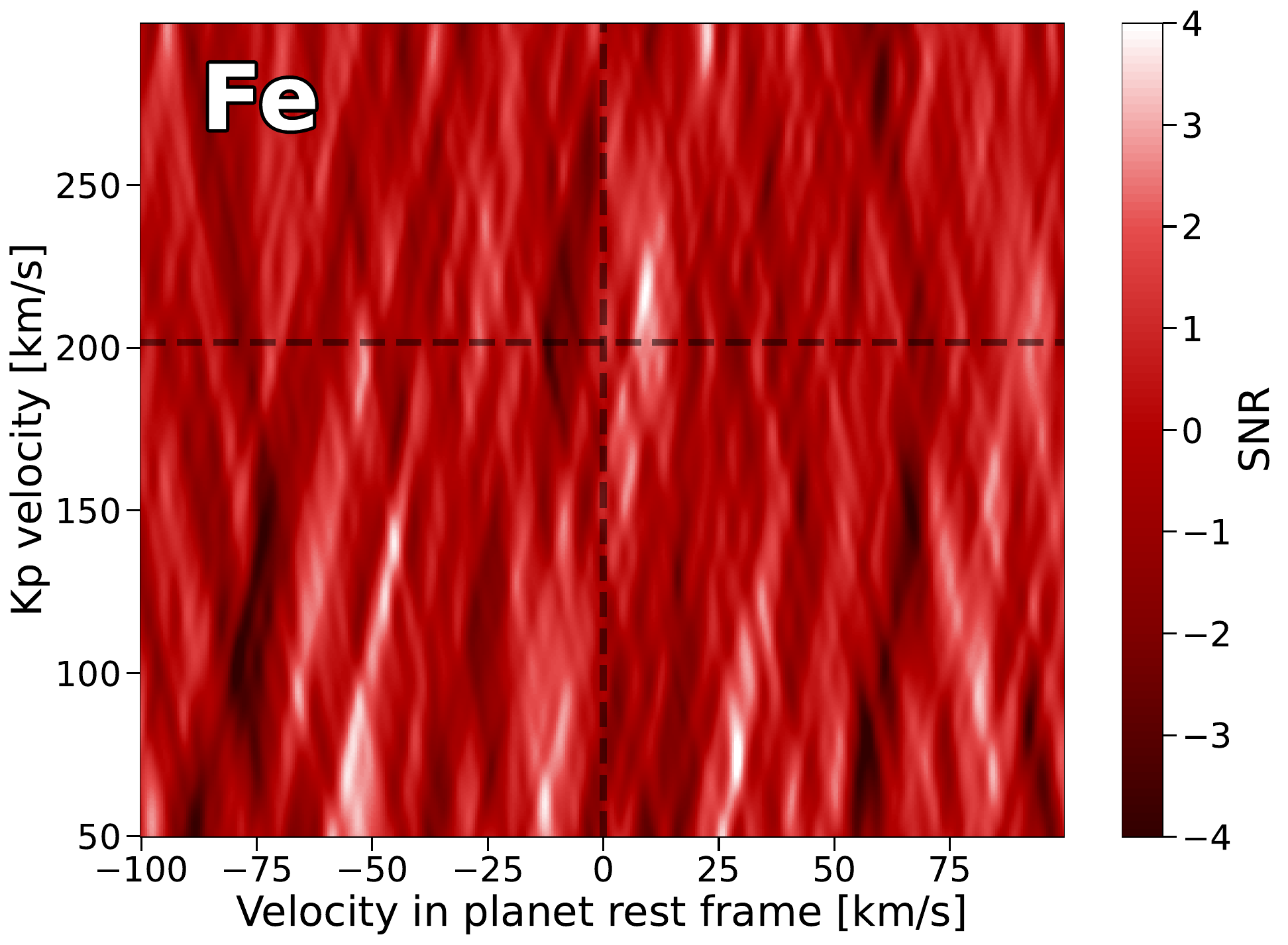}
    \end{minipage}

    \caption{Cross-correlation trail map (left panel) and $K_\mathrm{p} - v_\mathrm{sys}$ map (right panel) of NIRPS data, obtained with the reduction pipeline described in Subsection \ref{subsec:Data analysis}. Black dashed lines (right panel) denote the expected $K_\mathrm{p}$ determined from \cite{Prinoth+2023}, and $v_\mathrm{sys}$=0 km~s$^{-1}$. The colourmaps show a non-detection of Fe from two transits of WASP-189b observed with the NIRPS spectrograph. The observation gap occurred during the first transit on the night of 2023-04-24 is visible as a banded structure of higher scatter between phases 0.0 and 0.02 (left panel). 
    }
    \label{Figure:WASP189b_NIRPS_FeNonDetection}
\end{figure*}

The two-dimensional cross-correlation time series are transformed into velocity-velocity diagrams \citep[so called $K_\mathrm{p} - v_\mathrm{sys}$ maps, ][]{Brogi+2012} by phase-folding the planetary signal for different orbital configurations. The maximum S/N is obtained when the cross-correlation functions are co-added along the trail corresponding to the motion of the planet in the velocity space. Each trail is given by the Keplerian equation:
    \begin{equation}
    \label{eq:rv(t,Kp)_equation}
    rv(t, K_{p}) = K_{p} \sin \left( 2 \pi \phi(t) \right) + v_\mathrm{sys} ,
    \end{equation}
where $K_{p}$ is the semi-amplitude of the radial velocity of the planet; $\phi(t)$ is the planetary orbital phase, and $v_\mathrm{sys} = \Delta v_\mathrm{sys}$ is the systemic velocity (here set at zero km~s$^{-1}$ because spectra were already shifted to the stellar rest frame before cross-correlating). The systemic velocity is shifted 
as varying the $K_\mathrm{p}$ values within the range from 50 km~s$^{-1}$ to 300 km~s$^{-1}$ in steps of 0.5 km~s$^{-1}$. We choose a wide range for the orbital radial velocity of the planet around the expected velocity \citep[$K_\mathrm{p}$ = 200.7 km~s$^{-1}$, ][]{Prinoth+2023} following the method described in \cite{Brogi+2012}.  
Cross-correlation time series and $K_\mathrm{p} - v_\mathrm{sys}$ maps computed with Fe template for HARPS and NIRPS are shown in Figures \ref{Figure:WASP189b_HARPS_FeDetectionDopplerShadowCorrected},~\ref{Figure:WASP189b_NIRPS_FeNonDetection}, and~\ref{Figure:WASP189b_HARPS_FeDetectionDopplerShadowCorrected_Blueshift_DELTAkp}, respectively. 
To calculate the S/N, (i) we select a ``noise boxy region'' in the velocity space far away from where the planetary signal is expected, and free of artificial features and residuals from both telluric and Rossiter-McLaughlin corrections (between $K_\mathrm{p}\in$ [$-75$, $-25$] km~s$^{-1}$ and $v_\mathrm{sys}\in$ [150, 300] km~s$^{-1}$ for both HARPS and NIRPS, i.e. the cyan dashed box in Figure~\ref{Figure:WASP189b_HARPS_FeDetectionDopplerShadowCorrected}) to estimate the 1-$\sigma$ noise level; (ii) we calculate the standard deviation of this noise box region, and eventually (iii) we compute the ratio between in-transit cross-correlation values summed up and the noise estimated according to point i) and ii).
When computing cross-correlations from NIRPS spectra, the cross-correlation maps exhibit a non-negligible continuum, unrelated to any planetary signal. This is likely a consequence of the method used to compute the cross-correlation function (Equation~\ref{eq:CCF_function}), which is statistically more akin to a cross-variance rather than Pearson's correlation coefficients \citep{Rodgers+1988_PearsonCoeff}. In Pearson's correlation, the data and template are mean-subtracted, effectively removing any baseline offset. Thus, we fit a spline function for each time step (namely, to each row of the cross-correlation map) to model the continuum level and subtract it. 
Instead of fitting a single global polynomial (which might poorly capture local variations or overfit the data), splines fit low-degree polynomials in each region between the knots. We set the knots at $-100$, $-50$, 0, 50, and 100 km~s$^{-1}$, and choose a third order polynomial to fit the regions.
Once the spline is fitted to each row (for each time slice), the resulting fitted values represent the "continuum" or broad underlying shape of the cross-correlation. 
This continuum is eventually subtracted from the original cross-correlation values for each velocity and time step. This effectively removes the broad trends, leaving behind the smaller, more significant cross-correlation features (such as dips from spectral lines).
%

The iron signature we significantly detect in the optical with HARPS (S/N $\sim$ 5.5 at $K_\mathrm{p,detection}$ = 223.5 km~s$^{-1}$) appears blueshifted in the $K_\mathrm{p} - v_\mathrm{sys}$ map (Figure \ref{Figure:WASP189b_HARPS_FeDetectionDopplerShadowCorrected}). Since we expect the planetary signal to be at zero-velocity in the planet's rest frame, we measure the offset of the signature by fitting a Gaussian function to the peak in the S/N versus velocity 
(Figure~\ref{Figure:WASP189b_HARPS_FeDetectionDopplerShadowCorrected_Blueshift_DELTAkp}). We find the neutral iron feature to be blueshifted by $-5.54~\pm~$0.44~km~s$^{-1}$. We compute the uncertainty associated with the velocity shift as the square root of the variance.  
In their survey of ultra-hot gas giants, \cite{Gandhi+2023} noted a net and ubiquitous blueshift of the order of 1-7 km~s$^{-1}$ in ultra-hot Jupiter atmospheres, including WASP-189b. Furthermore,~\cite{Prinoth+2022, Prinoth+2023} mentioned non-significantly but blueshifted signatures indicative of day-to-night winds and flows \citep{Seidel+2020, Seidel+2021, Seidel+2023} in the atmosphere of WASP-189b. Hence, our finding is consistent with previous literature results. Moreover, we build the $\Delta K_\mathrm{p} - v_\mathrm{sys}$ diagram in the case of the neutral iron detection with HARPS (Figure~\ref{Figure:WASP189b_HARPS_FeDetectionDopplerShadowCorrected_Blueshift_DELTAkp}). According to \cite{Wardenier+2023}, the $\Delta K_\mathrm{p}$ is calculated as $\Delta K_\mathrm{p}$ = $K_\mathrm{p} - K_\mathrm{p, measured}$~, being $K_\mathrm{p}$ = 200.7~$\pm$~4.9~km~s$^{-1}$ the expected semi-amplitude of the radial velocity of the planet \citep{Prinoth+2023}, and $K_{\mathrm{p, measured}} = K_{\star}~(M_\star/M_p)\simeq$
 194.7 $\pm$ 2.2\footnote{error bars are calculated as square root of the sum of squares: $\sigma_{K_p} = K_p \times \sqrt{\left(\frac{1}{3}\frac{\sigma_{M_\star}}{M_\star}\right)^2 +  \left(\frac{1}{3}\frac{P}{\sigma_{P}}\right)^2}
$.}~m~s$^{-1}$ the measured velocity. Values for $K_\star$, $M_\star$, and $M_p$ can be found in Table~\ref{Table1:StarPlanetParameters}. Thus, we calculate a $\Delta K_\mathrm{p} \simeq 6$~km~s$^{-1}$. In this context,~$\Delta K_\mathrm{p} - v_\mathrm{sys}$ maps help to compare with model predictions because the sign of $\Delta K_\mathrm{p}$ depends on the three dimensional distribution of a species across the atmosphere \citep{Wardenier+2023}. Indeed, \cite{Wardenier+2023} argue that $\Delta K_\mathrm{p}$ results positive in case of signals dominated by the trailing limb in the first half of the transit and by the leading limb in the second half. $\Delta K_\mathrm{p} >$ 0 might result surprising given the discussion in \cite{Wardenier+2023}, especially if we consider the opposite conclusion driven by the more significant atomic iron detection in \cite{Prinoth+2022, Prinoth+2023}. However, it is worth mentioning that the value of $\Delta K_\mathrm{p}$ we estimate is within the error bar, considering the uncertainty of $\pm$13~m~s$^{-1}$ on $K_\star$ by \cite{Anderson+2018}.

Apart from neutral Fe in the optical, we do not detect any of the chemical species aforementioned neither in HARPS, nor in NIRPS transmission spectrum of WASP-189b.  However, we stress that these non-detections are not due to these species being absent in the near-infrared from the atmosphere of WASP-189b, neither are due to a lack of sensitivity (Section~\ref{subsubsec:Injection-recovery results}). The cross-correlation trail maps and $K_\mathrm{p} - v_\mathrm{sys}$ showing NIRPS and HARPS non-detections are reported in Appendix~\ref{Appendix-Sec:NIRPS non-detections in the transit spectrum of WASP-189b} and~\ref{Appendix-Sec:HARPS non-detections in the transit spectrum of WASP-189b} together with upper limits (Table~\ref{Table:NIRPS_upper_limits},~\ref{Table:HARPS_upper_limits}). We estimate the upper limits assuming a detection threshold of 3$\times$1-$\sigma$, where 1-$\sigma$ is the continuum noise level from the boxy region used to calculate the S/N, namely the cross-correlation value from that specific region in the velocity space. It is worth specifying that the line-contrast upper limits as calculated in this work can only be obtained with the choice of cross correlation in this manuscript, as its strength directly maps to the average line strength.

\subsection{Atmospheric forward modelling}
\label{subsec:Atmospheric modelling}
We use \texttt{SCARLET} \citep{BennekeSeager2012, BennekeSeager2013, Benneke2015, Benneke+2019a, Pelletier+2021} 
to generate transmission spectra of WASP-189b. 
SCARLET models use molecular cross sections for H$_2$O~\citep{polyansky_exomol_2018}, OH~\citep{rothman_hitemp_2010}, CO~\citep{rothman_hitemp_2010, li_rovibrational_2015}, and TiO~\citep{mckemmish_exomol_2019} while atomic cross sections are from the VALD database~\citep{ryabchikova_major_2015} and continuum cross sections for H$^{-}$ (bound-free and free-free) are from \cite{Gray_2021}.  Molecular and atomic cross sections are computed using \texttt{HELIOS-K}~\citep{grimm_helios-k_2015, grimm_helios-k_2021} and chemical equilibrium abundances are calculated using \texttt{FastChem}~\citep{stock_fastchem_2018, stock_fastchem_2022}. We assume the thermal structure to be isothermal and the atmosphere to be cloud-free, which is a reasonable assumption considering the high temperature of ultra-hot gas giant atmospheres \citep{Sing+2016, Helling+2021}. We expect the iron signal to only probe the dayside of WASP-189b \citep{Wardenier+2023}. We compute all models at a spectral resolution of $250\,000$ and convolve the spectra generated by the instrumental function to match NIRPS HE mode spectral resolution of $R \sim 80\,000$ and HARPS HAM mode spectral resolution of $R \sim 120\,000$. 

For the cross-correlation, we generate a WASP-189b transmission template assuming a 1 $\times$ solar metallicity atmosphere at a temperature of 3\,000~K \citep[based on][]{Anderson+2018}, including opacity contributions from Fe, Ti, V, Mn, Mg, Ca, Cr, Ni, Y, Ba, Sc, C, Na,  Fe$^+$, Ti$^+$, TiO, H$_2$O, CO, OH, and H$^{-}$. 
Contributions for all atomic, ionic, and molecular species of interest to the transmission spectrum are reported in Figures \ref{Figure:WASP189b_SCARLET models atoms} and \ref{Figure:WASP189b_SCARLET models molecules and ions} in the Appendix \ref{Appendix-Sec:SCARLET atmospheric modelled spectra of WASP-189b}.


%

\label{subsec:Results and discussion:Cross-correlation analysis}

\subsubsection{The role of hydride}
\begin{figure}[!h]
   \centering
  \includegraphics[width=\columnwidth]{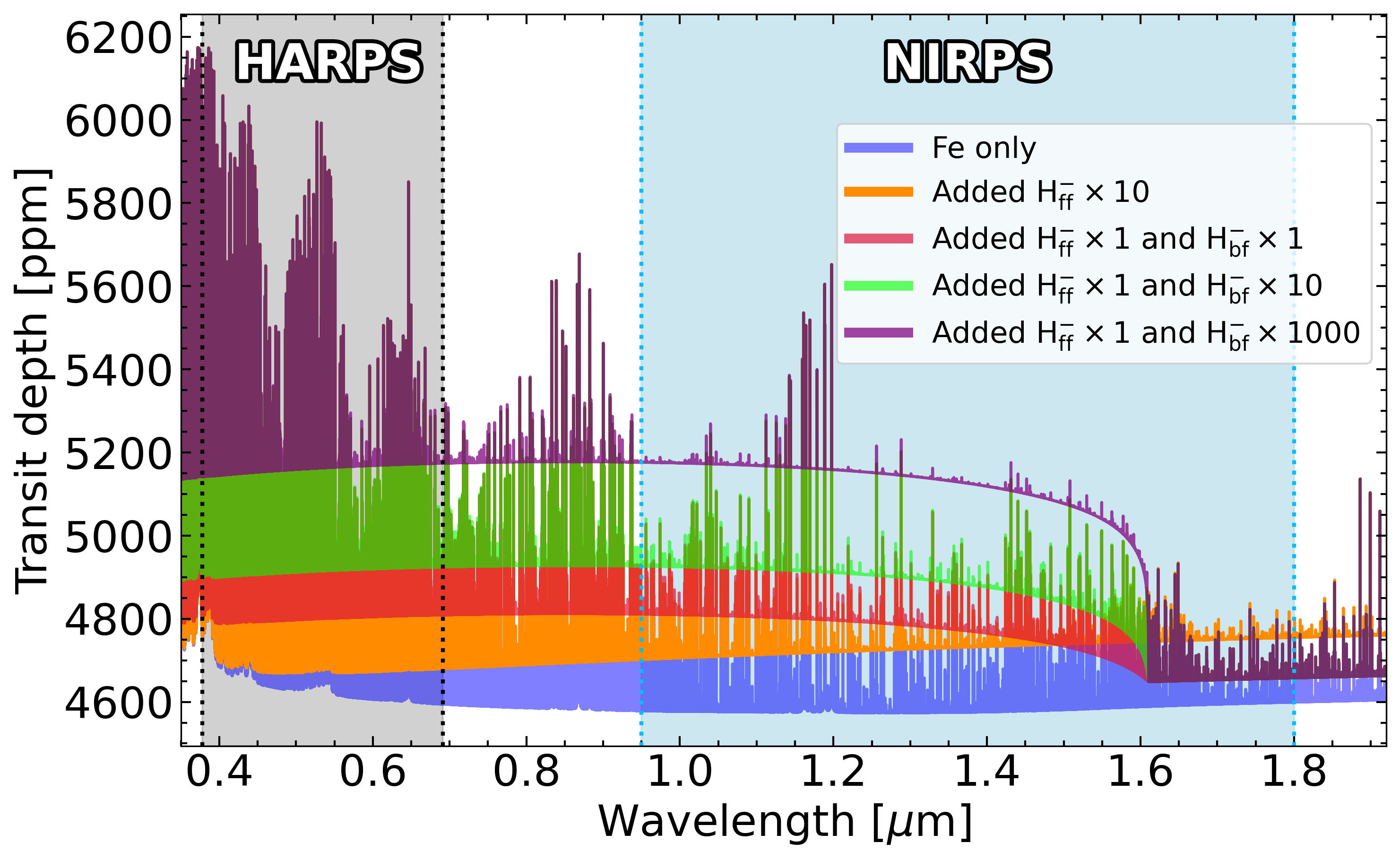}
   \caption{Iron atmospheric model spectrum as a function of wavelength including contributions given by the hydride as a source of continuum absorption. In blue the atmospheric model for Fe, with no absorber added. In orange, the template spectrum including H$^{-}$ free-free as source of continuum opacity, boosted by a factor of 10. In red, the spectrum including a solar amount of H$^{-}$ bound-free. In the green and purple spectra, the bound-free contribution is boosted by a factor of 10 and 1\,000, respectively. The purple spectrum includes contributions by H$_2$O beyond 1.6 $\mu$m. The grey and cyan coloured regions indicate HARPS and NIRPS coverage, respectively. It is worth mentioning that HARPS and NIRPS cover a similar fractional bandpass. We intentionally boost the continuum level differentiating the two main contributions to illustrate how the Fe spectrum changes accordingly.}
   \label{Figure:Fe+hydrideComparison}
\end{figure}

While ultra-hot Jupiters such as WASP-189b are likely too hot to form clouds on their dayside or terminator regions, hydrogen anions act as an important source of continuum opacity (Section \ref{sec:Introduction}). The impact of the hydride on the transmission spectra must therefore not be neglected. 
The equations to estimate the bound-free and the free-free transitions are implemented in \texttt{SCARLET} following \cite{Gray_2021}. 
Given the absorption coefficients\footnote{$a_{0}, a_{1}, a_{2},  a_{3}, a_{4}, a_{5}, a_{6}, f_{0}(\lambda), f_{1}(\lambda)$, and $f_{2}(\lambda)$ are the polynomial coefficients, and $\theta\propto 1/T$ indicates the temperature at which the cross-section is computed.} :
\begin{align}
   \alpha_{BF}(\lambda) = \left[ a_{0} + a_{1}\lambda + a_{2}\lambda^{2} + a_{3}\lambda^{3} + a_{4}\lambda^{4} + a_{5}\lambda^{5} + a_{6}\lambda^{6} \right], \\  
   \alpha_{FF}(\lambda) = \left[ 10^{f_{0}\lambda + f_{1}\lambda\log_{10}\theta + f_{2}\log_{10}^{2}\theta} \right], 
\end{align}
\cite{Gray_2021} calculates the H$^-$ bound-free and free-free cross-sections as:
\begin{align}
    \sigma_{BF}(\lambda) = \frac{10^{-17}}{10^{4}} \alpha_{BF}(H^{-}), \label{eq:cross-sections_BF}\\
    \sigma_{FF}(\lambda) = \frac{10^{-26}}{10^{3}} \alpha_{FF}(H^{-}).
    \label{eq:cross-sections_FF}
\end{align}
Figure \ref{Figure:Fe+hydrideComparison} shows how an example transmission spectrum generated by \texttt{SCARLET} is shaped by varying the abundance of hydride relative to Fe. The template spectrum includes Fe, TiO, OH, and contributions given by the hydride as a source of continuum absorption.
Figure \ref{Figure:Fe+hydrideComparison} is made with the purpose of visualising how the continuum absorption by the H$^{-}$ dampens the contrast of the neutral iron spectral lines along the whole range of wavelengths with respect to the continuum level. Specifically, the strength of iron lines depends on the relative abundance of hydride in the atmosphere of the planet at a given temperature and scale height. As the hydride abundance increases (holding the Fe abundance fixed), the contrast of Fe spectral lines decreases. 
Critically, the opacity contribution from H$^{-}$ can change the relative strength of spectral features in the optical and near-infrared (around 1.6~$\mu$m), in contrast to other continuum opacity sources (e.g. a grey cloud deck).
The simultaneous wavelength coverage of the bound-free and free-free absorption continua provides an opportunity to better characterise the role of these two hydride contributions in ultra-hot atmospheres.
While the cross-correlation method may be only weakly sensitive to continuum effects on the template, since it relies more on line positions, atmospheric retrievals are well-suited to constrain abundance ratios based on the relative contribution of different spectral features \citep[e.g.][]{brogi_retrieving_2019}.

\subsubsection{Injection-recovery results}
\label{subsubsec:Injection-recovery results}
\begin{figure*}[h!]
    \centering
    \begin{minipage}{0.48\hsize}  
        \centering
        \textsf{\textbf{\large NIRPS - injection-recovery test}}  
        \vspace{0.01em}  
        \includegraphics[width=\linewidth]{Figures/WASP189b_NIRPS_KpVsysMapFeNonDetection.pdf}
    \end{minipage}
    \hspace{0.1em}  
    \begin{minipage}{0.48\hsize}  
        \centering
        \textsf{\textbf{\large NIRPS - injection-recovery test}}  
        \vspace{0.01em}  
        \includegraphics[width=\linewidth]{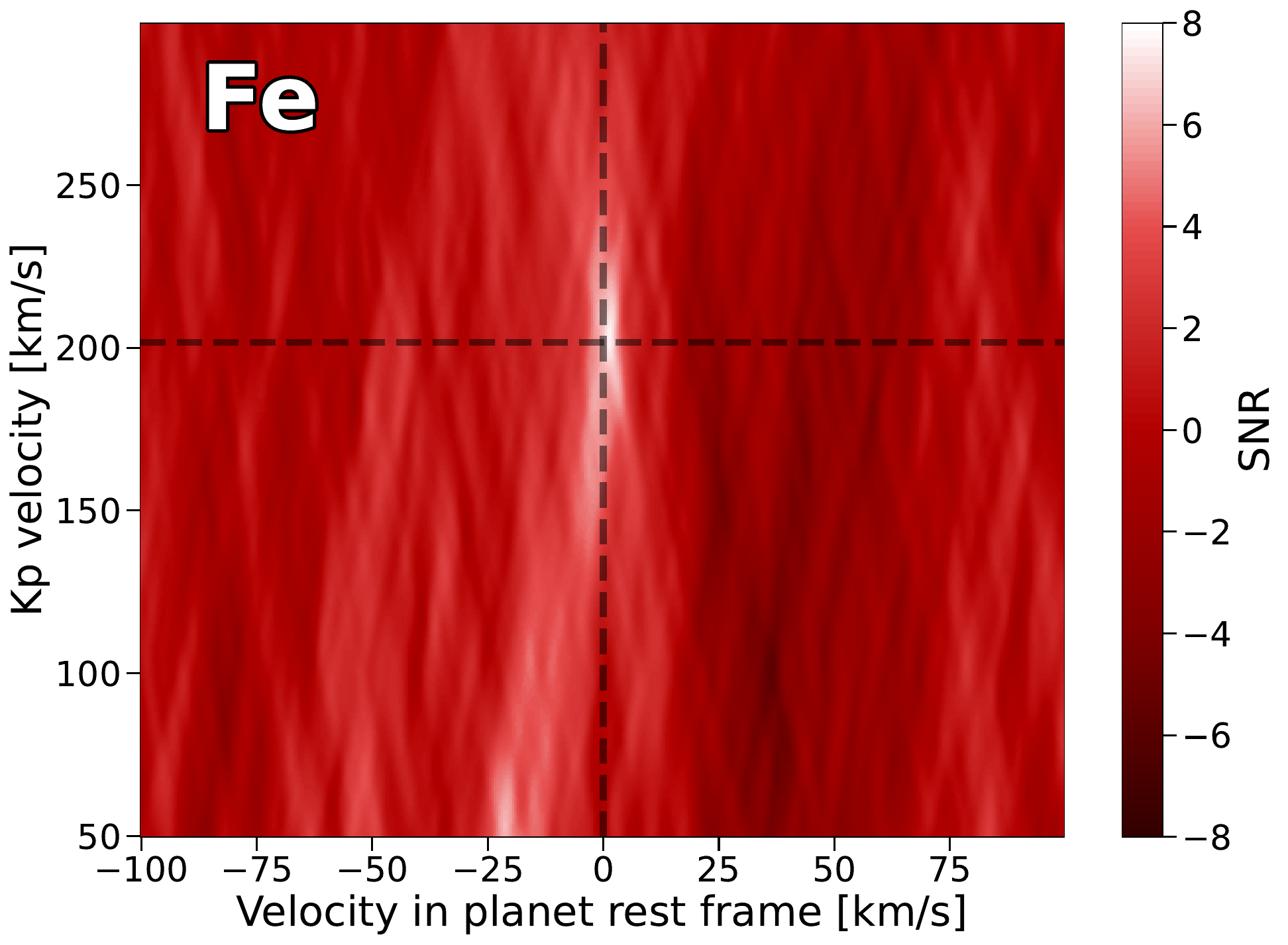}
    \end{minipage}

    \caption{$K_\mathrm{p} - v_\mathrm{sys}$ map showcasing the injection-recovery test on NIRPS data. Left panel: the injected iron atmospheric model includes contributions from the H$^-$ as a source of continuum opacity; the map shows no detection of atomic Fe in the near-infrared data of WASP-189b, consistent with non-injected results. Right panel: the injected iron atmospheric model does not include contributions from the hydride; in this case, the map shows a clear detection of neutral Fe in the near-infrared data. Black dashed lines denote the expected $v_\mathrm{sys}$ and $K_\mathrm{p}$ determined from \cite{Prinoth+2023}. 
    }
    \label{Figure:WASP189b_NIRPS_InjectionRecovery}
\end{figure*}

We use the synthetic iron atmospheric spectrum of WASP-189b 
as forward-model to characterise the underlying planetary signal. In order to quantify the expected amplitude of the atmospheric absorption, we inject the iron atmospheric model into the NIRPS spectra after they have been corrected for tellurics, but before  $\mathfrak{R}(\lambda)$ is calculated (see Subsection \ref{subsubsec:Transmission spectroscopy}). 
For cleaner results, we inject the model into non-BERV corrected spectra. The injected data are then processed following the same reduction pipeline, described in Subsection \ref{subsec:Data analysis}, as the non-injected spectra, using the known system parameters listed in Table \ref{Table1:StarPlanetParameters}. 
Following Subsection \ref{subsubsec:Cross-correlation functions of other atoms and molecules}, we compute the cross-correlation between the injected data and the modelled spectrum.
We showcase two scenarios: (1) we cross-correlate the NIRPS transmission spectrum with the template model described in Subsection~\ref{subsec:Atmospheric modelling}, considering only contributions given by atomic Fe spectral lines, and (2) we cross-correlate the data with the full template model, thus taking into account the hydride as source of continuum opacity.
The results of the injection-recovery test in Figure \ref{Figure:WASP189b_NIRPS_InjectionRecovery} demonstrate that we are able to recover the injected neutral iron-only signal at a signal-to-noise ratio of $\sim$8, 
indicating that the NIRPS data are sufficiently sensitive to detect a pure iron template. However, when contributions by H$^-$ are also included, the injection-recovery test results in a non-detection of neutral iron. 
Therefore, we argue that the signal strength from a pure Fe template does not match the observations. Other opacity sources, likely hydride, could play a role in dampening the neutral iron signal such that it is not observed in near-infrared data. 
A similar injection-recovery test, but where the template is injected in emission instead of absorption (to cancel out the real neutral Fe absorption signal from WASP-189b's atmosphere) in the HARPS data, finds that the Fe + H$^{-}$ template matches well the amplitude of the observed signal (see complementary Figure \ref{Figure:WASP189b_HARPS_negative_injection} in Appendix \ref{Appendix-Sec:HARPS_negative_injection_test}).


\subsection{Atmospheric retrievals}
\label{subsec: Atmospheric retrieval results}
   \begin{figure*}
   \centering
   \includegraphics[width=\textwidth]{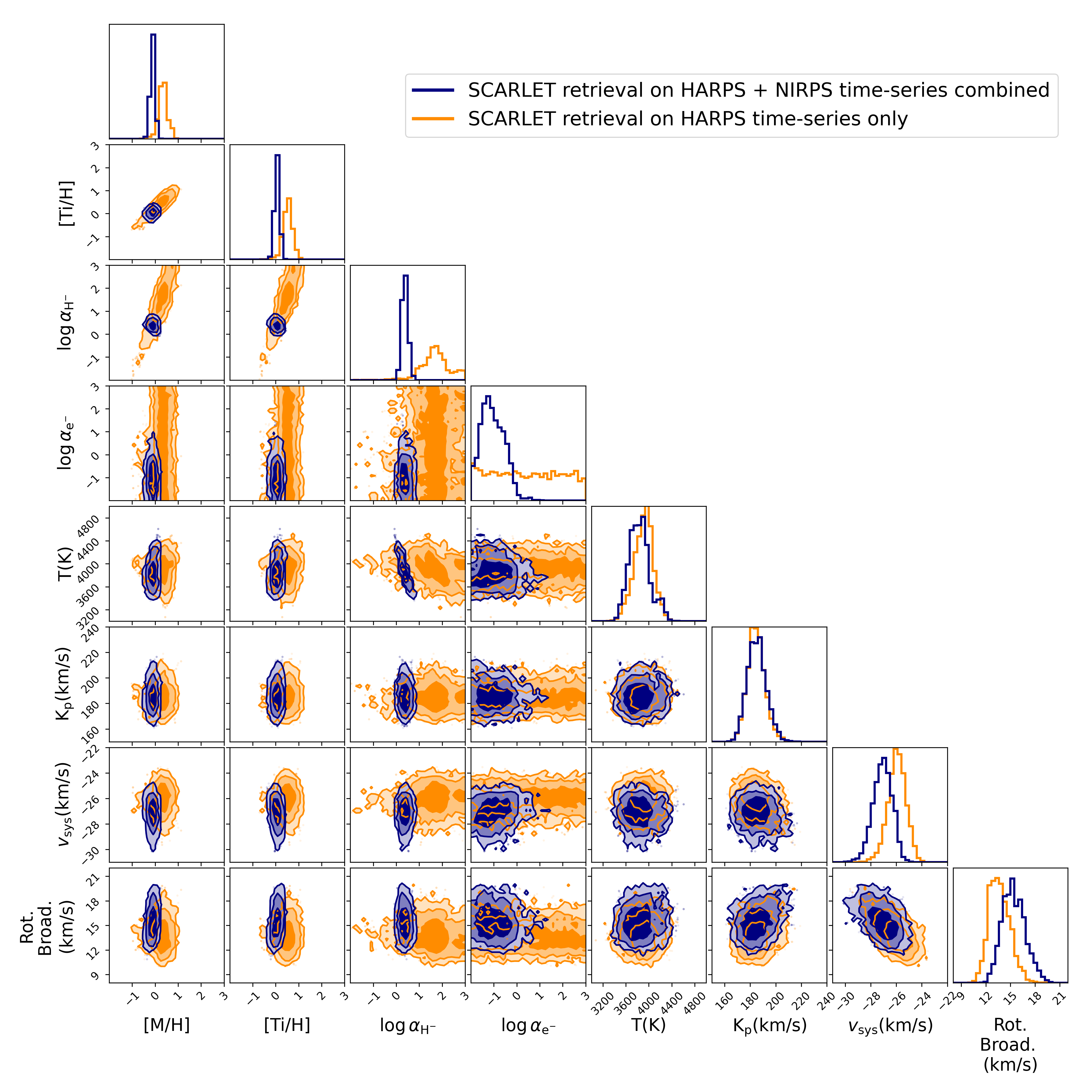}
      \caption{
      Retrieved constraints on the atmospheric and orbital properties obtained from the observed two transits of WASP-189b. The orange shows the retrieval considering only the HARPS data, while the blue shows with the inclusion of NIRPS.  Here the overall metallicity, as well as the titanium-only metallicity are well constrained to be near-solar.  Meanwhile the H$^-$ abundance is slightly above expectations from solar model in chemical equilibrium. The inclusion of the NIRPS data is particularly helpful in constraining the H$^-$ abundance, as well as setting a strict upper limit on the free electron density, which is unconstrained from the HARPS-only retrieval.      
      }\label{Figure:WASP189b_HARPSonly_Retrieval_CornerPlot}
   \end{figure*}

\begin{figure}
\centering
   \includegraphics[trim={0cm 0 0cm 0}, clip, width=\columnwidth]
   {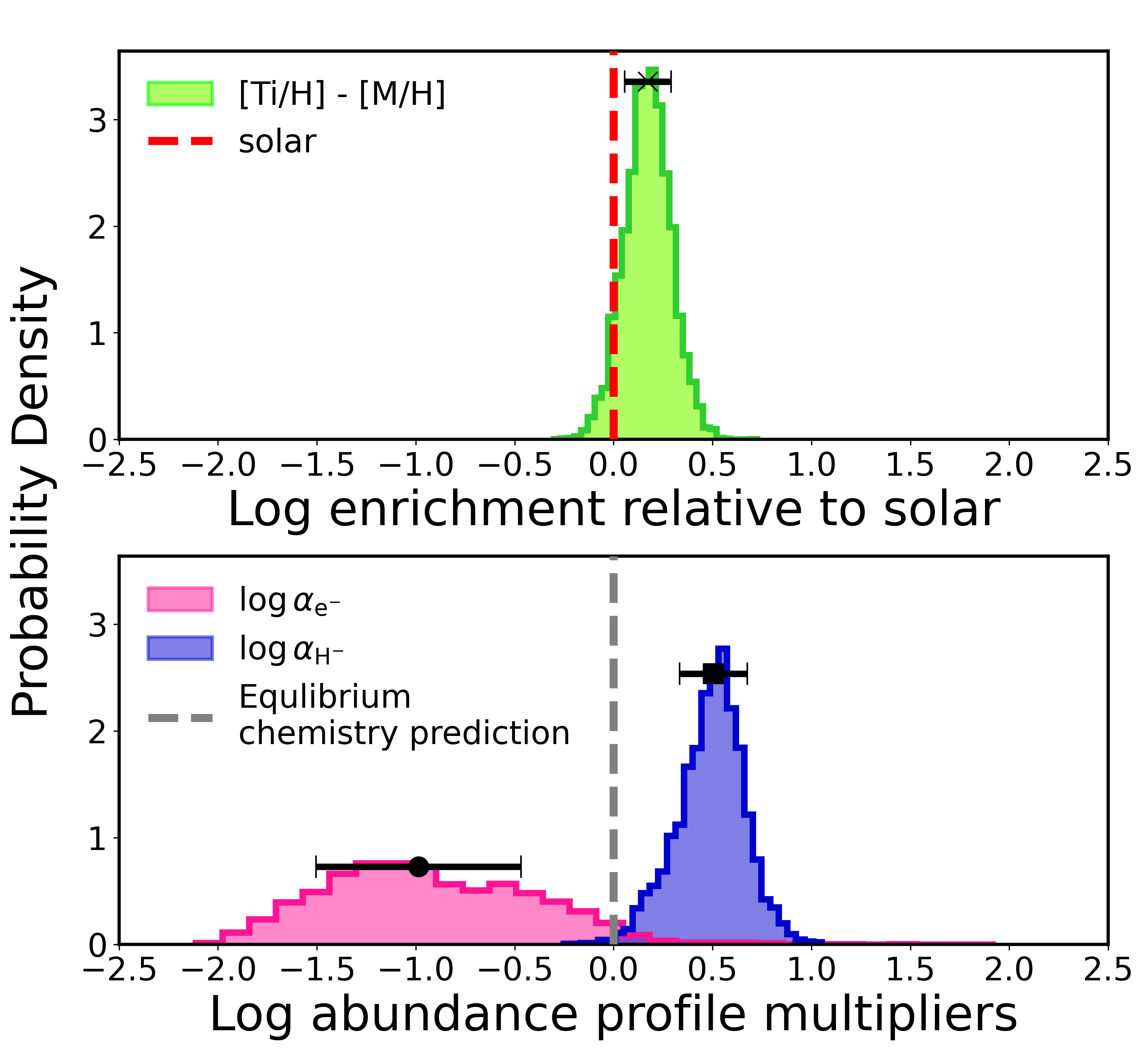}
      \caption{Ratio of the titanium metallicity relative to the overall metallicity (green distribution, cross marker) as well as the hydride (blue distribution, squared marker) and free electron (red distribution, dotted marker) abundance multipliers retrieved from the joint HARPS + NIRPS retrieval. The top panel shows the enrichment relative to solar (red dashed line) in logarithmic scale for titanium. The bottom panel shows the abundance profile for H$^-$ (bound-free) and free electrons multipliers as predicted by chemical equilibrium chemistry with \texttt{FastChem}. Here we see that the H$^-$ (bound-free) retrieved abundance is about 0.5 dex higher than that of the metals, as predicted by equilibrium chemistry (grey dashed line). This indicates that, while species such as Fe have very strong lines in the optical, others that only have weaker spectral features may be harder to detect with transmission spectroscopy on ultra-hot Jupiters like WASP-189b. The error bars, which are shown as black markers, were calculated as the standard deviation around the median value of each distribution.
      }\label{Figure:WASP189b_NIRPS+HARPS_Retrieval_H-/Fe_distribution}
\end{figure}
\begin{table}[!h]
    \centering
    \def\arraystretch{1.2}
    \caption{WASP-189b \texttt{SCARLET} atmospheric posteriors from the retrieval analysis on the HARPS and NIRPS time series combined.}
    \begin{tabularx}{\linewidth}{c X c}
\toprule
\toprule
       Parameters & Description & Value \\
    \midrule
    \([M/H]\) & Overall log metallicity & $-$0.11$^{+0.9}_{-0.9}$ \\
    \([Ti/H]\) & Log titanium metallicity & 0.06$^{+0.10}_{-0.11}$ \\
    \(\log\alpha_{\mathrm{H}^{-}}\) & Log hydrogen anions abundance multiplier & 0.38$^{+0.13}_{-0.12}$\\
    \(\log\alpha_{\mathrm{e}^{-}}\) & Log free electrons abundance multiplier  & $-$1.09$^{+0.59}_{-0.46}$ \\
    T [K] \ & Atmospheric temperature & 3\,832$^{+172}_{-164}$ \\
    $\mathrm{K}_\mathrm{p}$ [km/s] & Planet velocity semi-amplitude & 185.0$^{+7.5}_{-7.1}$\\
    $\mathrm{V}_\mathrm{sys}$ [km/s] & Systemic velocity & $-$27.06$^{+0.75}_{-0.78}$\\
    FWHM [km/s] & Rotational broadening full width at half maximum & 36.05$^{+3.53}_{-3.06}$ \\ 
    \bottomrule
    \bottomrule
    \label{Table:retrieval results best-fit}
    \end{tabularx}
\end{table}

%

In order to move beyond the cross-correlation method and infer posterior probabilities on atmospheric parameters of interest, we must use an adequate high-resolution likelihood prescription \citep[e.g.][]{brogi_retrieving_2019, gibson_detection_2020, gibson_relative_2022}.  
Currently, the presented method to clean the data and remove contributions from the host star and Earth's atmosphere to reveal the planetary signal relies on using a telluric model as well as a Doppler shadow correction (Sections \ref{subsubsec:Cleaning steps} and \ref{subsubsec:Removal of the Doppler shadow}).  
However, while the telluric correction is done at the spectral level, the average effect of the Doppler shadow is instead removed from the cross-correlation function. 
It is therefore not so obvious how to compute a likelihood, which is typically done at the spectral level, when including such a Doppler shadow correction.
Nevertheless, this step cannot be neglected as removing contributions from the star is necessary given that the stellar photosphere also contains Fe, which could otherwise bias any inference of planetary atmospheric properties.

As accurately correcting for the Doppler shadow distortion on every stellar line residual at the spectral level would require a perfectly accurate model of the host star photosphere and is beyond the scope of the paper, we opt instead to proceed forward and retrieve the atmospheric properties of WASP-189b using the methodology presented by \cite{Pelletier+2021, Pelletier+2023}.
For this we now re-analyse the data, but instead starting from the telluric uncorrected order-by-order data products for both HARPS and NIRPS and using a Principal Component Analysis (PCA)-based approach to detrend the data of telluric and stellar contributions.
Indeed while the cross-correlation maps calculated from the NIRPS DRS-corrected data still show some H$_2$O residuals, even in the case where tellurics are masked aggressively (Figure~\ref{Figure:NIRPS_non_detections}), we find that this data driven method is better at removing tellurics, albeit at the cost of potentially distorting the atmospheric trace of WASP-189b (we refer the reader to Appendix~\ref{Appendix-Sec:TelluricCorrectionsComparison} for a detailed discussion).
However, while PCA can alter and even remove an underlying planetary signal~\citep{brogi_retrieving_2019, cheverall_feasibility_2024}, this effect is less severe for short orbital period planets where atmospheric absorption lines are rapidly accelerating, such as in the case of WASP-189b observed in transmission. Nevertheless, reproducing the effect of the detrending procedure on the spectral line shapes and contrasts of the underlying planetary signal is critical in order to accurately infer properties of its atmosphere. As such, all likelihood calculations in the retrieval are made using such a `processed’ model that mimics the effect of PCA on spectral features~\citep{Pelletier+2021, Pelletier+2023}.  
We opt to remove 10 principal components, although we tested removing anywhere between 3 and over 20 components, finding overall consistent results. We further discard from the analysis all exposures where the planetary trace overlaps with the Doppler shadow to avoid contamination from any potential remaining stellar residuals.
Using this PCA-based approach, we similarly find an Fe absorption signature in the HARPS data (see Appendix Figure~\ref{Figure:PCA_Fe_CCF}) but do not detect an equivalent signature in the NIRPS data.  This is not necessarily surprising given that Fe has much stronger spectral features in the optical than in the near-IR (Figure~\ref{Figure:Fe+hydrideComparison}).  Nevertheless, the presence of numerous Fe lines of various strengths spread across the HARPS and NIRPS wavelength ranges allows us to probe the effect of various continuum opacity sources.  For example, the pseudo-continuum from TiO will affect the line contrast of Fe spectral features in and out of TiO absorption bands. Meanwhile, H$^-$ (bound-free) will instead dampen all Fe lines below $\sim$1.5~$\mu$m roughly uniformly, while H$^-$ (free-free) will preferentially weaken lines at longer wavelengths.

We use the high-resolution atmospheric retrieval framework detailed in \cite{Pelletier+2021, Pelletier+2023, Pelletier+2024, bazinet_subsolar_2024} which uses \texttt{SCARLET} to produce transmission spectra, the CCF-to-log likelihood prescription of \cite{gibson_detection_2020}, and \texttt{emcee}~\citep{foreman-mackey_emcee_2013} as a sampler. We run a retrieval first on the HARPS data alone, and then a retrieval also including the NIRPS data, for both transits simultaneously. Other than abundance parameters, the retrieval fits for the atmospheric temperature, the velocity parameters $K_\mathrm{p}$ and $v_\mathrm{sys}$, as well as a free broadening parameter. The temperature is assumed to be uniform across the atmosphere (isothermal).  $K_\mathrm{p}$ and $v_\mathrm{sys}$ are fitted rather than fixed as the observed position of the atmospheric signature of WASP-189b can differ from the values predicted for a uniform and static atmosphere.  We also include an extra Gaussian broadening parameter as the observed signal is more extended (blurred) in $K_\mathrm{p}-v_\mathrm{sys}$ space than expected from the instrumental resolution only.

As thermal dissociation and ionisation are important in ultra-hot Jupiter atmospheres, especially at sub-millibar pressures probed by transmission spectroscopy, abundances of species can vary by orders of magnitude in pressure.  As such, we opt against so called `free retrievals' that parametrise the abundance of elements as uniform-in-pressure and instead compute abundance profile using \texttt{FastChem}, which takes into account ionisation and dissociation.  
However, in order to give the model some flexibility in exploring non-equilibrium scenarios, we fit for both an overall metallicity ([M/H]) controlling the abundances refractory elements 
as well as a separate metallicity for Ti-bearing species ([Ti/H]).  Here [M/H] refers to the $\log_{10}$ metallicity relative to solar and controls the abundance of all non-Ti bearing species included in the retrieval (Fe, Fe$^{+}$, Cr, Mg, V, Mn, Ca, Ni, Na, H$_2$O, CO, and OH), with a value of [M/H] = 0.5 corresponding to an atmospheric composition for WASP-189b being $10^{0.5}$$\times$ that of the Sun~\citep{asplund_chemical_2009}.  Meanwhile, [Ti/H] controls the abundances of TiO, Ti, and Ti$^{+}$.
The motivations for fitting [Ti/H] separately is because titanium species are often observed to be underabundant in ultra-hot Jupiter atmospheres, likely due to nightside cold-trapping~\citep{hoeijmakers_hot_2020, hoeijmakers_mantis_2024, merritt_non-detection_2020, maguire_high-resolution_2023, Gandhi+2023}, in which case the TiO abundance can be greatly over-predicted by models \citep[e.g.][]{Pelletier+2023, Pelletier+2024}. 
Similarly, the opacity contributions of H$^-$ (bound-free and free-free) will strongly depend on the availability of electrons in the atmosphere of WASP-189b, which is dependent on the ionisation of metals \citep{Gray_2021}. 
Given the plausibility that non-equilibrium processes play a role in affecting the abundances of H$^-$ and e$^-$, we also fit for $\log_{10}$ multiplicative factors ($\log\alpha_{\mathrm{H}^{-}}$ and $\log\alpha_{\mathrm{e}^{-}}$) to both the hydride and e$^-$ abundance profiles predicted by \texttt{FastChem}.  A value of $\log\alpha_{\mathrm{H}^{-}}$ = 1 would therefore correspond to the H$^-$ abundance being 10$\times$ higher at all pressures than expected from equilibrium chemistry for the given composition (set by [M/H] and [Ti/H]) and temperature structure.


Our retrievals therefore have a total of eight parameters (Table~\ref{Table:retrieval results best-fit}), with the overall results shown in Figure~\ref{Figure:WASP189b_HARPSonly_Retrieval_CornerPlot}.  We find the overall metallicity [M/H] (mostly driven by Fe) to be well consistent with solar.  Intriguingly, our retrieval provides a bounded constraint on the titanium metallicity [Ti/H] despite the lack of a cross-correlation detection of any Ti-bearing species.  Rather, this constraint is driven by the retrieval preferring to add TiO bands as a source of pseudo-continuum.  Interestingly, the retrieved [Ti/H] abundance comes out to be nearly the same as the overall metallicity.  While normally one should be sceptical of retrieved posteriors on species not detected in cross-correlation, in this particular case TiO is known from higher S/N data to be present in the transmission spectrum of WASP-189b~\citep{Prinoth+2022, Prinoth+2023}.  Nevertheless, we do not consider this a robust detection of TiO and rather simply take it as further evidence that models of WASP-189b including TiO in its atmosphere are preferred.

The retrieval also constrains the abundance of H$^-$, finding this source of opacity to be more prevalent than predicted from chemistry predictions. This may indicate that the continuum set by H$^-$ is more pronounced than expected.  We note, however, that an over abundance of H$^-$ could also be the result of other sources of opacity in the atmosphere of WASP-189b unaccounted for in our models (e.g. metal oxides like VO, FeO, MgO, CaO, etc. or metal hydrides like FeH, MgH, SiH, AlH, CaH, etc.).  The abundance of free electrons is unsurprisingly not well constrained when including HARPS only, with a much stricter upper limit obtained when including NIRPS.  Overall, even though the cross-correlation detection of Fe is mostly driven by the HARPS data, the inclusion of NIRPS with a high sensitivity across near-infrared wavelengths is highly constraining, especially for H$^-$ and e$^-$.  The temperature is found to be around 3\,800~K, which is similar to what was found by \cite{Gandhi+2023}.
Moreover, we retrieve a rotational broadening value of $\simeq$36~km~s$^{-1}$, about one order of magnitude higher than the equatorial rotational velocity of the planet estimated to be of $\simeq$3~km~s$^{-1}$\footnote{$v_\mathrm{eq.~rot.}$ = $\omega~\times~R_\mathrm{p}=(2\pi/P)~\times~R_\mathrm{p}$, where $\omega=2\pi/P$ is the angular velocity, $P$ is WASP-189b orbital period, and $R_\mathrm{p}$ is the planet radius.}. Indeed, transmission data are more sensitive to atmospheric dynamics, hence we could expect a rotational broadening parameter $\not\approx$ planet rotation rate. We surmise that the blurred atomic iron signal detected in HARPS optical data, and, consequently, the high retrieved rotational broadening parameter, are linked to the combination of the planet's rotation rate, day-to-night winds \citep[known in ultra-hot gas giants,][]{Seidel+2021}, and super rotating jets, all acting to contribute various blue and red Doppler shifts which all together blur the signal. What found is consistent with \cite{Prinoth+2022}, who also explain the broadening of the spectral lines to be due to super-rotational winds and flows causing an exacerbation of the red- and blueshifts of the limbs.

Broadly speaking, all of our retrieved parameters are fairly consistent with the atmosphere of WASP-189b having a solar-like composition, with the exception that the H$^-$ bound-free contribution (in the form of the hydrogen anion abundance) appears to be elevated relative to what equilibrium chemistry predicts by about 0.5 dex (Figure~\ref{Figure:WASP189b_NIRPS+HARPS_Retrieval_H-/Fe_distribution}).  Interestingly, the H$^-$ continuum being slightly higher than expected has little implications for the detectability of metals with very deep absorption lines in the optical, as evident by the wealth of species that have already been detected in transmission on WASP-189b~\citep[e.g.][]{Prinoth+2023}.  However, this could mean that other elements or molecules that only have weaker spectral features (even if there are many), such as H$_2$O, may be harder to detect in ultra-hot Jupiter atmospheres using transmission spectroscopy.





\section{Conclusions}
\label{sec:Conclusions}
This study presents an in-depth analysis of WASP-189b's atmospheric composition using high-resolution transmission spectra from two transit time series gathered simultaneously with HARPS and NIRPS. A key finding is the detection of atomic Fe in the optical at a maximum S/N of $\sim$5, accompanied by its absence in the near-infrared wavelength regime. We show that the lack of detections in the near-infrared could be due to the opacity contribution given by the hydride bound-free transitions that dampens the relative contrast of spectral lines. Atmospheric retrievals, run on both HARPS only and HARPS+NIRPS datasets combined, indicate that the hydride-to-Fe ratio exceeds the equilibrium model predictions by about 0.5 dex, hinting at an elevated abundance of electrons that are bound to hydrogen atoms in the upper atmosphere. We find the overall metallicity to be consistent with solar, with our retrieval also constraining the titanium metallicity despite the lack of a cross-correlation detection of any Ti-bearing species. Although the retrieved [Ti/H] is consistent with [M/H], we do not consider this to be a robust detection of any Ti-bearing species and rather simply take it as evidence that WASP-189b's atmospheric models including TiO in its atmosphere are preferred. This can also mean that we are sensitive the pseudo-continuum of TiO, but not its spectral lines.

These results highlight the challenges posed by the hydride continuum opacity when detecting metals and other species in transmission in the near-infrared. The high-resolution data from NIRPS, even in case of non-detections, offer valuable constraints on the atmospheric opacity and composition of WASP-189b. The gathered NIRPS high-resolution spectra allow us to probe subtle opacity effects from the hydride continuum in the near-infrared, advancing our understanding of metal and molecular abundances in the challenging environments of ultra-hot Jupiter atmospheres.
Extending observations into the mid-far-infrared (\textit{K}, \textit{L}, \textit{M} photometric bands beyond 2~$\mu$m) could help overcome and mitigate the limitations imposed by the hydride continuum in transmission, allowing for detections of water (H$_2$O), carbon monoxide (CO), trihydrogen cation (H$_3^+$), and other species that may be present but remain undetectable at shorter wavelengths.  

Additionally, the findings of this study complement recent CRIRES+ $K$-band dayside observations of WASP-189b showing cross-correlation signals of CO and Fe, both blueshifted possibly due to day-to-night winds~\citep{Lesjak+2025}. Notably, the $K$-band extends further into the infrared compared to the reddest limit of NIRPS and lies beyond where H- bound-free interactions can absorb photons and is fully in the free-free regime. 
The detection of Fe emission in the $K$-band thus likely indicates that the atmosphere of WASP-189b is more transparent past $\sim$1.6\,$\mu$m, as suggested by the NIRPS data (Table~\ref{Table:retrieval results best-fit}). 
The dayside thermal emission viewing geometry also enables the CRIRES+ observations to probe deeper pressures of the atmosphere and detect Fe lines that would otherwise appear weaker in the transmission spectrum that probes lower pressure levels. In general, ultra-hot Jupiter atmospheres with important H- contributions may be easier to probe in emission rather than in transmission at infrared wavelengths.

\begin{acknowledgements}
      This work has been carried out within the framework of the NCCR PlanetS supported by the Swiss National Science Foundation under grants 51NF40\_182901 and 51NF40\_205606.\\
VV, DE  acknowledge support from the Swiss National Science Foundation for project 200021\_200726. The authors acknowledge the financial support of the SNSF.\\
RA  acknowledges the Swiss National Science Foundation (SNSF) support under the Post-Doc Mobility grant P500PT\_222212 and the support of the Institut Trottier de Recherche sur les Exoplan\`etes (IREx).\\
RA, \'EA, FBa, BB, RD, LMa, LB, CC, NJC, LD, AL \& TV  acknowledge the financial support of the FRQ-NT through the Centre de recherche en astrophysique du Qu\'ebec as well as the support from the Trottier Family Foundation and the Trottier Institute for Research on Exoplanets.\\
EC, SCB, ED-M, NCS, ARCS \& JGd  acknowledge the support from FCT - Funda\c{c}\~ao para a Ci\^encia e a Tecnologia through national funds by these grants: UIDB/04434/2020, UIDP/04434/2020.\\
XDu  acknowledges the support from the European Research Council (ERC) under the European Union’s Horizon 2020 research and innovation programme (grant agreement SCORE No 851555) and from the Swiss National Science Foundation under the grant SPECTRE (No 200021\_215200).\\
HC, ML \& BA acknowledge support of the Swiss National Science Foundation under grant number PCEFP2\_194576.\\
\'EA, FBa, RD \& LMa  acknowledges support from Canada Foundation for Innovation (CFI) program, the Universit\'e de Montr\'eal and Universit\'e Laval, the Canada Economic Development (CED) program and the Ministere of Economy, Innovation and Energy (MEIE).\\
SCB  acknowledges the support from Funda\c{c}\~ao para a Ci\^encia e Tecnologia (FCT) in the form of work of work through the Scientific Employment Incentive program (reference 2023.06687.CEECIND).\\
XB and JMA acknowledge funding from the European Research Council under the ERC Grant Agreement n. 337591-ExTrA.\\
XB, XDe, TF \& VY  acknowledge funding from the French ANR under contract number ANR\-18\-CE31\-0019 (SPlaSH), and the French National Research Agency in the framework of the Investissements d'Avenir program (ANR-15-IDEX-02), through the funding of the ``Origin of Life" project of the Grenoble-Alpes University.\\
The Board of Observational and Instrumental Astronomy (NAOS) at the Federal University of Rio Grande do Norte's research activities are supported by continuous grants from the Brazilian funding agency CNPq. This study was partially funded by the Coordena\c{c}\~ao de Aperfei\c{c}oamento de Pessoal de N\'ivel Superior—Brasil (CAPES) — Finance Code 001 and the CAPES-Print program.\\
BLCM  acknowledge CAPES postdoctoral fellowships.\\
BLCM  acknowledges CNPq research fellowships (Grant No. 305804/2022-7).\\
NBC  acknowledges support from an NSERC Discovery Grant, a Canada Research Chair, and an Arthur B. McDonald Fellowship, and thanks the Trottier Space Institute for its financial support and dynamic intellectual environment.\\
JRM  acknowledges CNPq research fellowships (Grant No. 308928/2019-9).\\
ED-M  further acknowledges the support from FCT through Stimulus FCT contract 2021.01294.CEECIND. ED-M  acknowledges the support by the Ram\'on y Cajal grant RyC2022-035854-I funded by MICIU/AEI/10.13039/50110001103 and by ESF+.\\
JIGH, RR, ASM, NN \& AKS  acknowledge financial support from the Spanish Ministry of Science, Innovation and Universities (MICIU) projects PID2020-117493GB-I00 and PID2023-149982NB-I00.\\
ICL  acknowledges CNPq research fellowships (Grant No. 313103/2022-4).\\
CMo  acknowledges the funding from the Swiss National Science Foundation under grant 200021\_204847 “PlanetsInTime”.\\
Co-funded by the European Union (ERC, FIERCE, 101052347). Views and opinions expressed are however those of the author(s) only and do not necessarily reflect those of the European Union or the European Research Council. Neither the European Union nor the granting authority can be held responsible for them.\\
This project has received funding from the European Research Council (ERC) under the European Union's Horizon 2020 research and innovation programme (project {\sc Spice Dune}, grant agreement No 947634). This material reflects only the authors' views and the Commission is not liable for any use that may be made of the information contained therein.\\
ARCS  acknowledges the support from Funda\c{c}ao para a Ci\^encia e a Tecnologia (FCT) through the fellowship 2021.07856.BD.\\
LD  acknowledges the support of the Natural Sciences and Engineering Research Council of Canada (NSERC) [funding reference number 521489] and from the Fonds de recherche du Qu\'ebec (FRQ) - Secteur Nature et technologies [funding file number 332355].\\
RLG  acknowledge CAPES graduate fellowships.\\
FG  acknowledges support from the Fonds de recherche du Qu\'ebec (FRQ) - Secteur Nature et technologies under file \#350366.\\
AL  acknowledges support from the Fonds de recherche du Qu\'ebec (FRQ) - Secteur Nature et technologies under file \#349961.\\
NN  acknowledges financial support by Light Bridges S.L, Las Palmas de Gran Canaria.\\
NN acknowledges funding from Light Bridges for the Doctoral Thesis "Habitable Earth-like planets with ESPRESSO and NIRPS", in cooperation with the Instituto de Astrof\'isica de Canarias, and the use of Indefeasible Computer Rights (ICR) being commissioned at the ASTRO POC project in the Island of Tenerife, Canary Islands (Spain). The ICR-ASTRONOMY used for his research was provided by Light Bridges in cooperation with Hewlett Packard Enterprise (HPE).\\
AP acknowledges support from the Unidad de Excelencia María de Maeztu CEX2020-001058-M programme and from the Generalitat de Catalunya/CERCA.\\
AKS  acknowledges financial support from La Caixa Foundation (ID 100010434) under the grant LCF/BQ/DI23/11990071.\\
TV  acknowledges support from the Fonds de recherche du Qu\'ebec (FRQ) - Secteur Nature et technologies under file \#320056.

\end{acknowledgements}

   \bibliographystyle{aa} 
   \bibliography{references}

\appendix
\section{\texttt{CONAN} prior and posterior distribution}
\label{Appendix-Sec:CONAN prior and posterior distribution}
\begin{table}[!h]
    \caption{\texttt{CONAN} prior and posterior distribution.}
    \begin{tabularx}{\linewidth}{XXX}
    \toprule
    \midrule
       Parameters  & Priors & Posteriors\\
       \hline
       & & \\
       Duration (T$_{14}$) [days]  & $\mathcal{N}$(0.1806, 0.01) & 0.179$^{+0.002}_{-0.002}$ \\
       
       Planet-to-star radius ratio (R$_{p}$/R$_{\star}$) & $\mathcal{U}$(0, 0.1) & 0.079$^{+0.003}_{-0.005}$ \\
       
       Impact parameter (b) & $\mathcal{N}$(0.478, 0.012) & 0.048$^{+0.01}_{-0.01}$ \\

       Mid-transit time (T$_{0}$) [BJD$_{TDB}$] & $\mathcal{N}$(2460100.6, 0.1) & 2460100.599$^{+0.001}_{-0.001}$\\

       Orbital period (P$_{\rm{orb}}$) [days] & 2.724033 &  2.724033 \\

       Eccentricity (e) & 0 & 0 \\

       Omega ($\omega$) [deg] & 90 & 90 \\

       q1$_{\rm{ExTrA}}$ & $\mathcal{N}$(0.147, 0.011) & 0.148$^{+0.011}_{-0.011}$ \\
       
       q2$_{\rm{ExTrA}}$ & $\mathcal{N}$(0.147, 0.042) & 0.152$^{+0.039}_{-0.041}$ \\

       q1$_{\rm{Euler}}$ & $\mathcal{N}$(0.432, 0.013) & 0.434$^{+0.013}_{-0.014}$\\
       
       q2$_{\rm{Euler}}$ & $\mathcal{N}$(0.253, 0.014) & 0.253$^{+0.012}_{-0.013}$\\

    \midrule
    \bottomrule  
    \end{tabularx}
    \label{Table:CONAN prior and posterior distribution parameters}
    \vspace{0.1cm}
    
    \small Notes: We assume a normal (i.e. Gaussian, $\mathcal{N}$) distribution of priors for the transit duration, the impact parameter, the mid-transit time, and for the quadratic limb darkening parameters for both Euler and ExTrA; and a uniform priors distribution ($\mathcal{U}$) for the planet-to-star radius ratio.
    \end{table}

\section{Gaussian fitting to the Doppler shadow}
\label{Appendix-Sec:DopplerShadowCorrection}
\begin{table}[!h]
    \caption{Initial guesses as to Gaussian fitting.}
    \begin{tabularx}{\linewidth}{ccccc}
    \toprule
    \midrule
         & $\mathrm{A}$ & $\mu~\mathrm{[km/s]}$ & $\sigma~\mathrm{[km/s]}$ & $\mathrm{c}$ \\
    \midrule
    1$\mathrm{^{st}}$ comp. & -0.0001 & 47 & 0.5 & 10$^{-5}$\\
    2$\mathrm{^{nd}}$ comp. & 0.0075 & 0 & 50 & 10$^{-5}$\\
    \midrule
    \bottomrule  
    \end{tabularx}
    \label{Table:GaussianPriors}
    \end{table}

\section{Hydrogen lines in the transit spectrum of WASP-189b}
\label{Appendix-Sec:Hydrogen lines in the transit spectrum of WASP-189b}
\begin{figure}[!h]
   \centering
   \includegraphics[trim={0.7cm 0 1.8cm 0}, clip, width=\columnwidth]{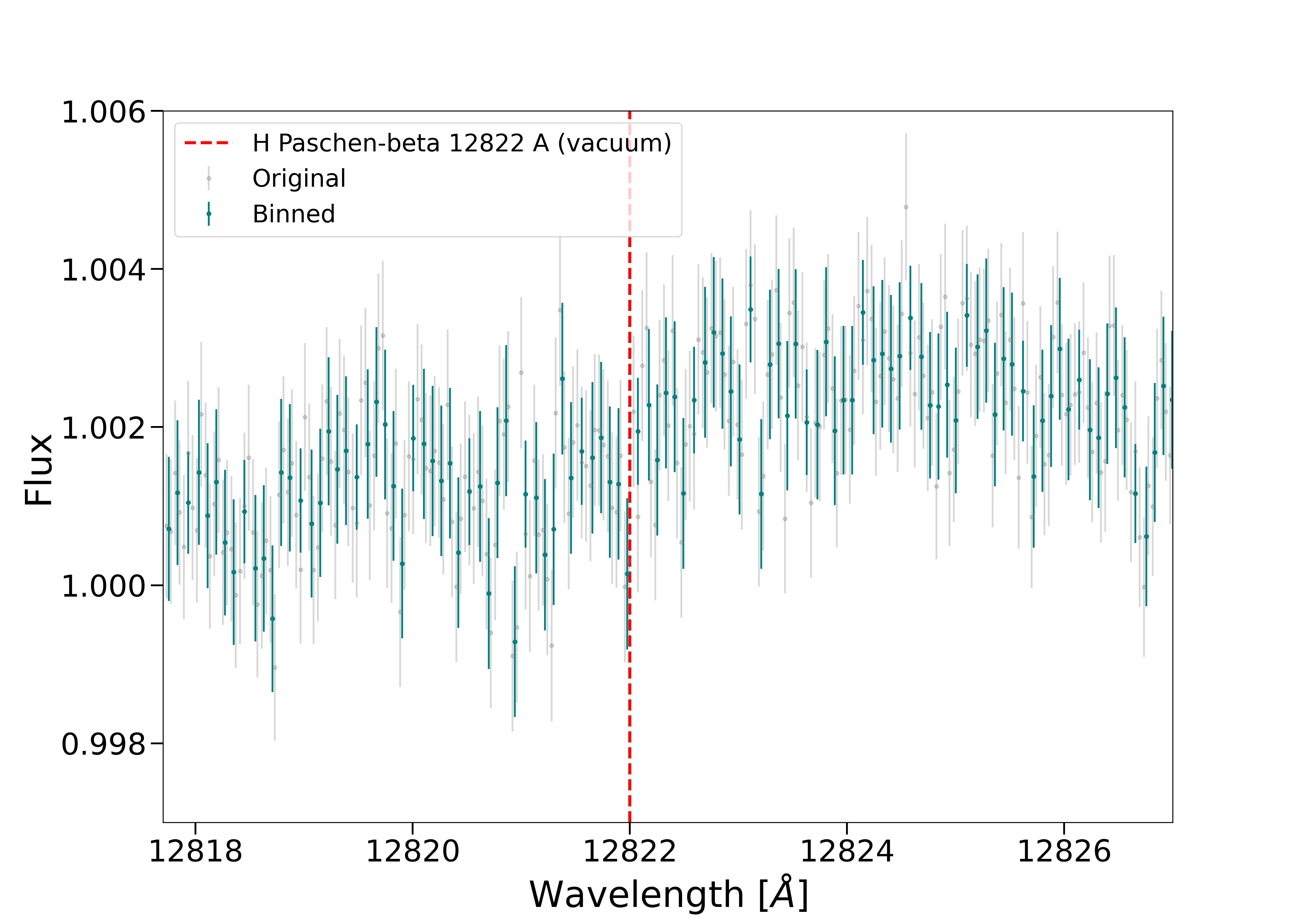}
   \caption{Transmission spectrum of WASP-189b around the hydrogen Paschen-$\beta$ line (dashed red line). No evidences of hydrogen spectral lines belonging to the Paschen series.}
   \label{Figure:WASP189b_TS@HPaschenBeta}
\end{figure}
\begin{figure}[!h]
   \centering
   \includegraphics[trim={0.7cm 0 1.8cm 0}, clip, width=\columnwidth]{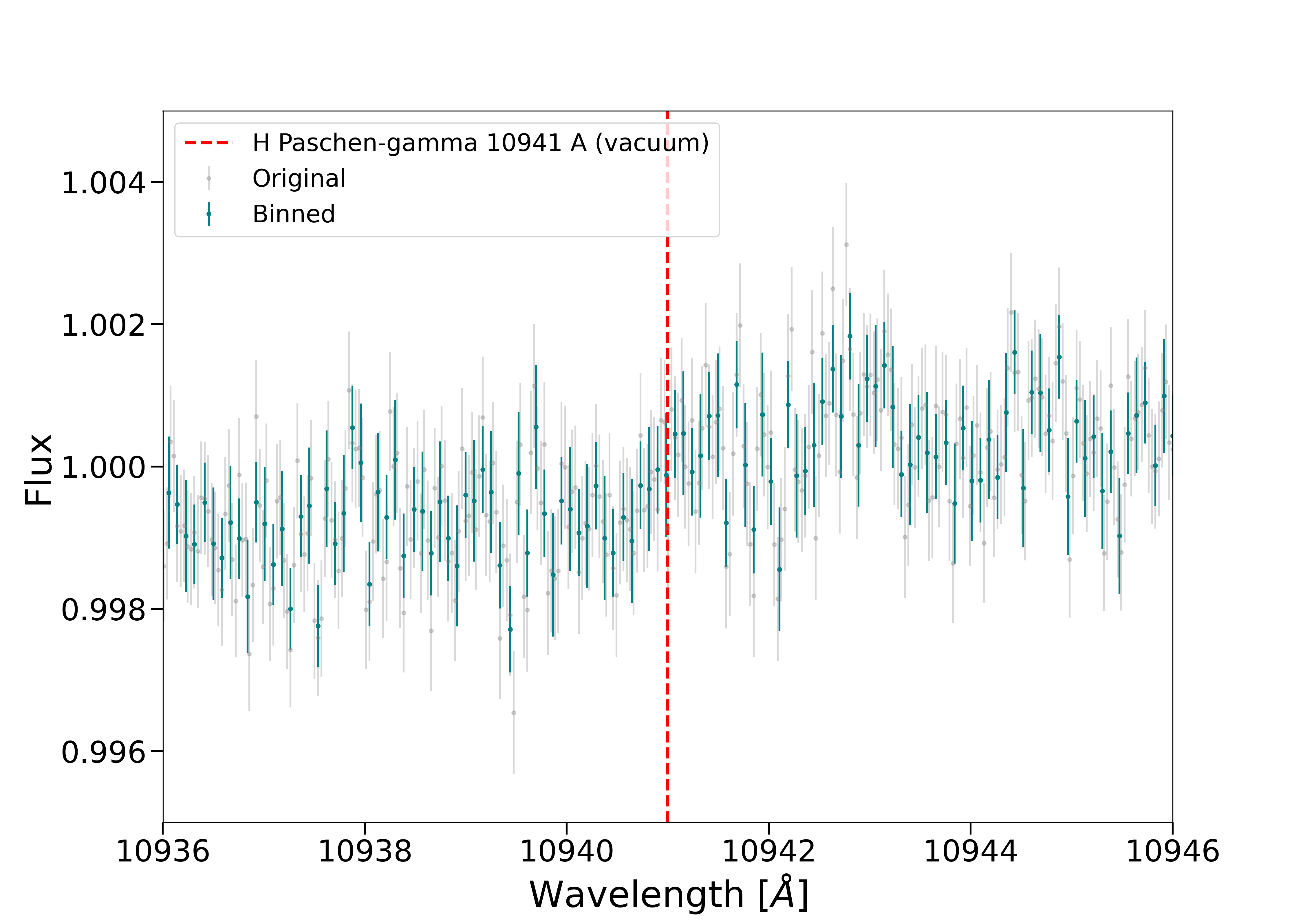}
   \caption{Transmission spectrum of WASP-189b around the hydrogen Paschen-$\gamma$ line (dashed red line). No evidences of hydrogen spectral lines belonging to the Paschen series.}
   \label{Figure:WASP189b_TS@HPaschenGamma}
\end{figure}

\section{\texttt{SCARLET} atmospheric modelled spectra of WASP-189b}
\label{Appendix-Sec:SCARLET atmospheric modelled spectra of WASP-189b}
\begin{figure*}[!h]
   \centering
   \includegraphics[width=0.97\columnwidth]{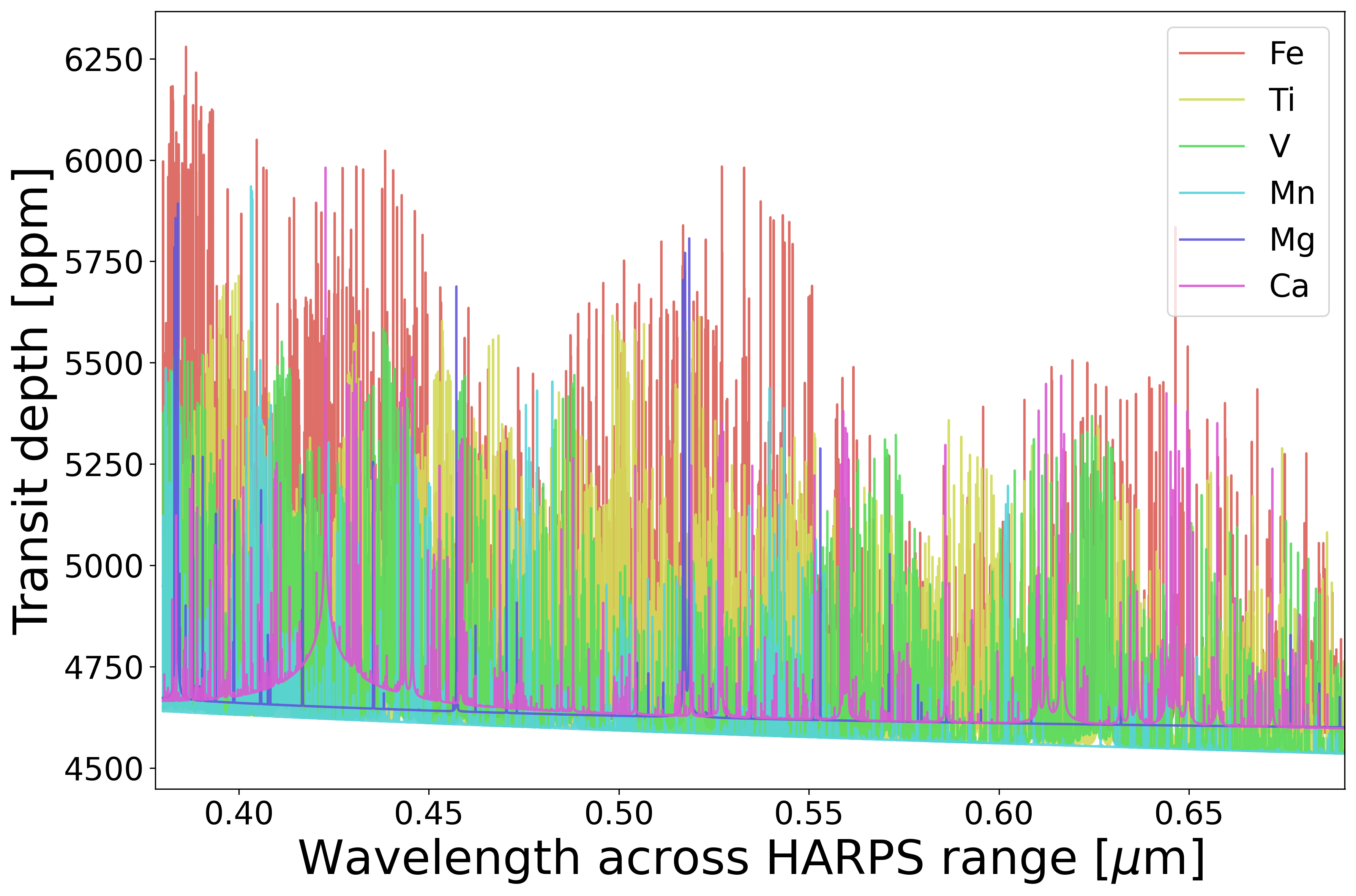}
    \includegraphics[width=0.97\columnwidth]{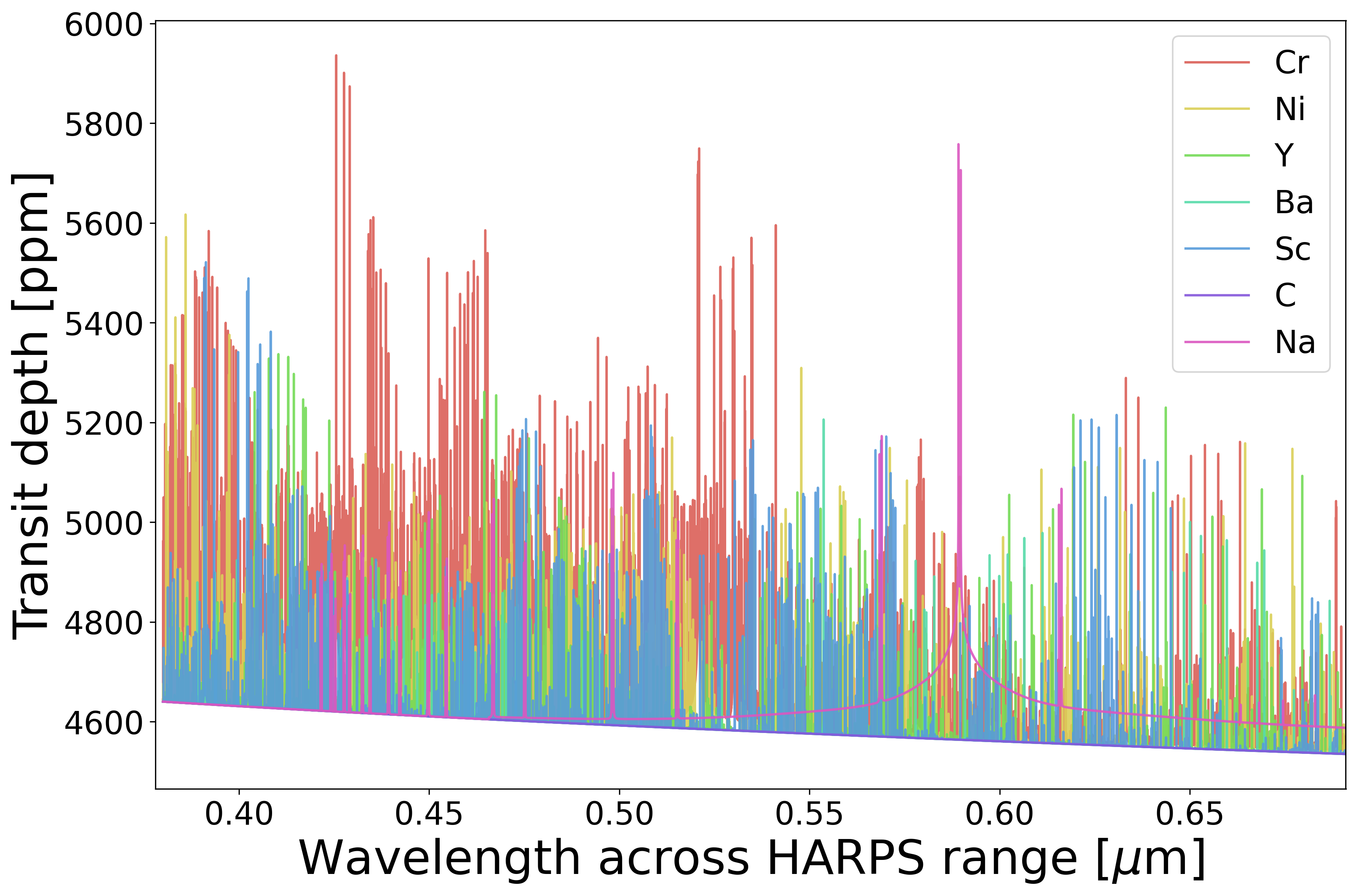}
    \includegraphics[width=0.97\columnwidth]{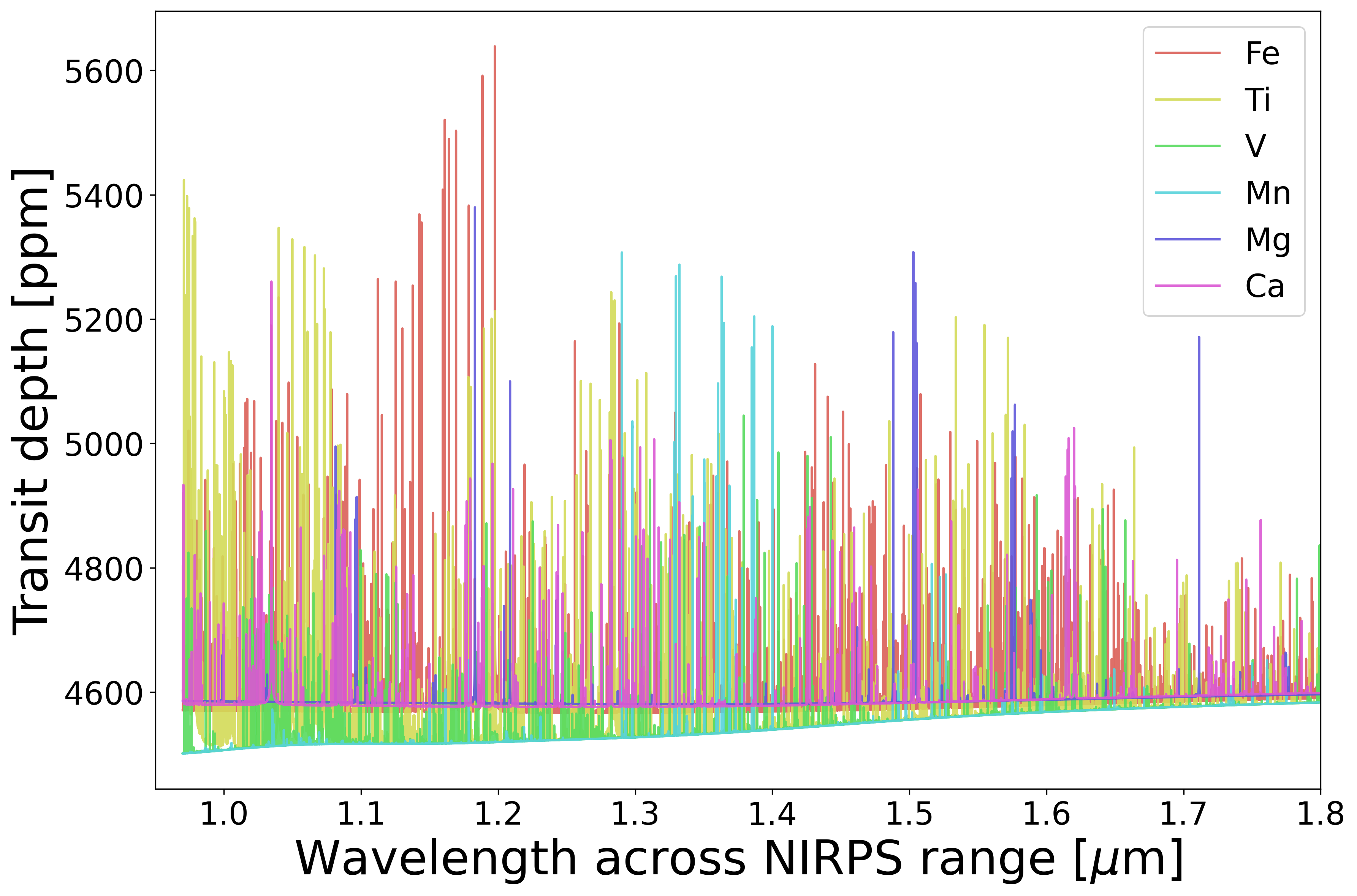}
    \includegraphics[width=0.97\columnwidth]{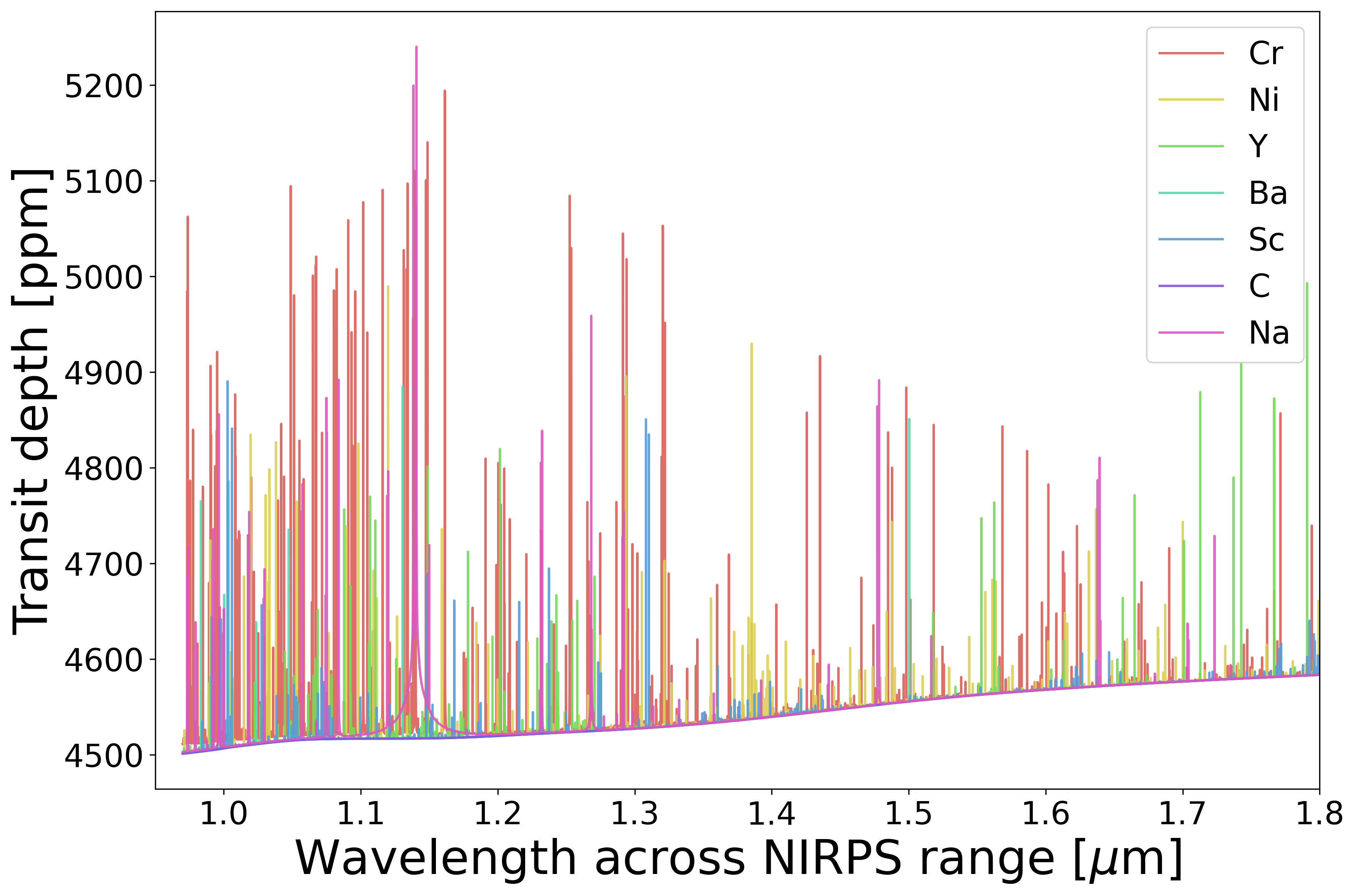}
   \caption{Full resolution \texttt{SCARLET} synthetic spectra of the atmosphere of WASP-189b for all the chemical species of interest (details in Section~\ref{subsec:Atmospheric modelling}). Top and bottom panels show all the atomic species in the HARPS and NIRPS coverage, respectively.}
   \label{Figure:WASP189b_SCARLET models atoms}
\end{figure*}

\begin{figure*}[!h]
   \centering
    \includegraphics[width=0.95\columnwidth]{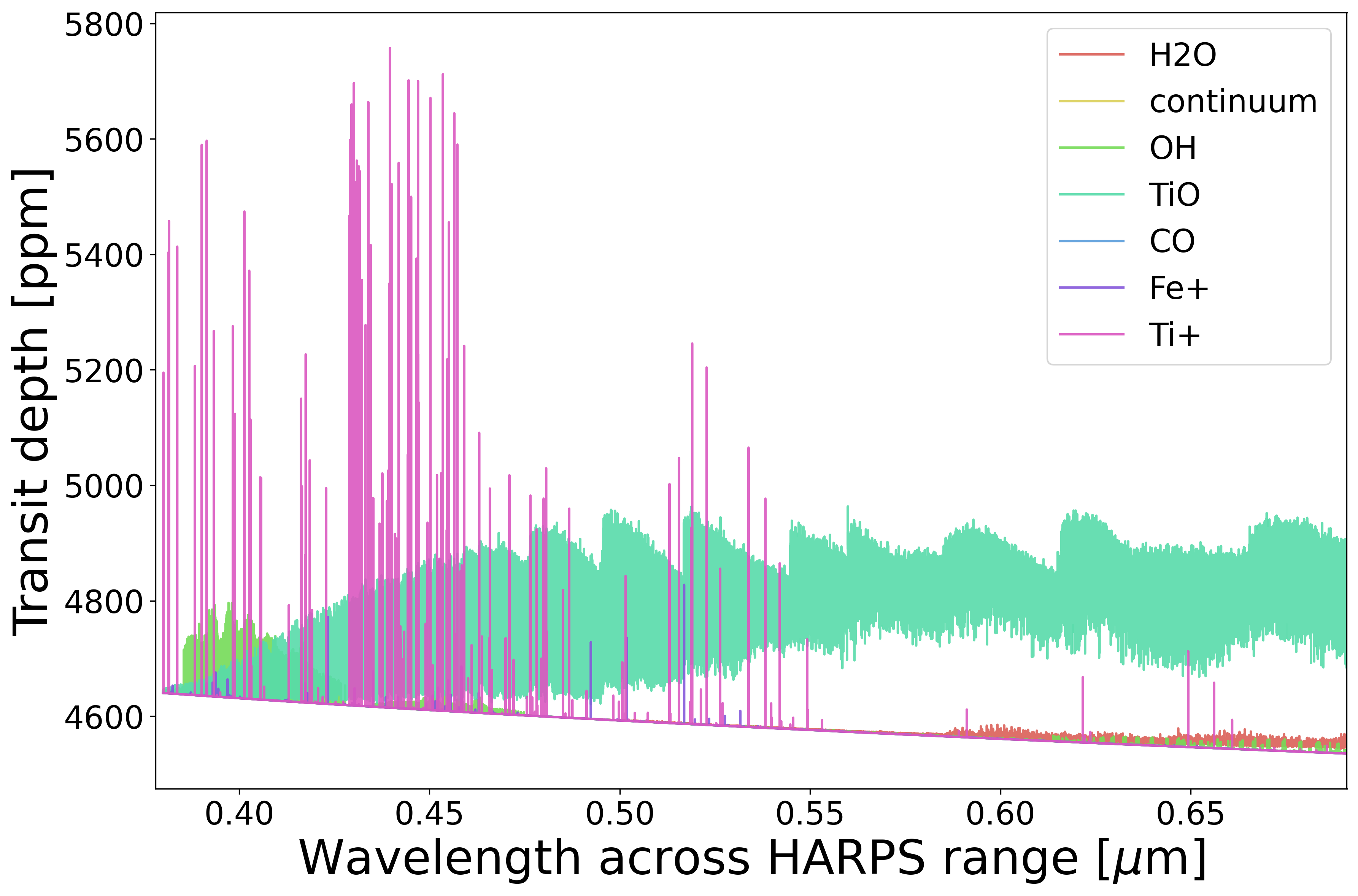}
   \includegraphics[width=0.95\columnwidth]{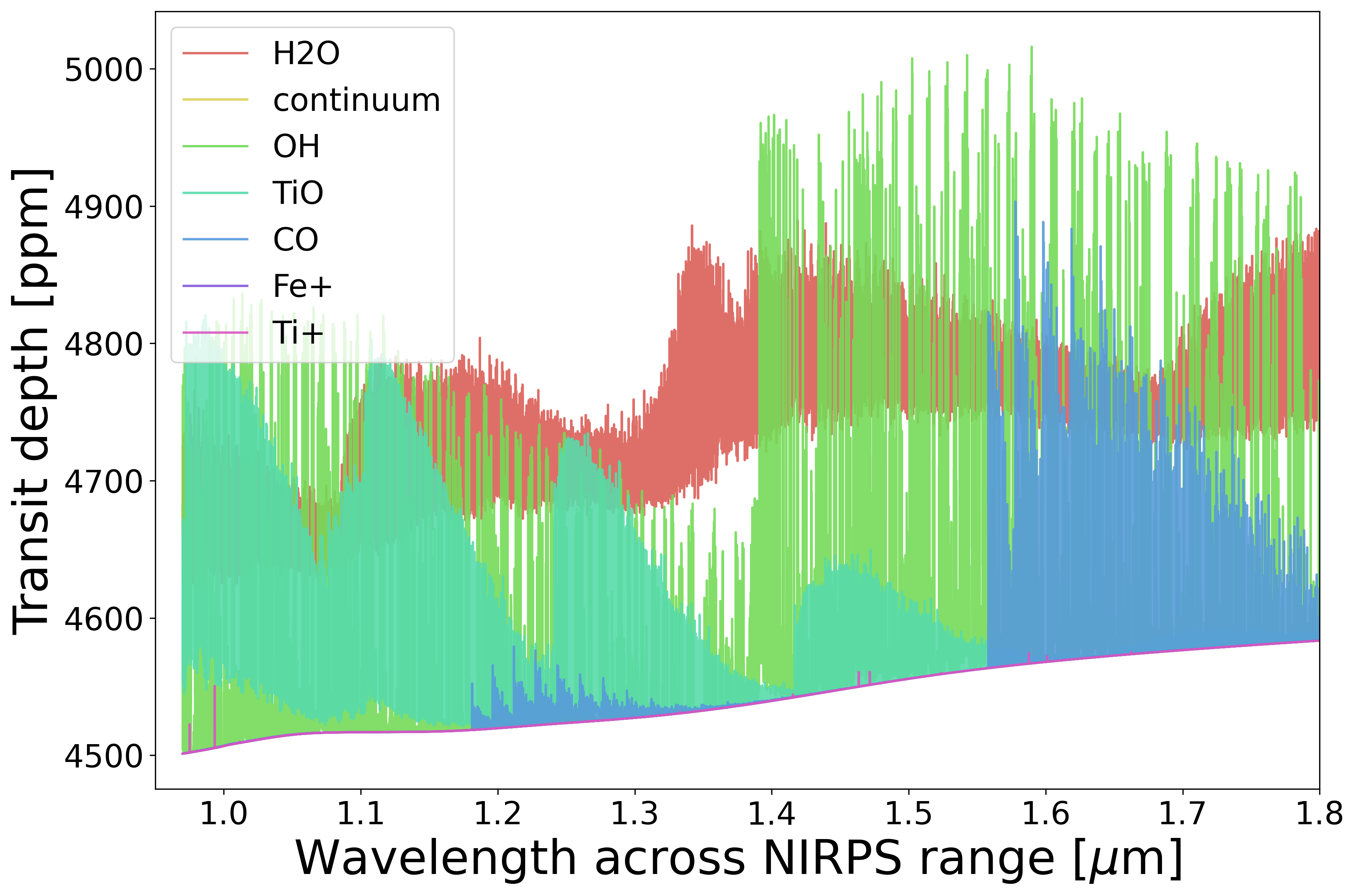}
   \caption{Full resolution \texttt{SCARLET} synthetic spectra of the atmosphere of WASP-189b for all the chemical species of interest (details in Section~\ref{subsec:Atmospheric modelling}). Left and right panels show the molecular and ionic species in the HARPS and NIRPS coverage, respectively.}
   \label{Figure:WASP189b_SCARLET models molecules and ions}
\end{figure*}

\clearpage
\section{Comparison between different telluric correction methods}
\label{Appendix-Sec:TelluricCorrectionsComparison}
\begin{figure}[!h] 
    \centering
    
    \textsf{\textbf{NIRPS - \texttt{molecfit} corrected}} 
    \includegraphics[width=\columnwidth]{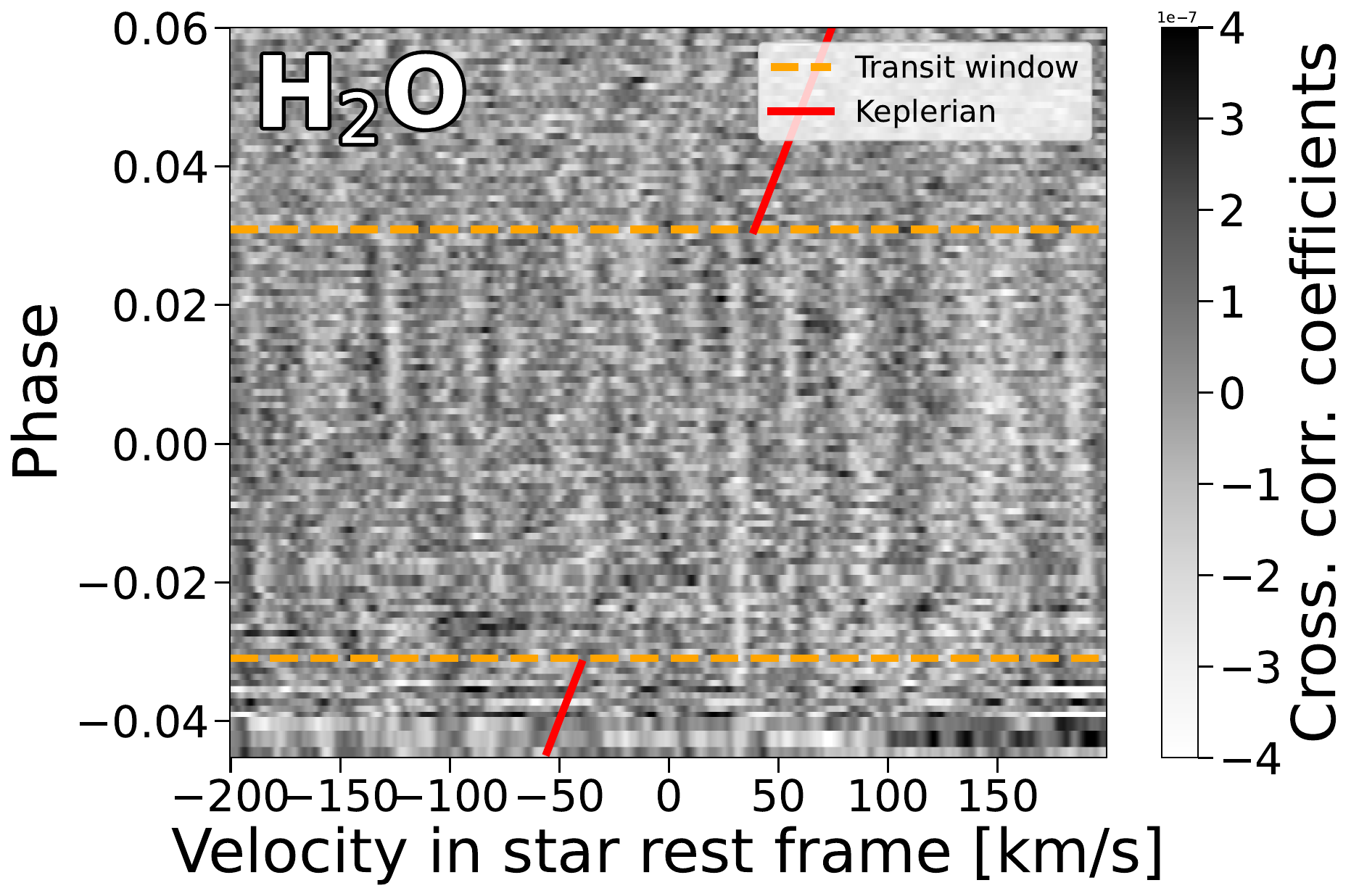}
    
    
    \textsf{\textbf{NIRPS - DRS 3.2.0 corrected}} 
    \includegraphics[width=\columnwidth]{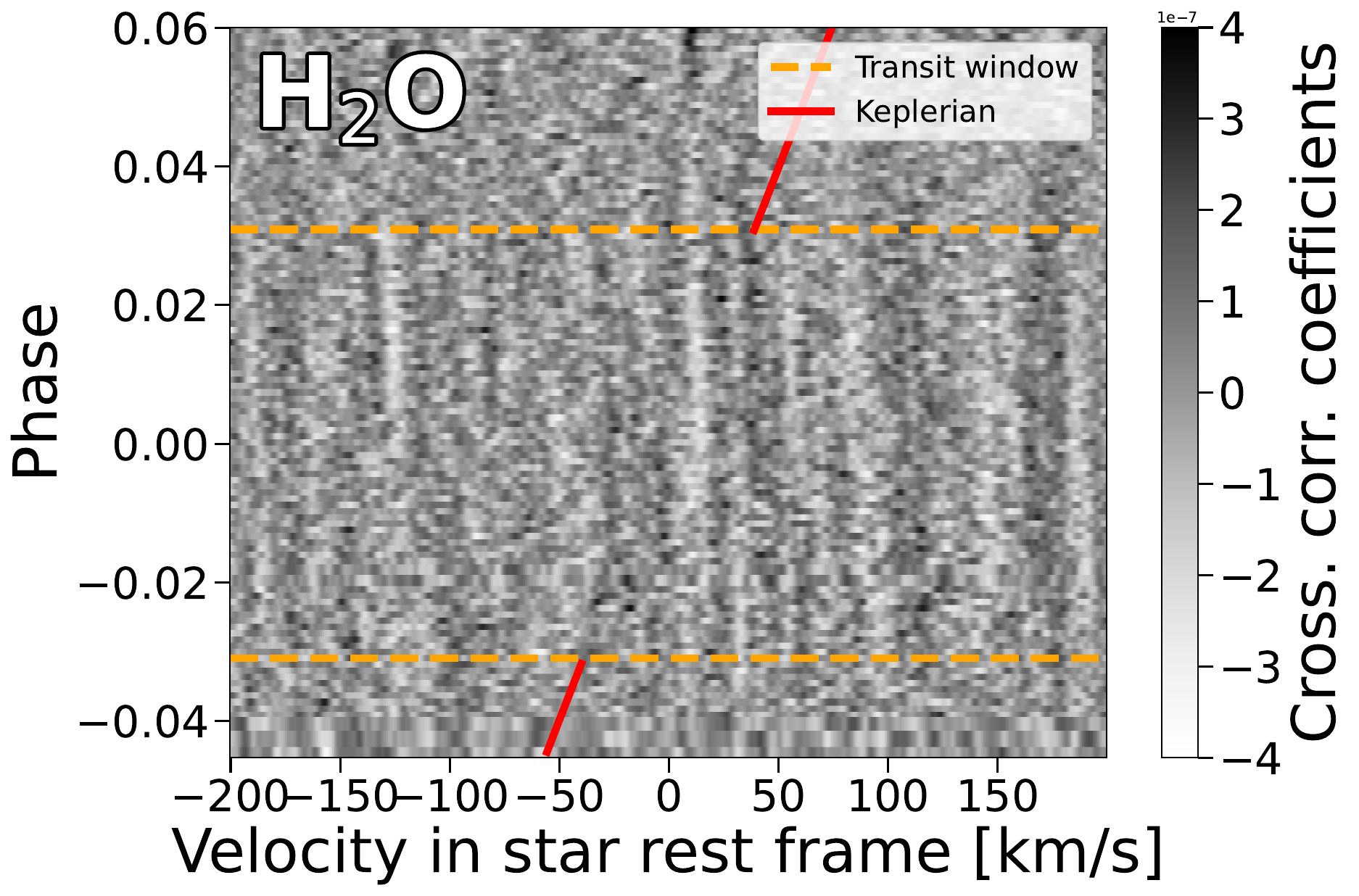}
    
    
    \textsf{\textbf{NIRPS - PCA corrected}} 
    \includegraphics[width=0.40\textwidth]{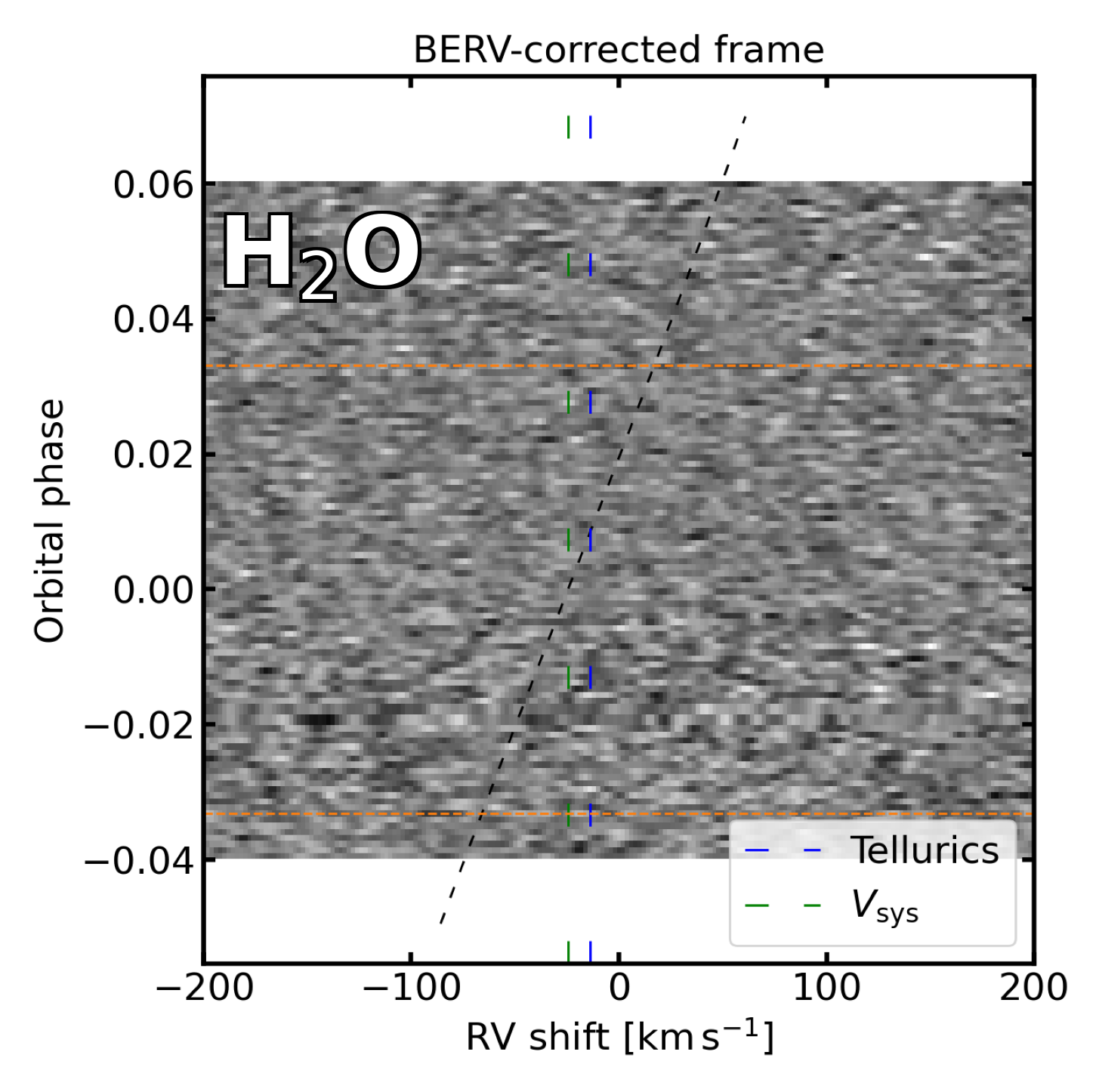}

    \caption{Top panel: Cross-correlation trail map for H$_2$O out of \texttt{molecfit} telluric corrected NIRPS data and a 90\% filtering. 
    Central panel: Cross-correlation trail map for H$_2$O out of DRS telluric corrected NIRPS data and a 90\% filtering (see Sec.~\ref{sec:Methods} and~\ref{subsubsec:Cross-correlation functions of other atoms and molecules} for a detailed description of the data analysis process). 
    Bottom panel: Cross-correlation trail map for H$_2$O out of PCA corrected NIRPS data.}
    \label{fig:tellcorrcomparison}
\end{figure}

We conduct additional tests to evaluate the quality of the telluric correction applied to near-infrared NIRPS data. Specifically, we compare the telluric correction performed using the \texttt{molecfit} tool against that implemented by the instrument's automated DRS 3.2.0 based on \cite{allart_automatic_2022}. To obtain a cleaner cross-correlation map for H$_2$O out of the DRS telluric corrected NIRPS data, we increase the filtering percentage. Despite the improvements, both water cross-correlation maps (focused on the second WASP-189b transit happened on 2023-06-04, top and central panels in Figure~\ref{fig:tellcorrcomparison}) still display residuals (as vertical stripes) highlighting that neither \texttt{molecfit} nor the DRS 3.2.0 correction are able to suppress telluric lines to a sufficient level. With remaining residuals at a greater level than the planetary signal, it is certain that this would bias inferred parameters from an atmospheric retrieval. In comparison, we find that the PCA method (see the bottom panel in Figure~\ref{fig:tellcorrcomparison}) is more effective at cleaning the data of water residuals. The persistence of telluric remnants above a comfortable level, even after processing NIRPS data with the DRS pipeline or correction using \texttt{molecfit}, inevitably guarantees contamination by water telluric residuals. Therefore, we proceed with the alternative PCA approach to clean NIRPS data, before they are digested by the atmospheric retrieval. The following cross-correlation trail maps for H$_2$O are useful tovisualise how different telluric correction approaches behave applied to near-infrared NIRPS data.

\clearpage
\section{HARPS negative injection test}
\label{Appendix-Sec:HARPS_negative_injection_test}
       \begin{figure}[!h]
    \includegraphics[width=\columnwidth]{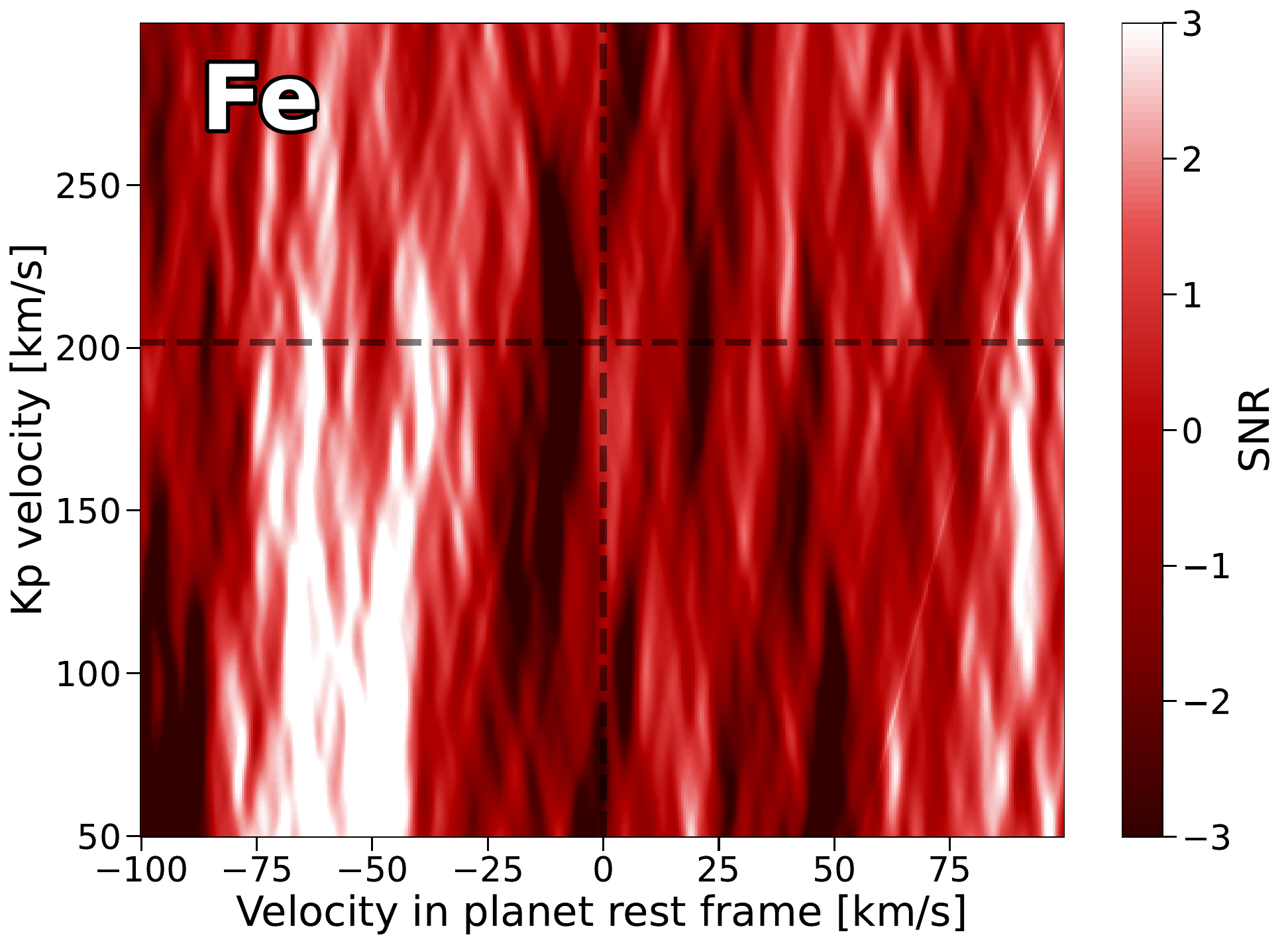}
            \caption{$K_\mathrm{p} - v_\mathrm{sys}$
              showcasing how the atomic Fe signal detected in HARPS data was cancelled out by the negative injection (in emission) of the fiducial model. This probe that the model used matches well the amplitude of the real signal (details in Section~\ref{subsubsec:Injection-recovery results}). Peaks at $\sim -$50 and +50 km~s$^{-1}$ are artefacts of the Rossiter-McLaughlin correction.}
            \label{Figure:WASP189b_HARPS_negative_injection}
   \end{figure}

\section{NIRPS non-detections in the transit spectrum of WASP-189b}
\label{Appendix-Sec:NIRPS non-detections in the transit spectrum of WASP-189b}
\begin{table}[!h]
    \caption{Line-contrast upper limits calculated for non-detections of listed neutral chemical species in the NIRPS transmission spectrum of WASP-189b.}
    \begin{tabularx}{\linewidth}{c c}
\toprule
\toprule
       Species & 3-$\sigma$ line-contrast upper limit (NIRPS)\\
    \midrule
    Fe & 3.1 $\times$ 10$^{-6}$\\
    Ti & 6.1 $\times$ 10$^{-6}$\\
    V  & 1.8 $\times$ 10$^{-6}$\\
    Mn & 2.0 $\times$ 10$^{-6}$\\
    Mg & 2.1 $\times$ 10$^{-6}$\\
    Ca & 2.4 $\times$ 10$^{-6}$\\
    Cr & 2.0 $\times$ 10$^{-6}$\\
    Ni & 7.6 $\times$ 10$^{-7}$\\
    Y  & 7.2 $\times$ 10$^{-7}$\\
    Ba & 4.7 $\times$ 10$^{-7}$\\
    Sc & 7.8 $\times$ 10$^{-7}$\\
    Na & 1.0 $\times$ 10$^{-6}$\\
    TiO & 4.9 $\times$ 10$^{-6}$\\
    H$_2$O & 2.6 $\times$ 10$^{-6}$\\
    CO & 3.6 $\times$ 10$^{-6}$\\
    OH & 1.3 $\times$ 10$^{-5}$\\
   
    \bottomrule
    \bottomrule
    \label{Table:NIRPS_upper_limits}
    \end{tabularx}
    \vspace{+0.1cm}
    
    \small Notes: Upper limits are extracted as described in sub-Section \ref{subsubsec:Velocity-velocity maps} (i.e. the standard deviation of cross-correlation values from a "noise boxy" continuum region selected away from we expect the planetary signal to pop out in the velocity space).
\end{table}

\begin{figure*}[!h]
   \centering
   \includegraphics[width=\columnwidth]{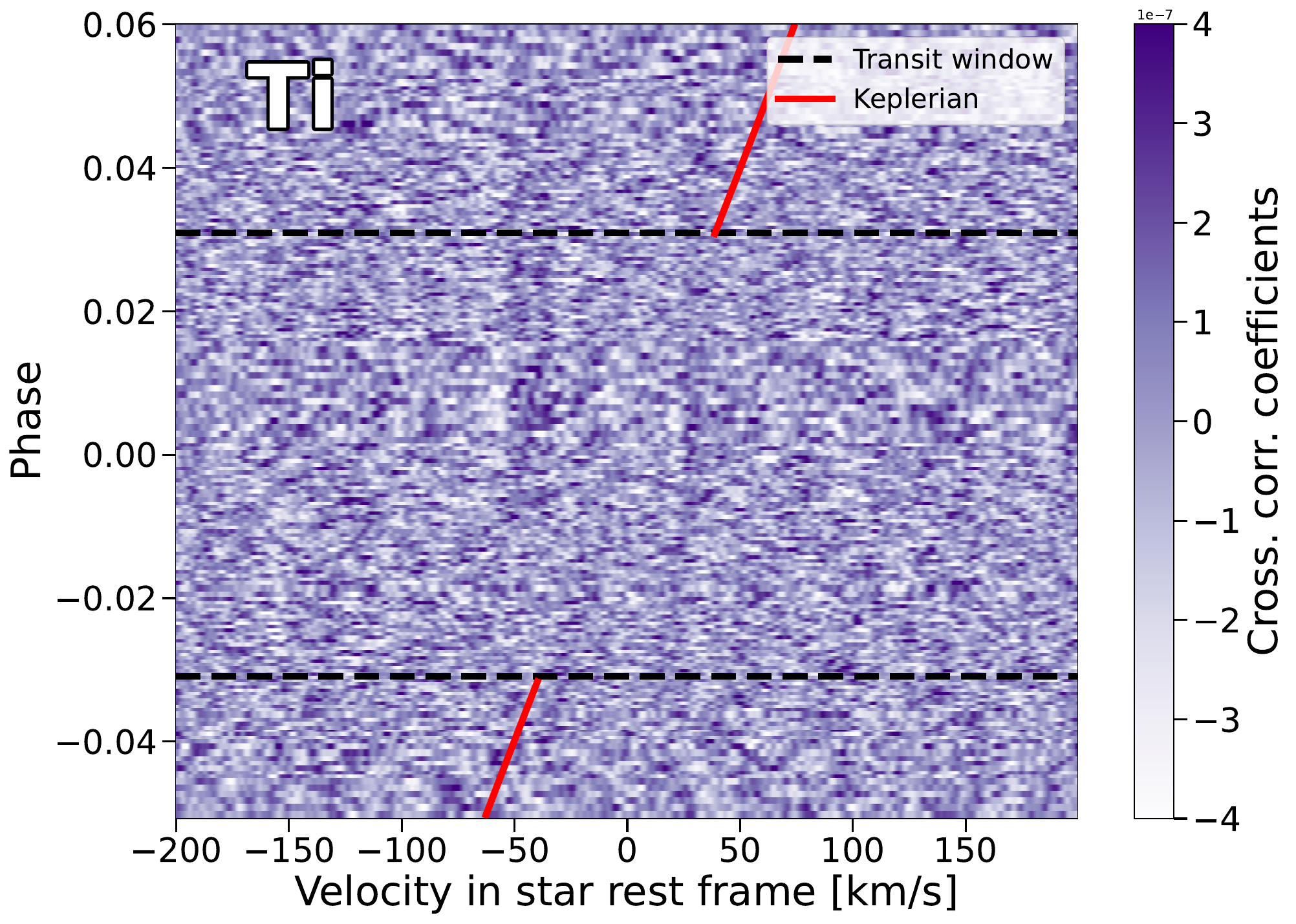}
    \includegraphics[width=\columnwidth]{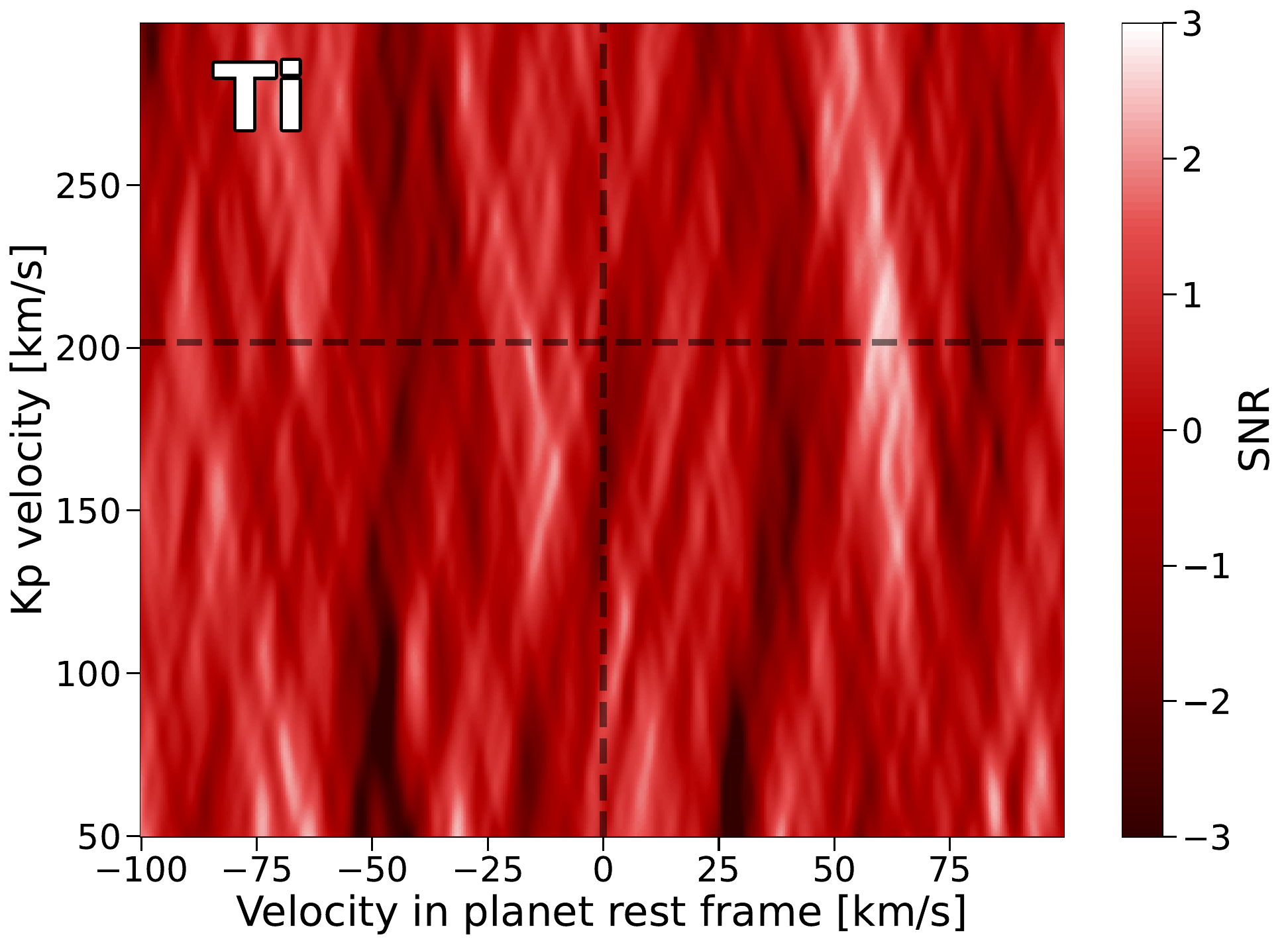}
   \includegraphics[width=\columnwidth]{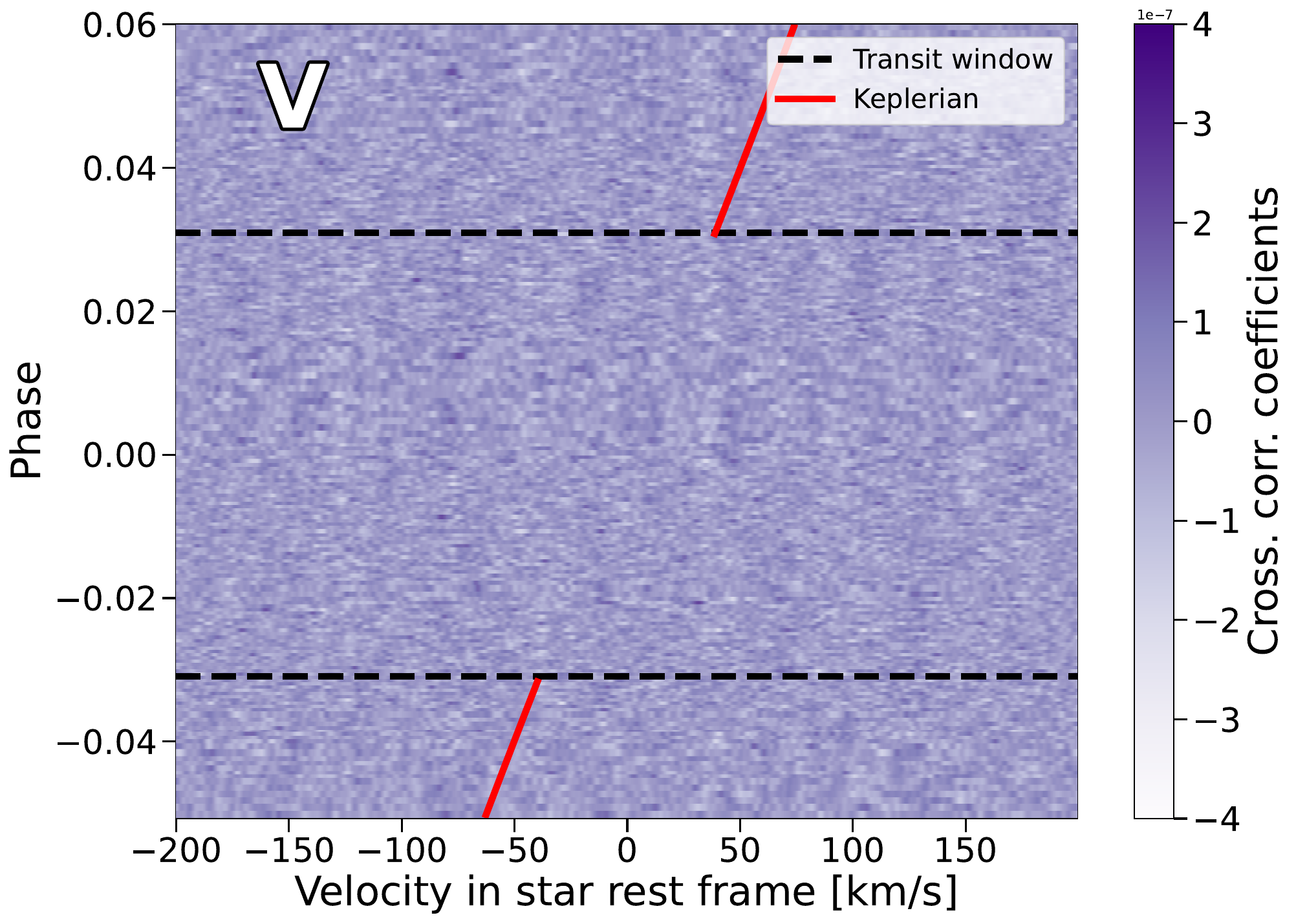}
    \includegraphics[width=\columnwidth]{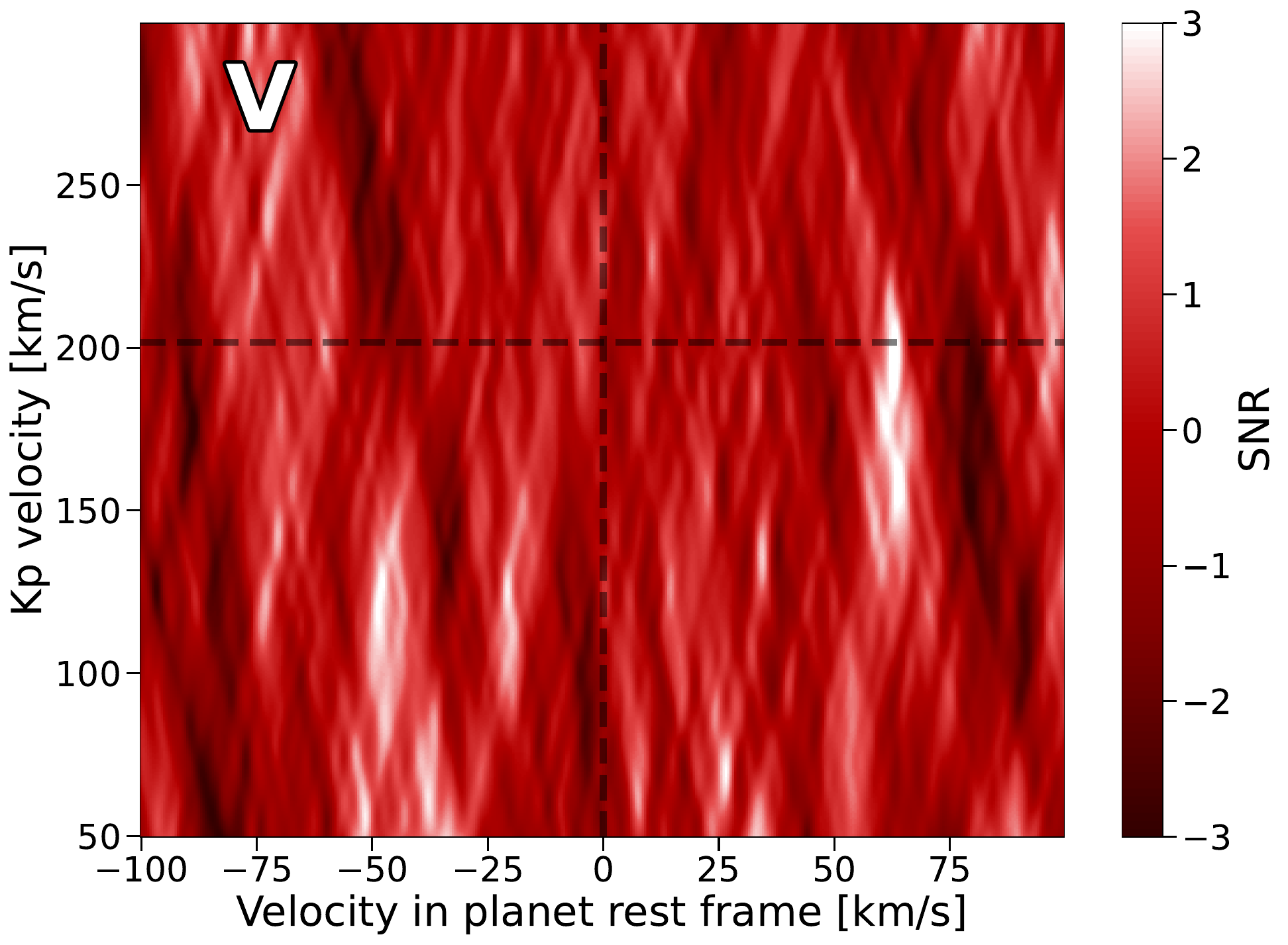}
   \includegraphics[width=\columnwidth]{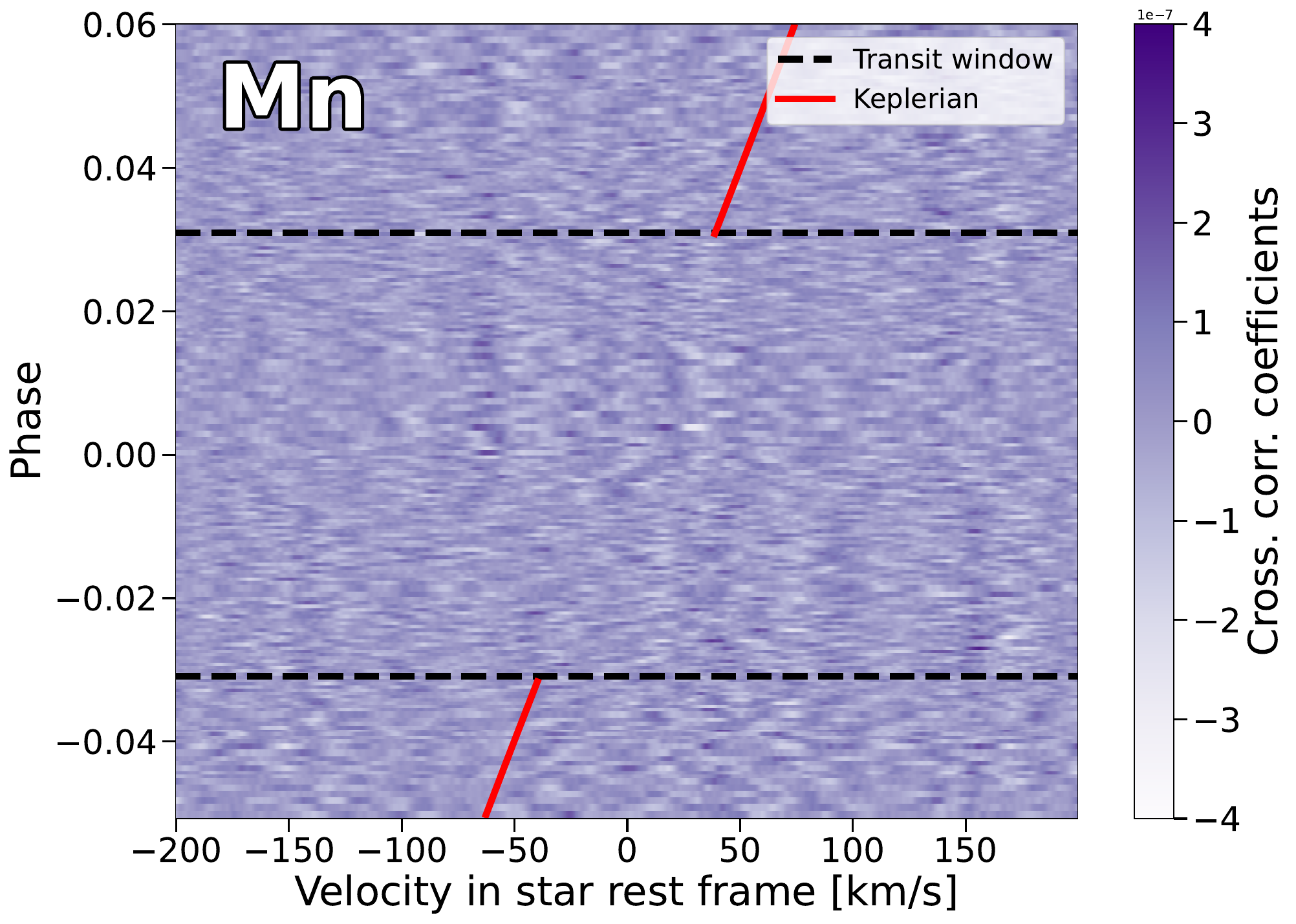}
    \includegraphics[width=\columnwidth]{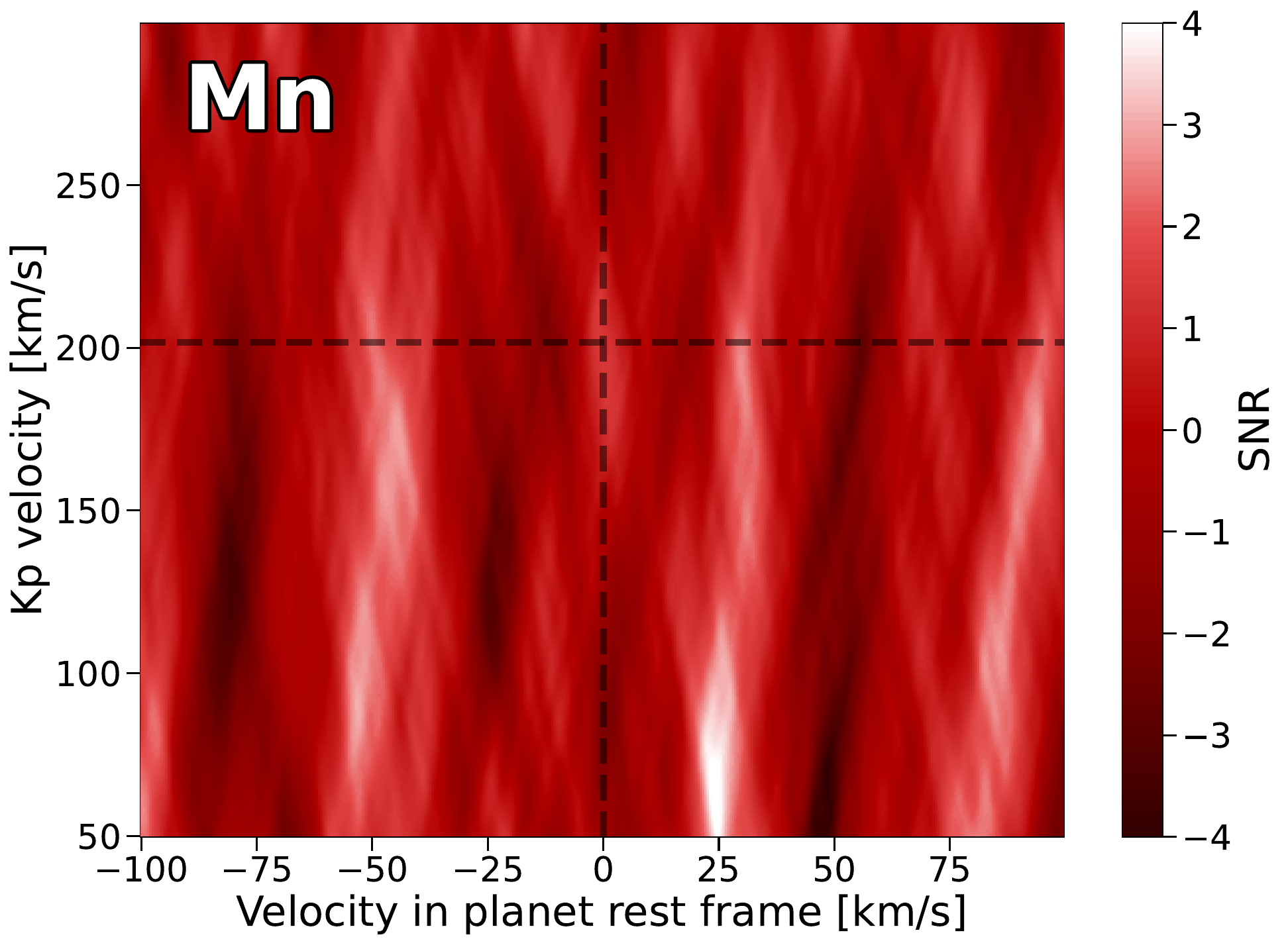}
    \caption{NIRPS cross-correlation results showing non-detected chemical species. Black dashed lines (right panel) denote the expected $K_\mathrm{p}$ determined from \cite{Prinoth+2023}, and $v_\mathrm{sys}=0$ ~km~s$^{-1}$. Left column: cross-correlation trail map. Right column: cross-correlation $K_\mathrm{p} - v_\mathrm{sys}$ map.}
   \label{Figure:NIRPS_non_detections}
\end{figure*}
\begin{figure*}[!h]
   \includegraphics[width=\columnwidth]{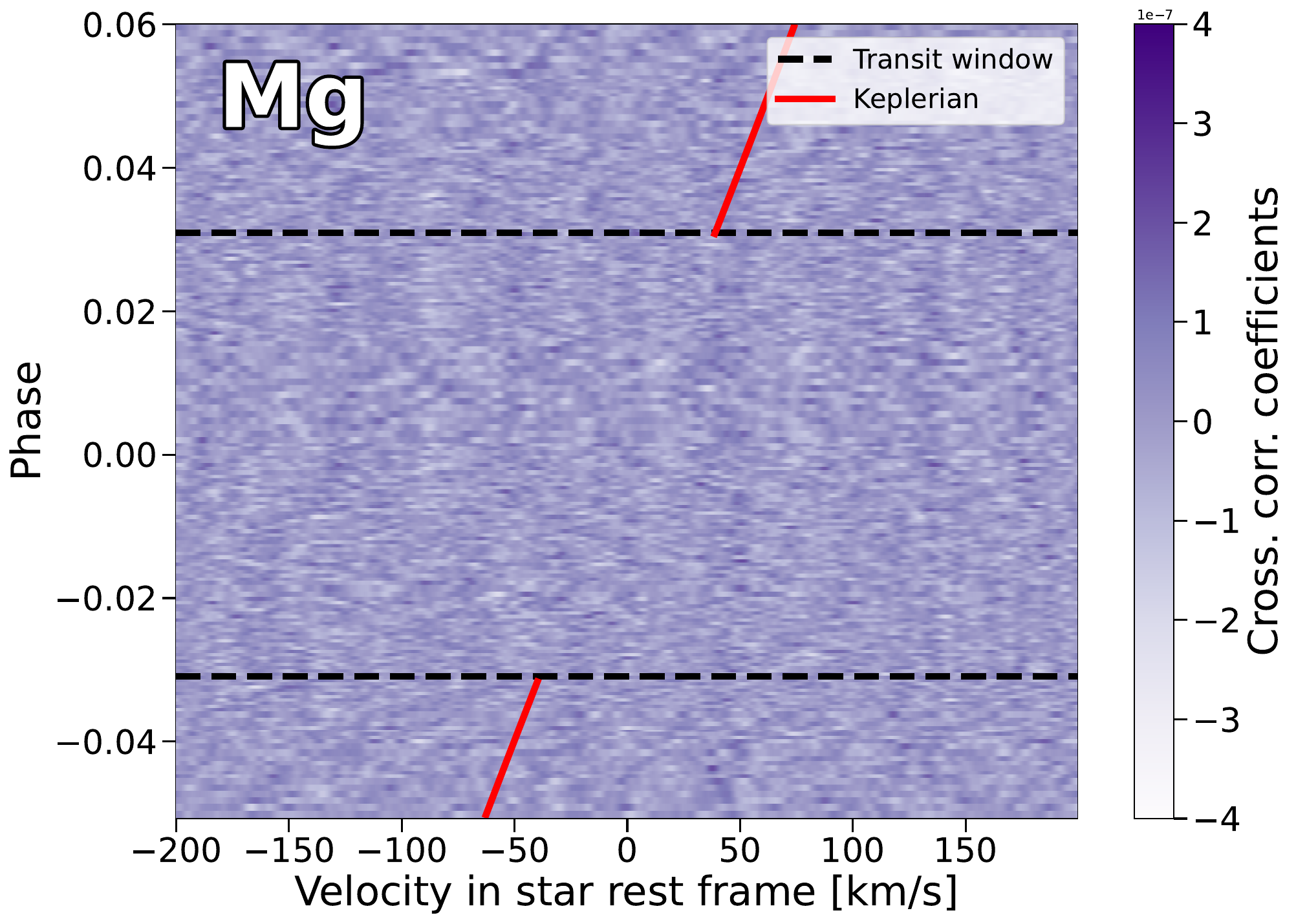}
    \includegraphics[width=\columnwidth]{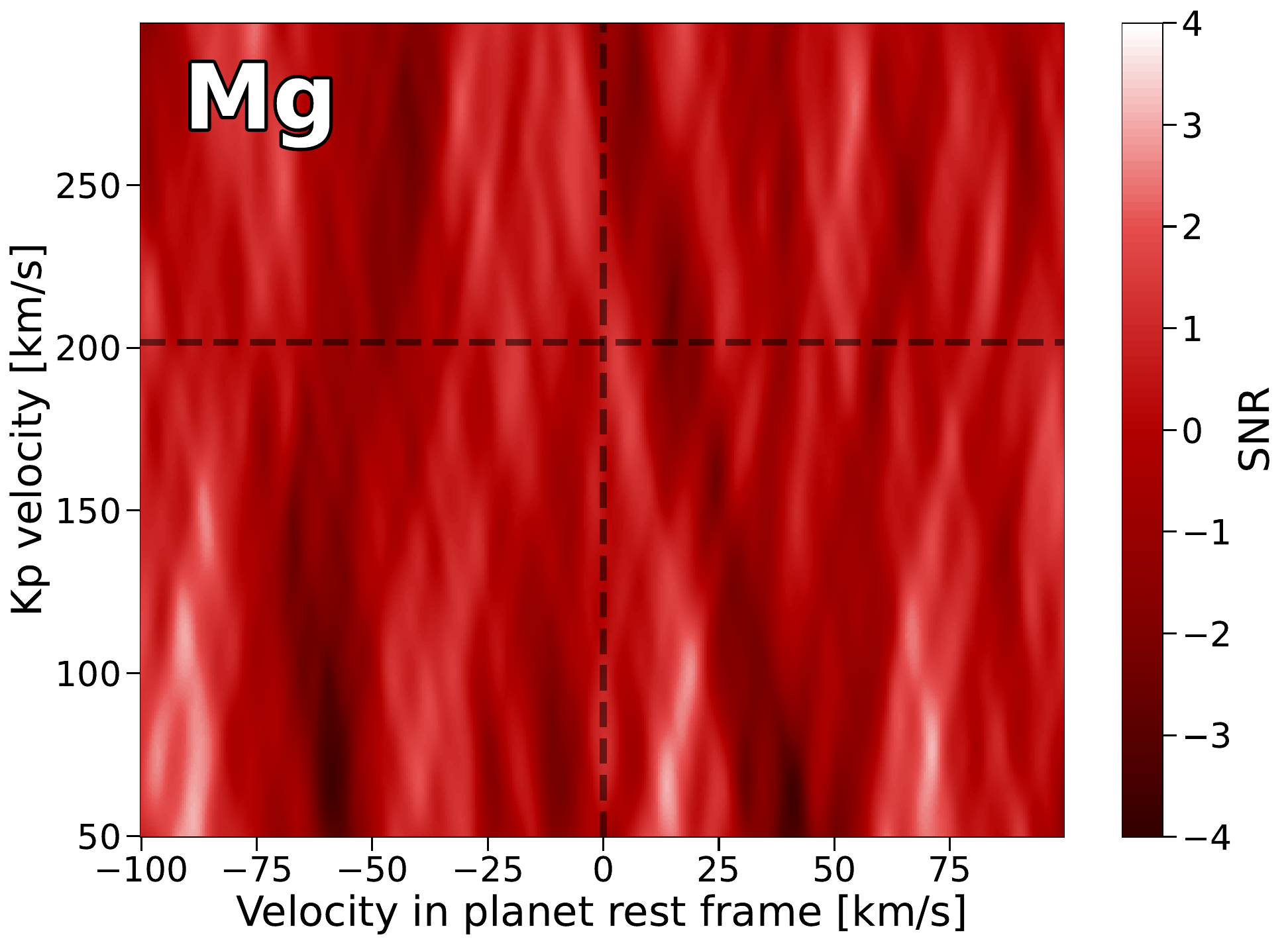}
   \includegraphics[width=\columnwidth]{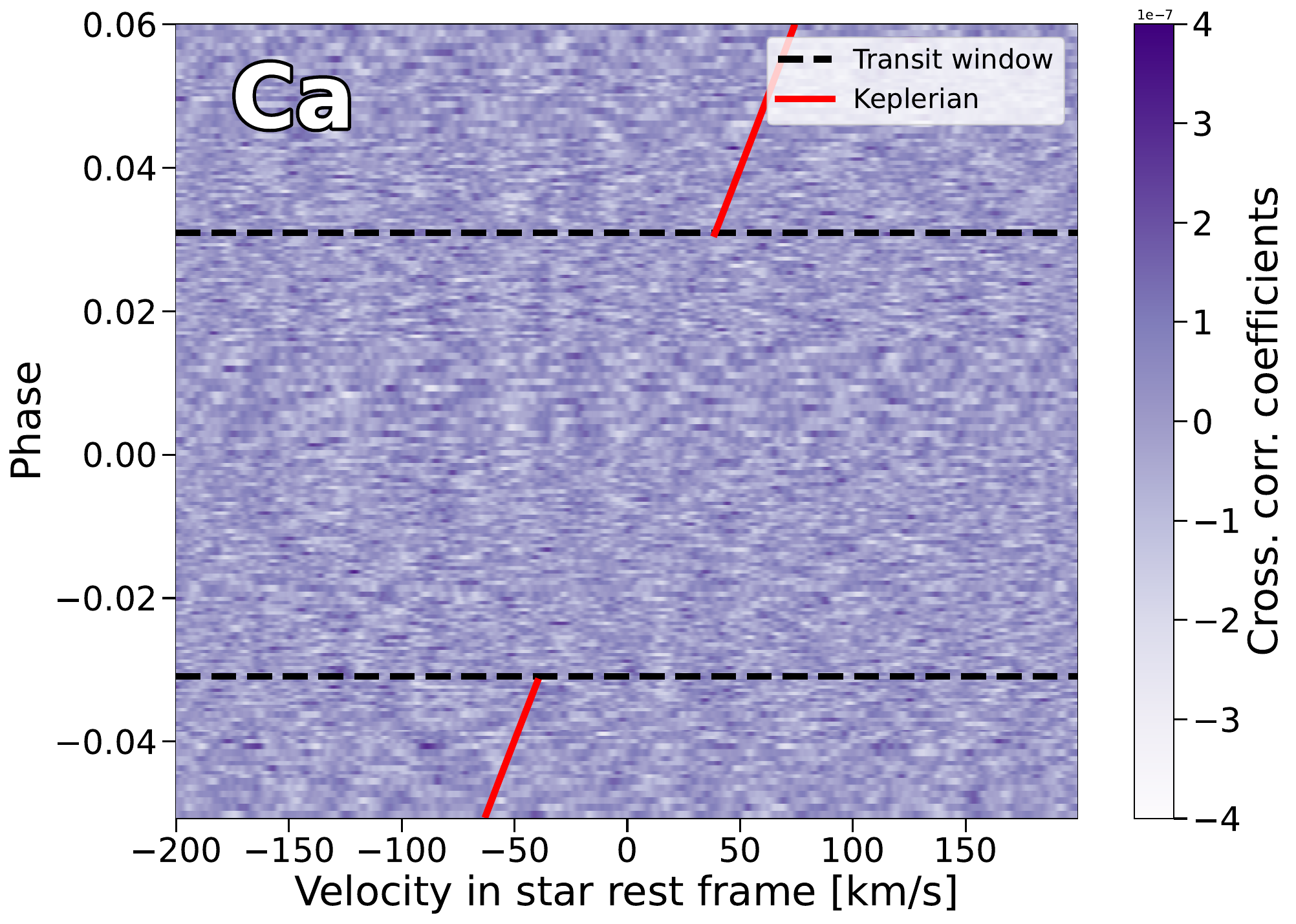}
    \includegraphics[width=\columnwidth]{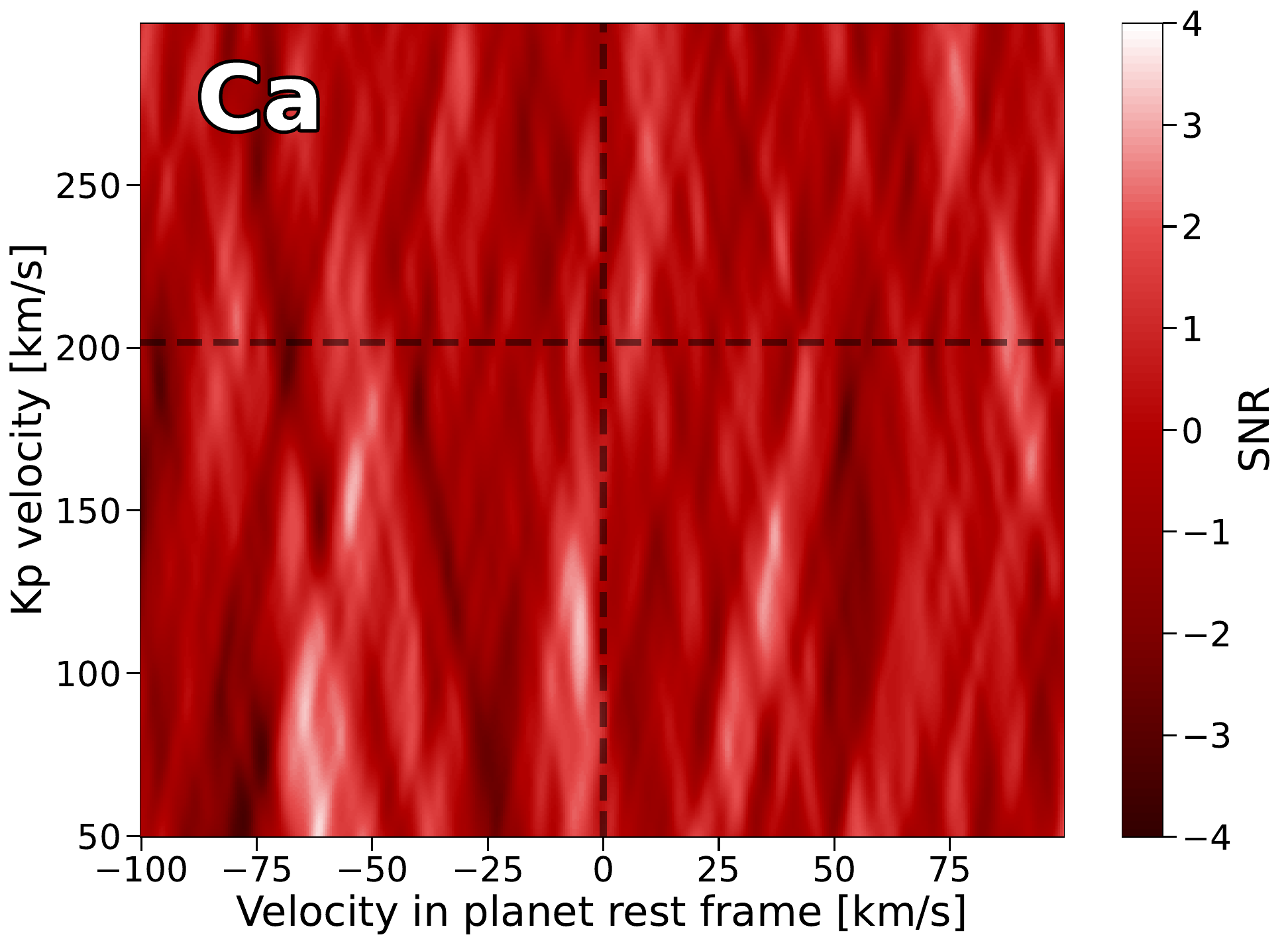} 
   \includegraphics[width=\columnwidth]{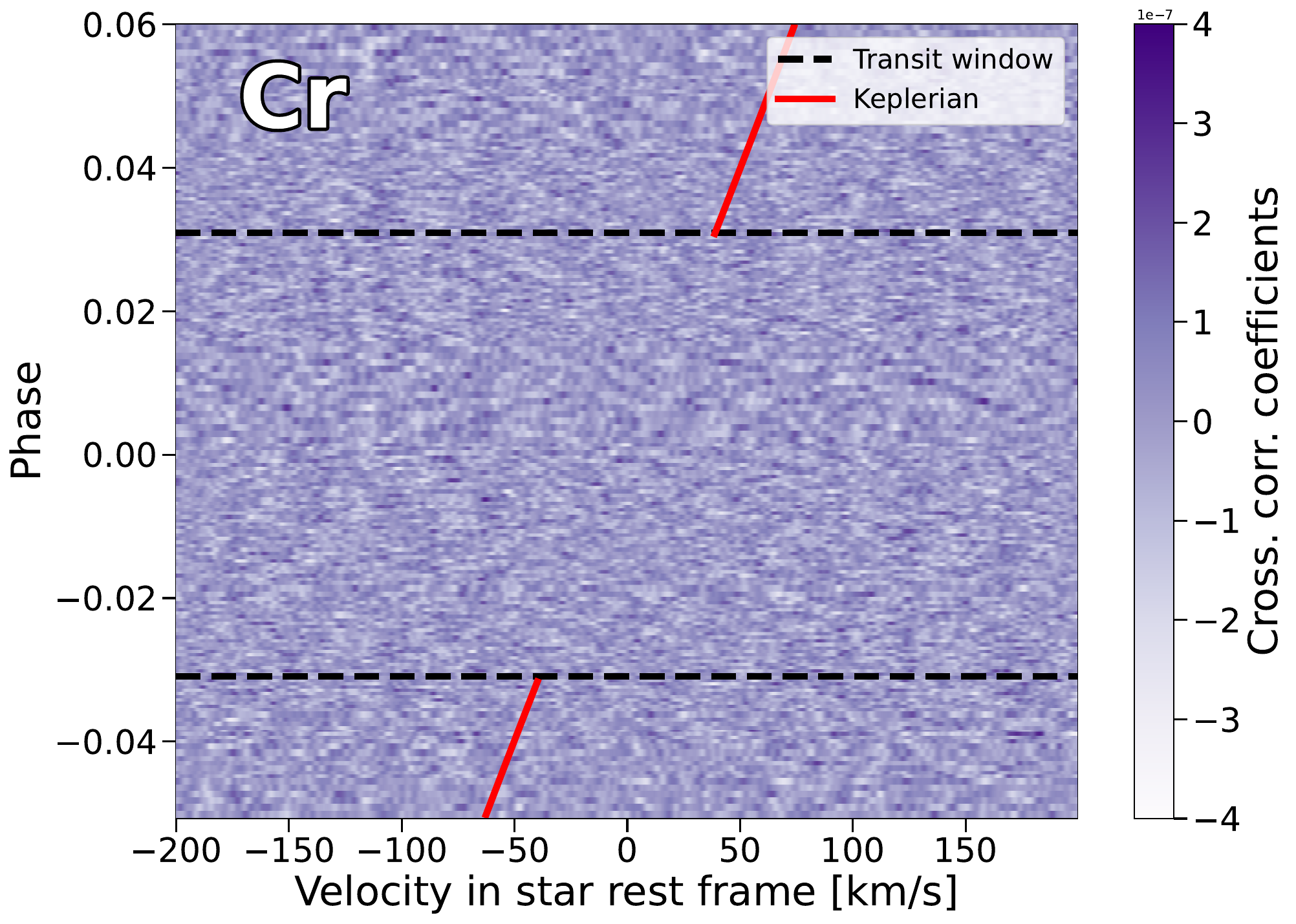}
   \hspace{+0.22cm}
    \includegraphics[width=\columnwidth]{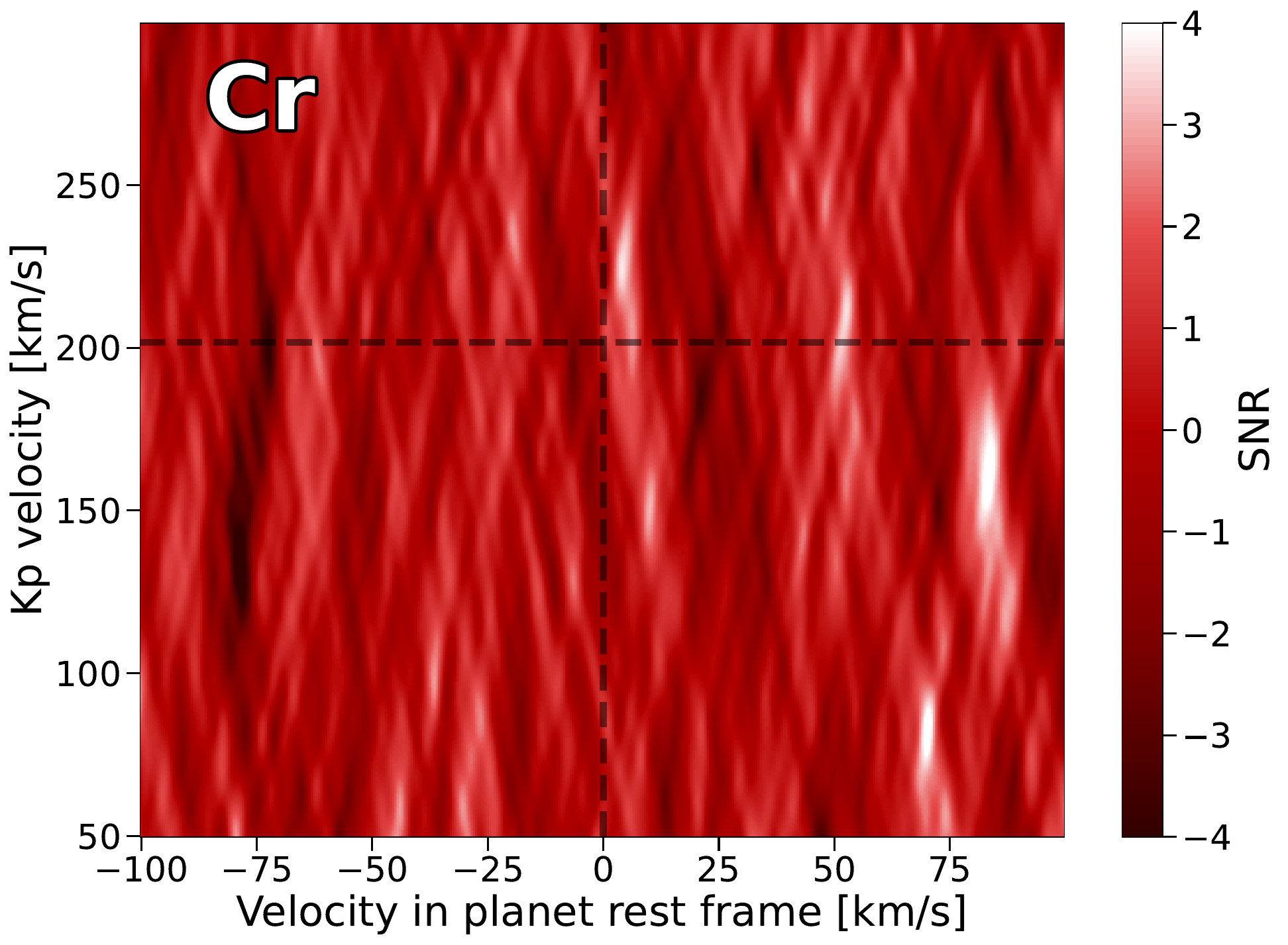}
    
   \caption{NIRPS non-detections,~\ref{Figure:NIRPS_non_detections} continued.}
\end{figure*}
\begin{figure*}[!h]
   \includegraphics[width=\columnwidth]{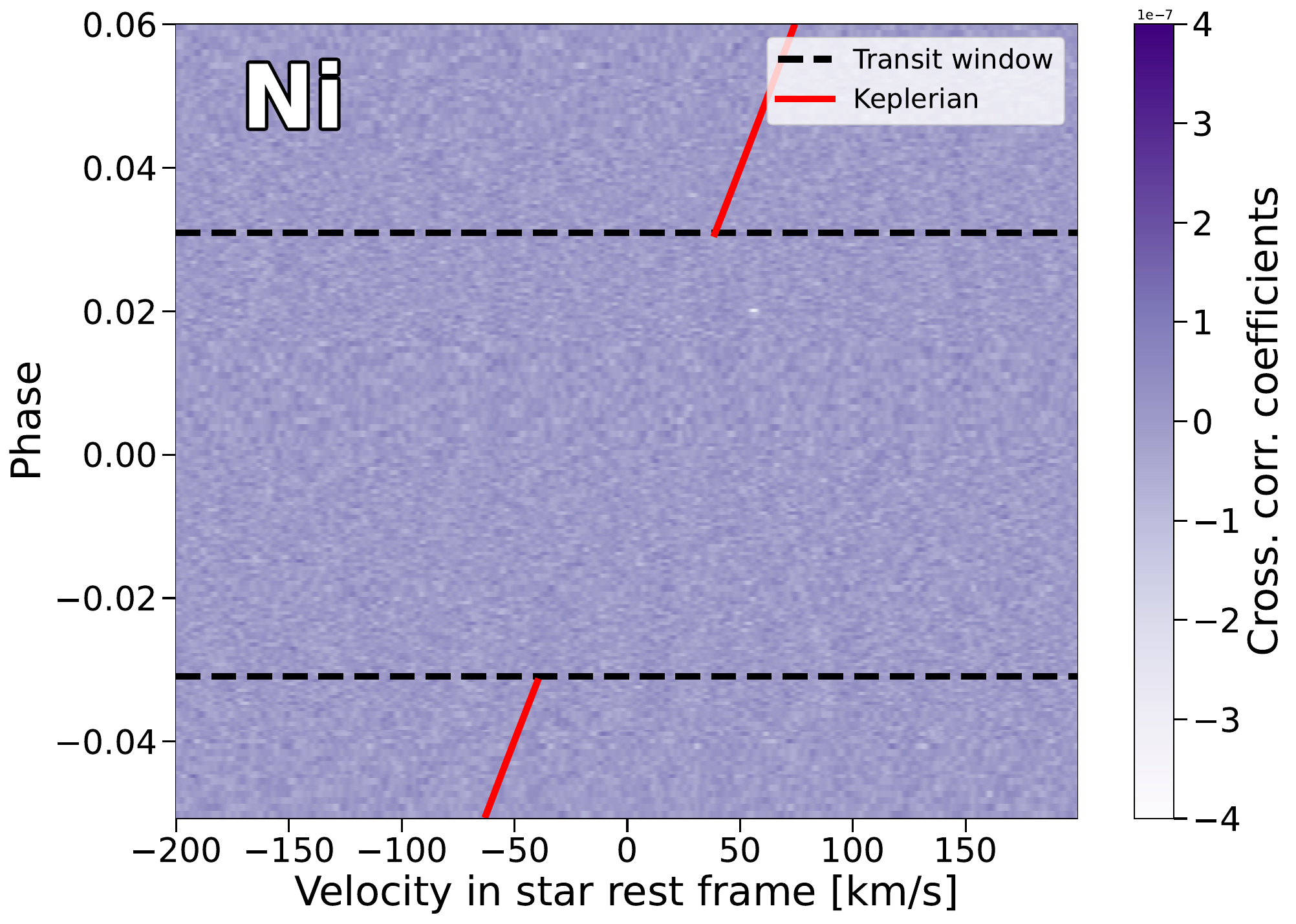}
    \includegraphics[width=\columnwidth]{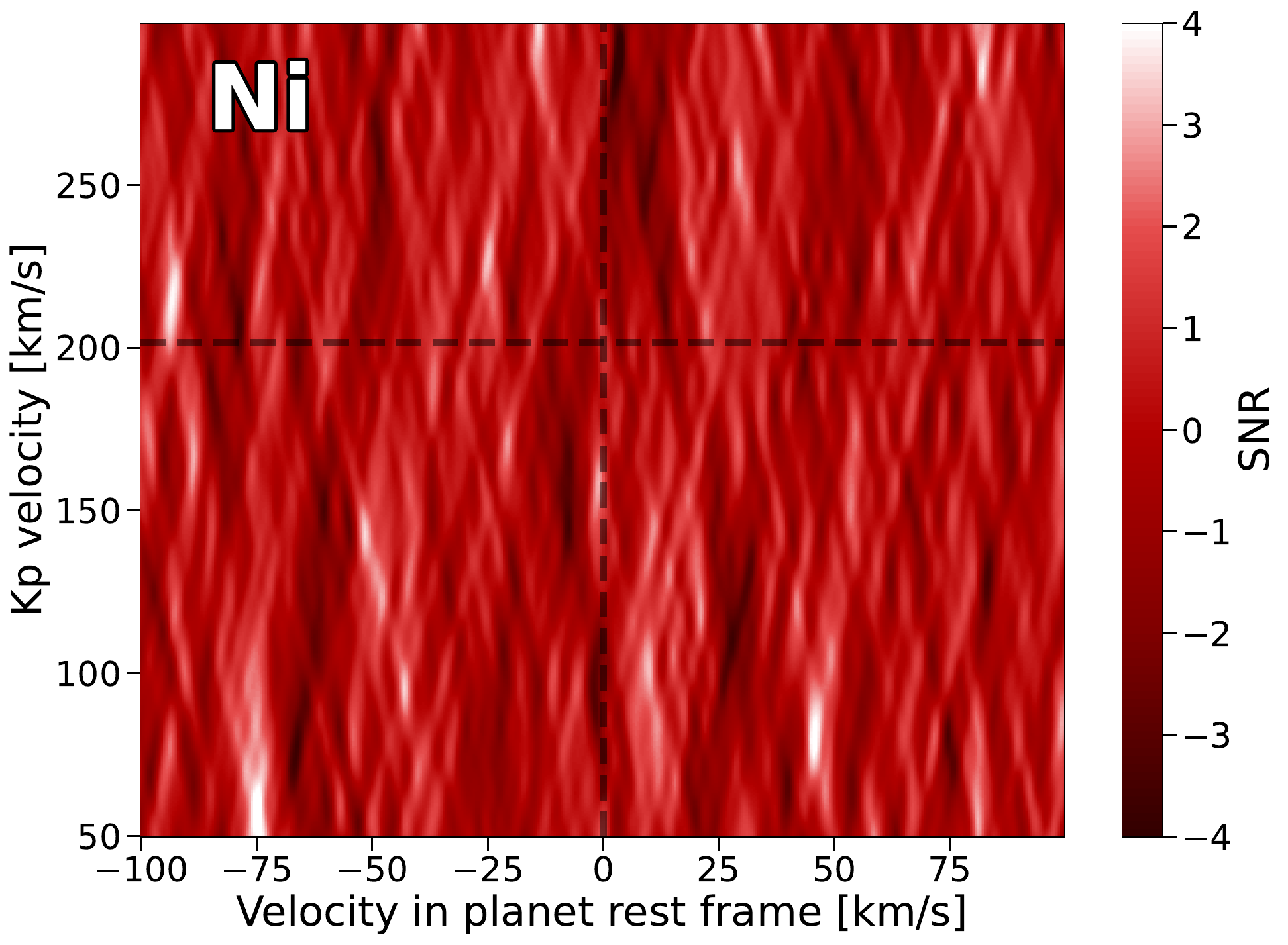}
   \includegraphics[width=\columnwidth]{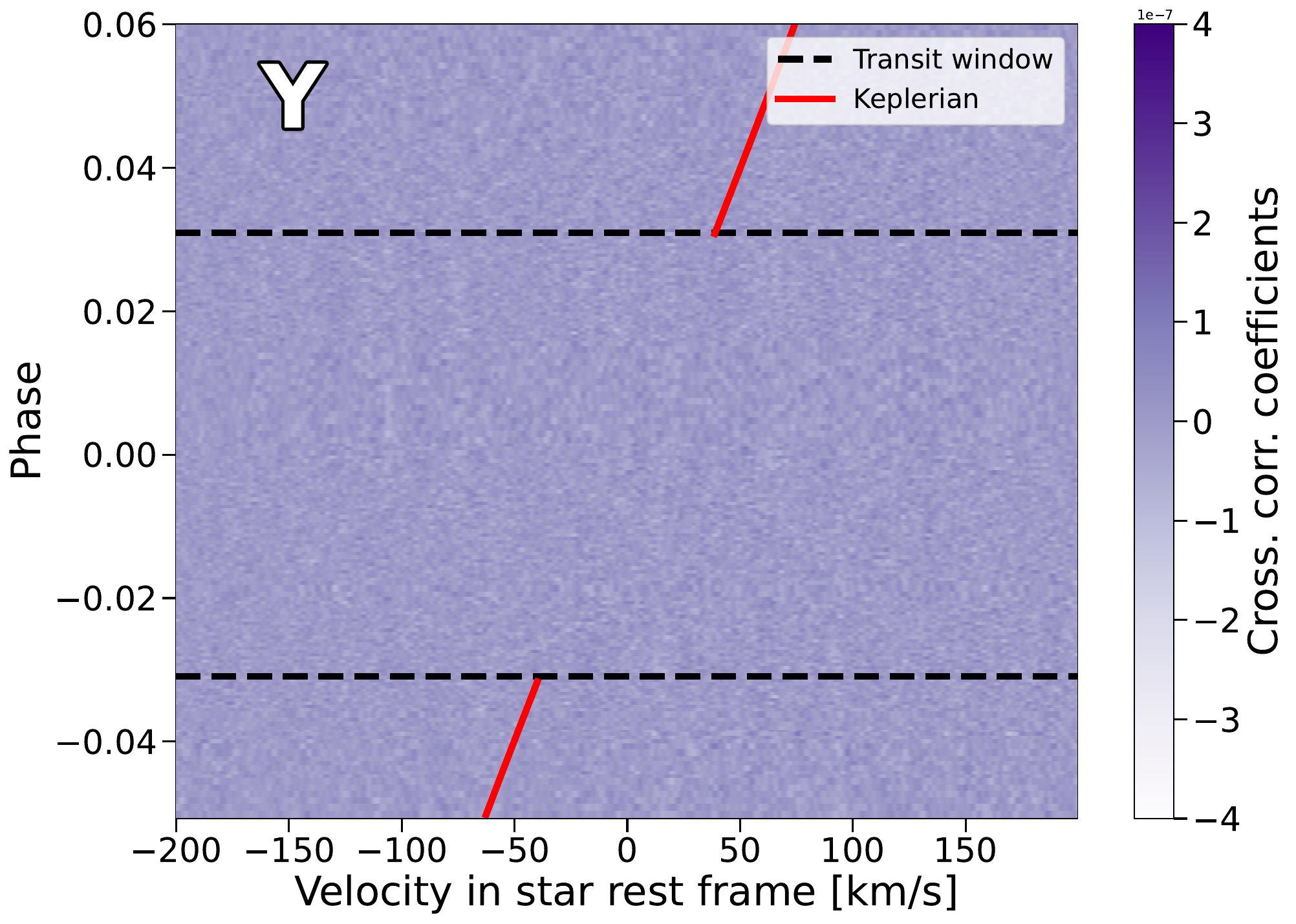}
   \includegraphics[width=\columnwidth]{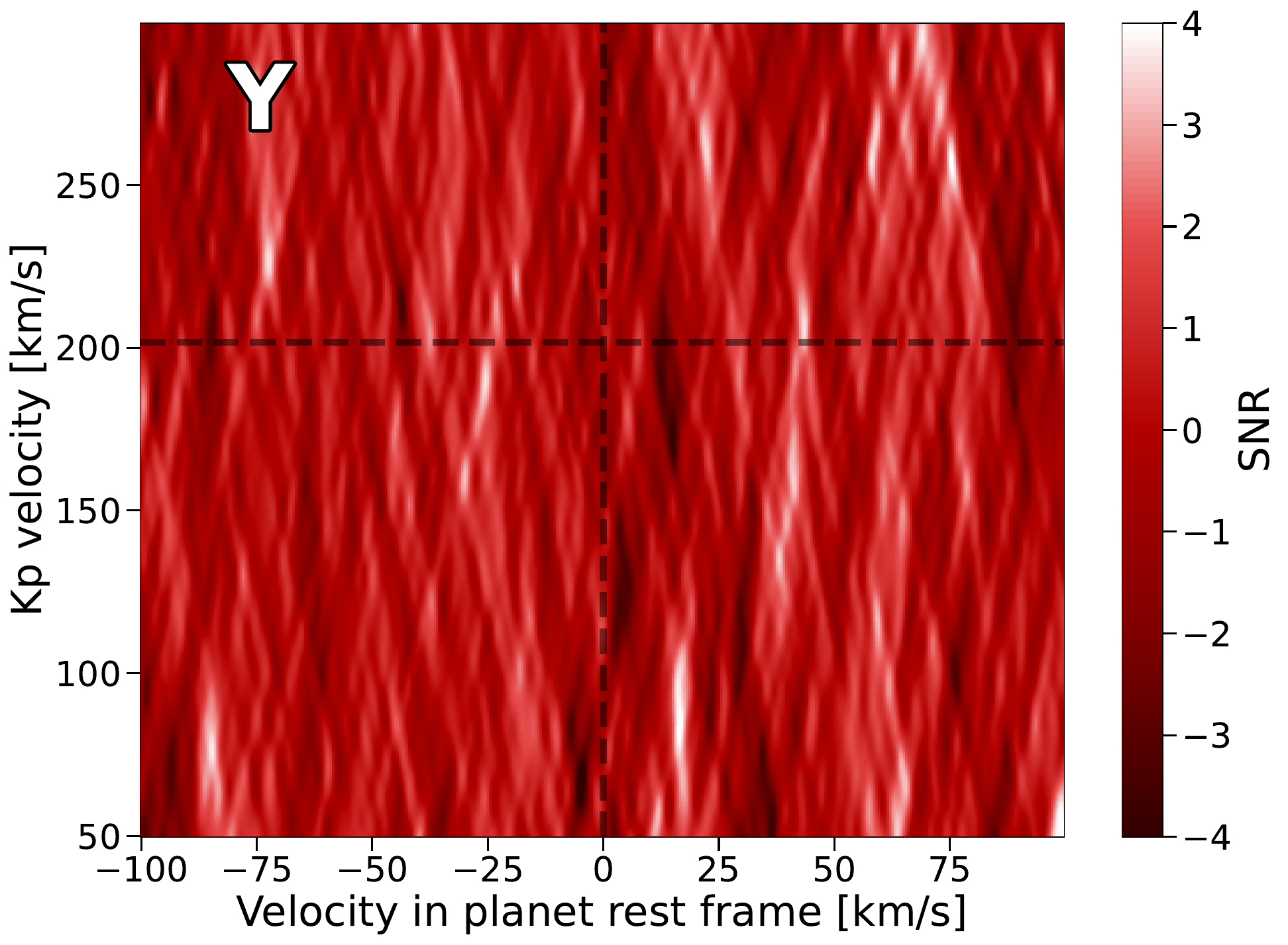} 
   \includegraphics[width=\columnwidth]{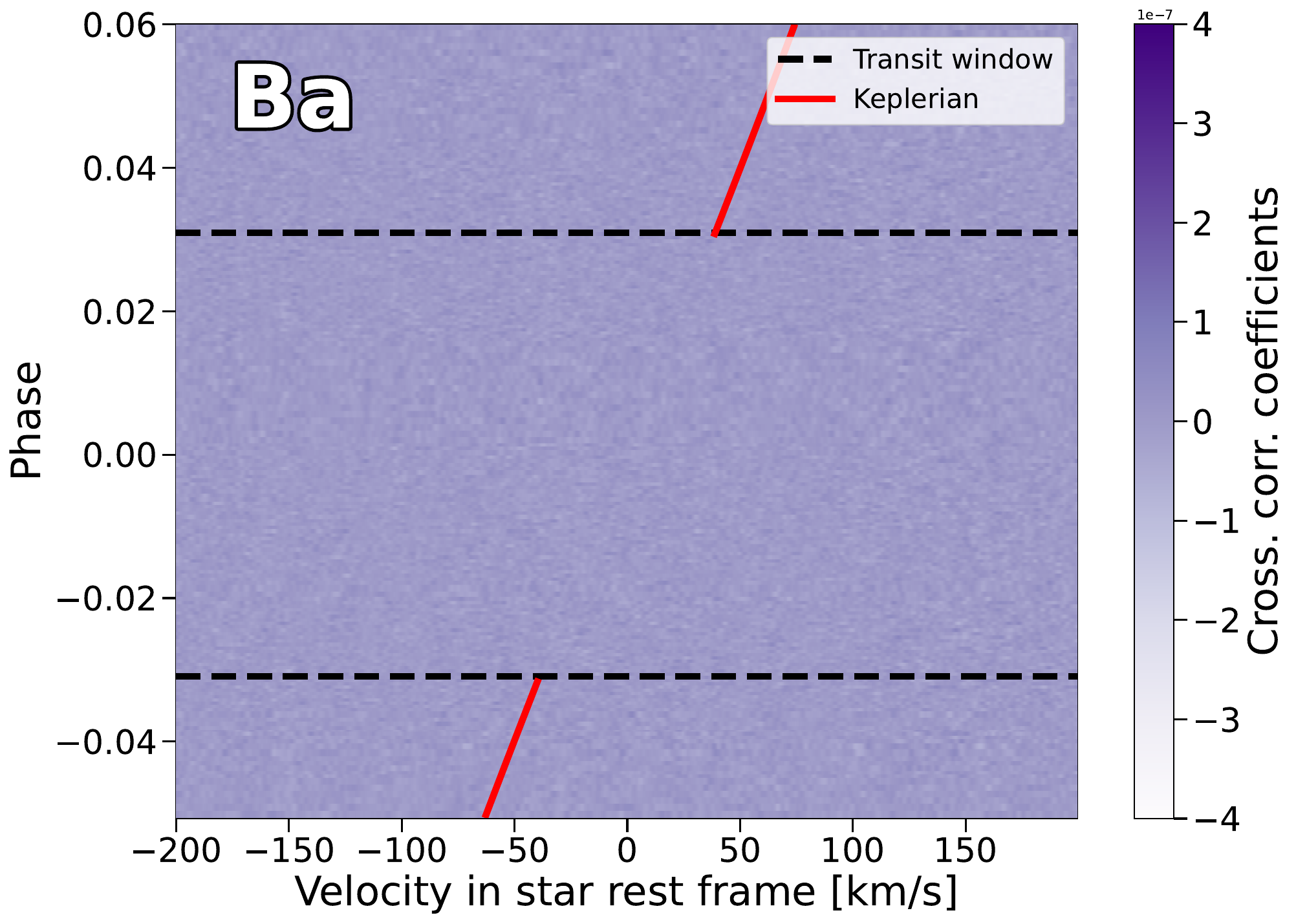}
   \hspace{+0.22cm}
    \includegraphics[width=\columnwidth]{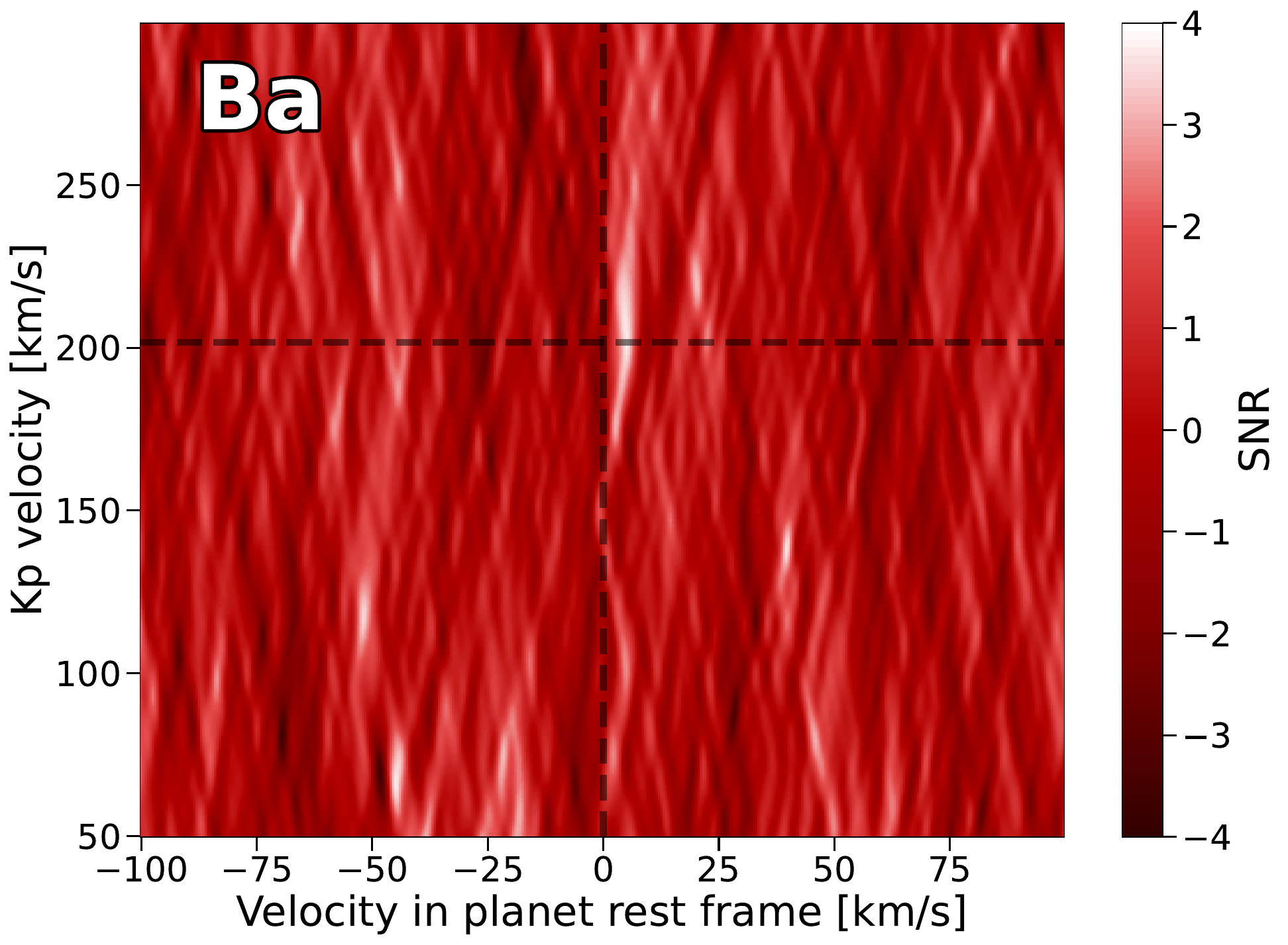} 
   \caption{NIRPS non-detections,~\ref{Figure:NIRPS_non_detections} continued.}
\end{figure*}
\begin{figure*}[!h]
   \includegraphics[width=\columnwidth]{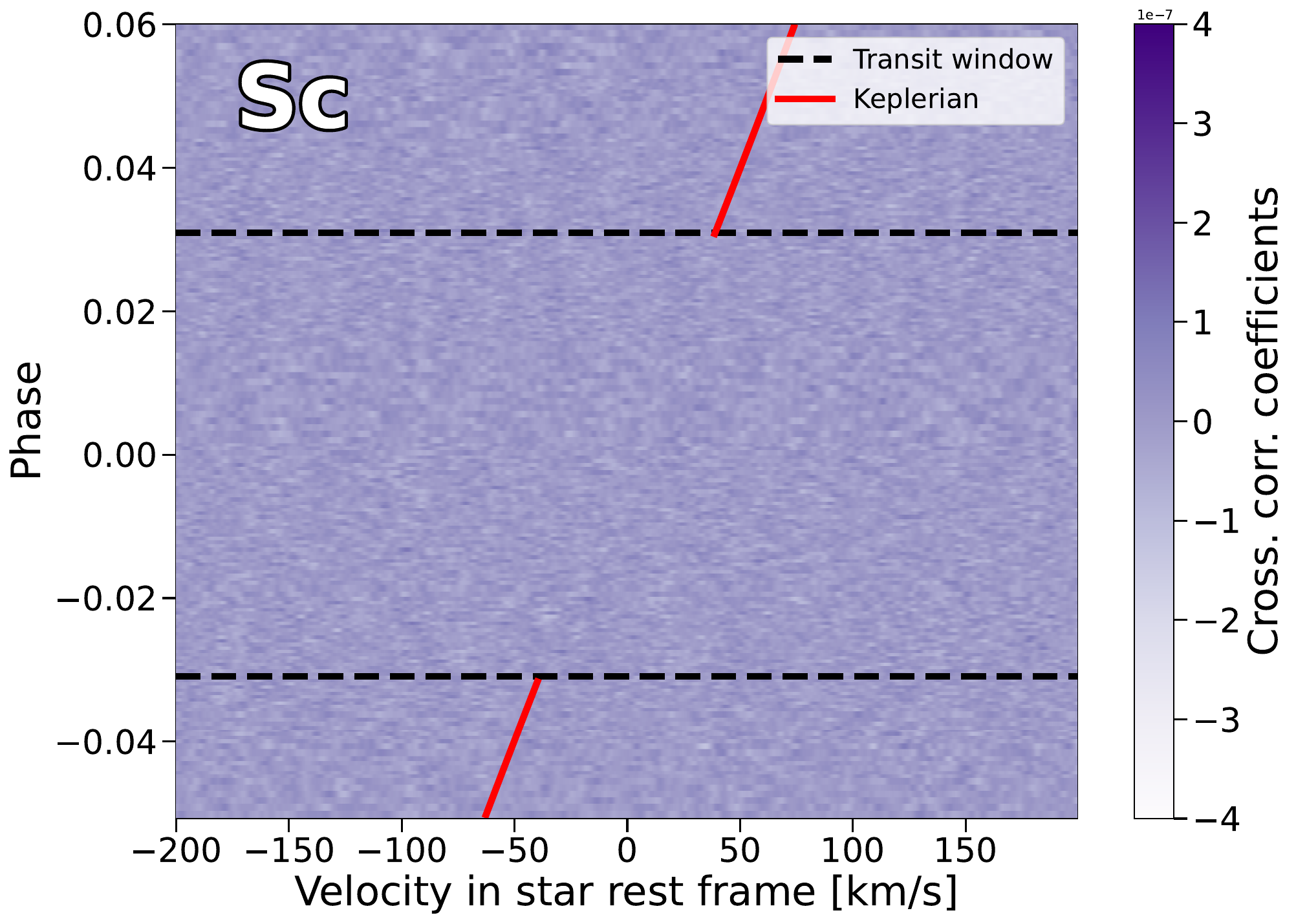}
    \includegraphics[width=\columnwidth]{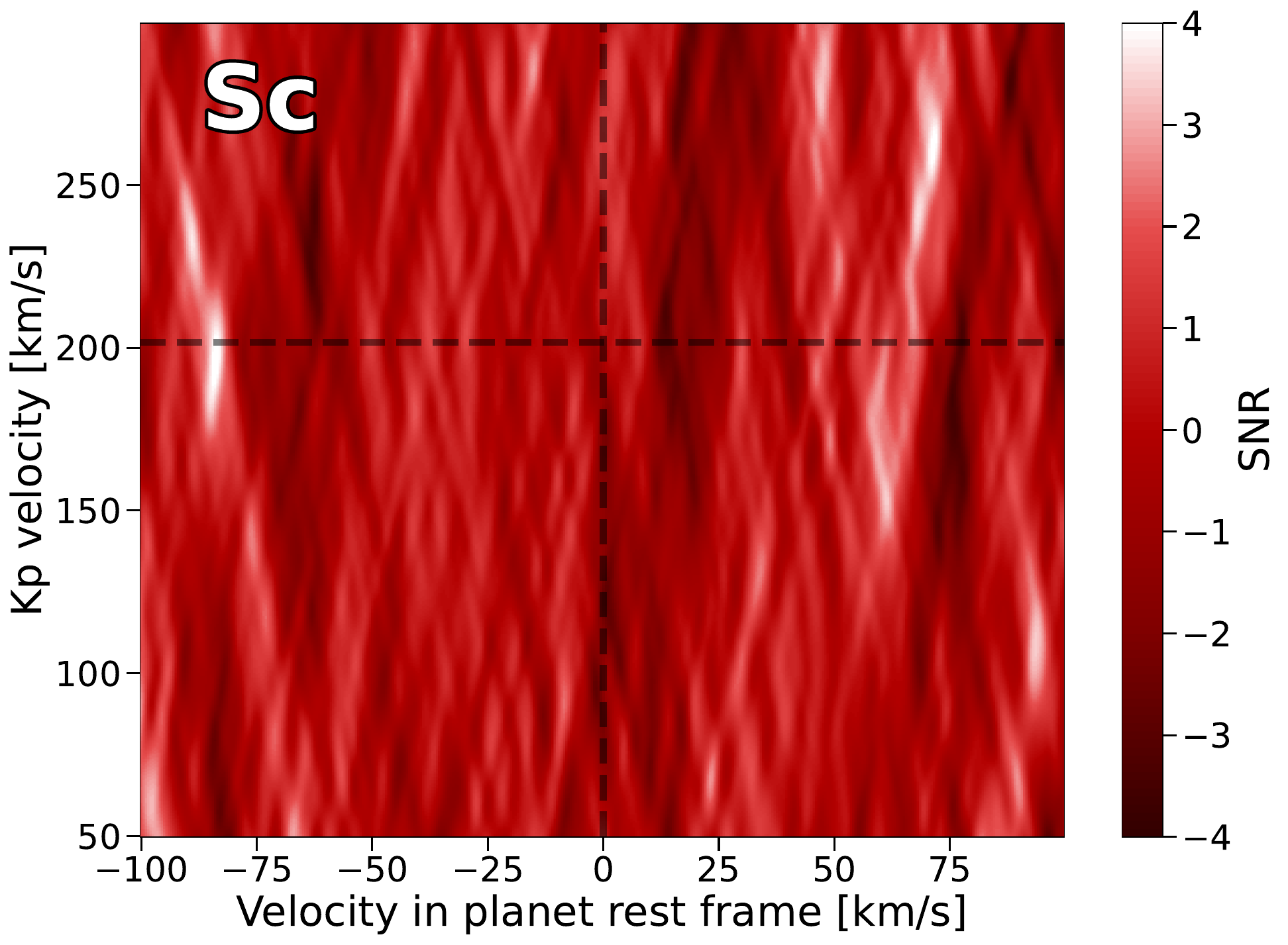}
   \includegraphics[width=\columnwidth]{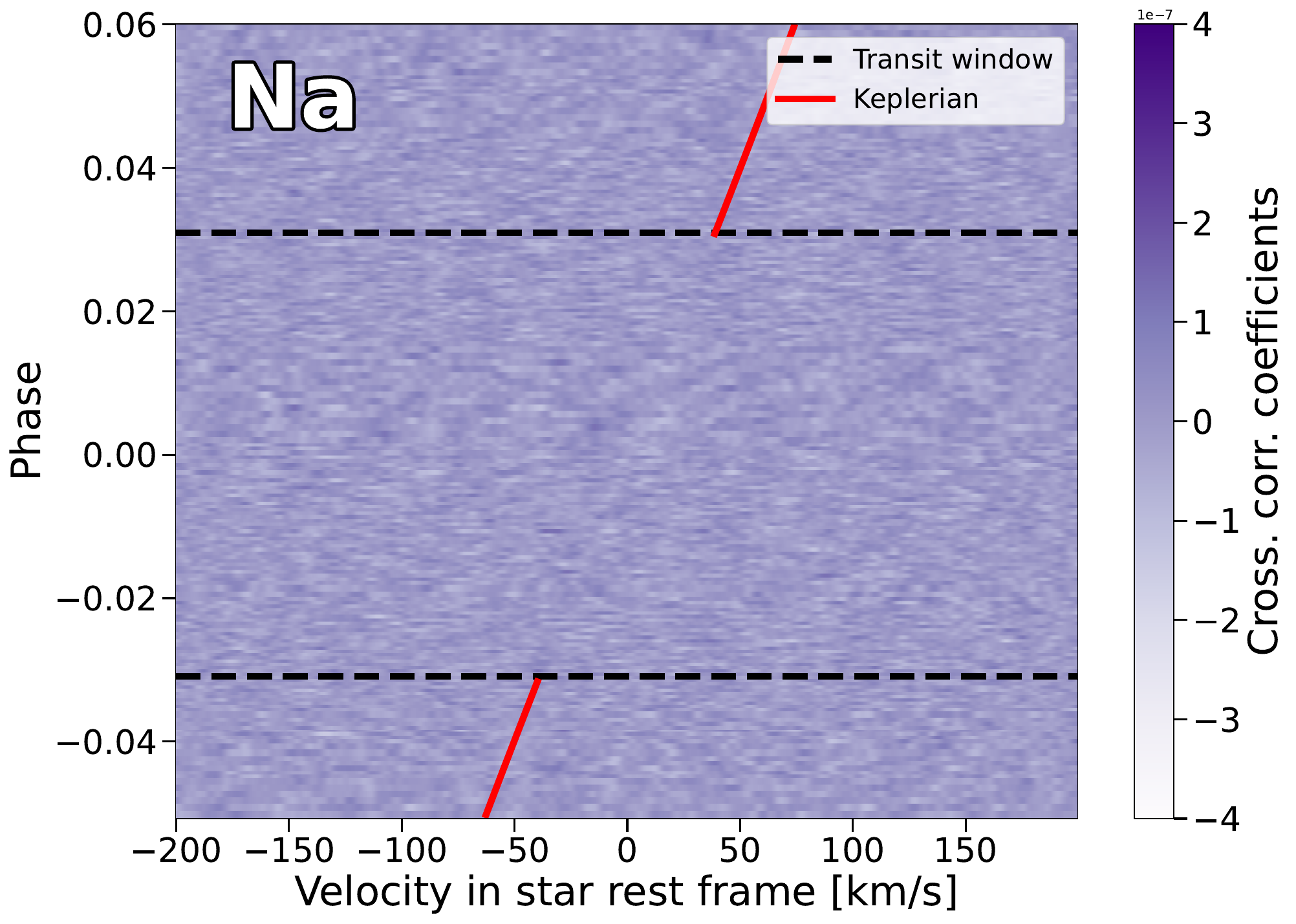}
    \includegraphics[width=\columnwidth]{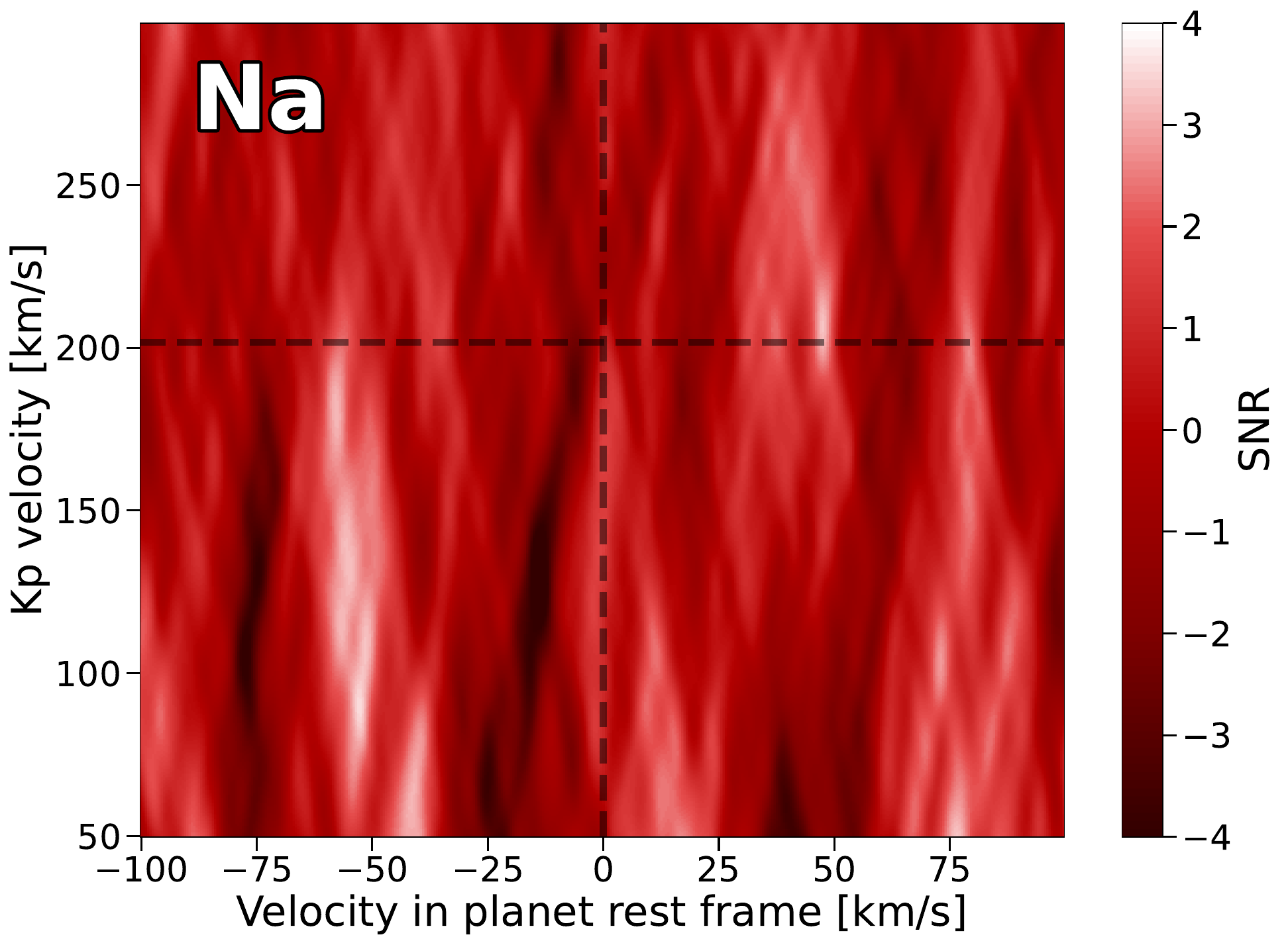}
   \includegraphics[width=\columnwidth]{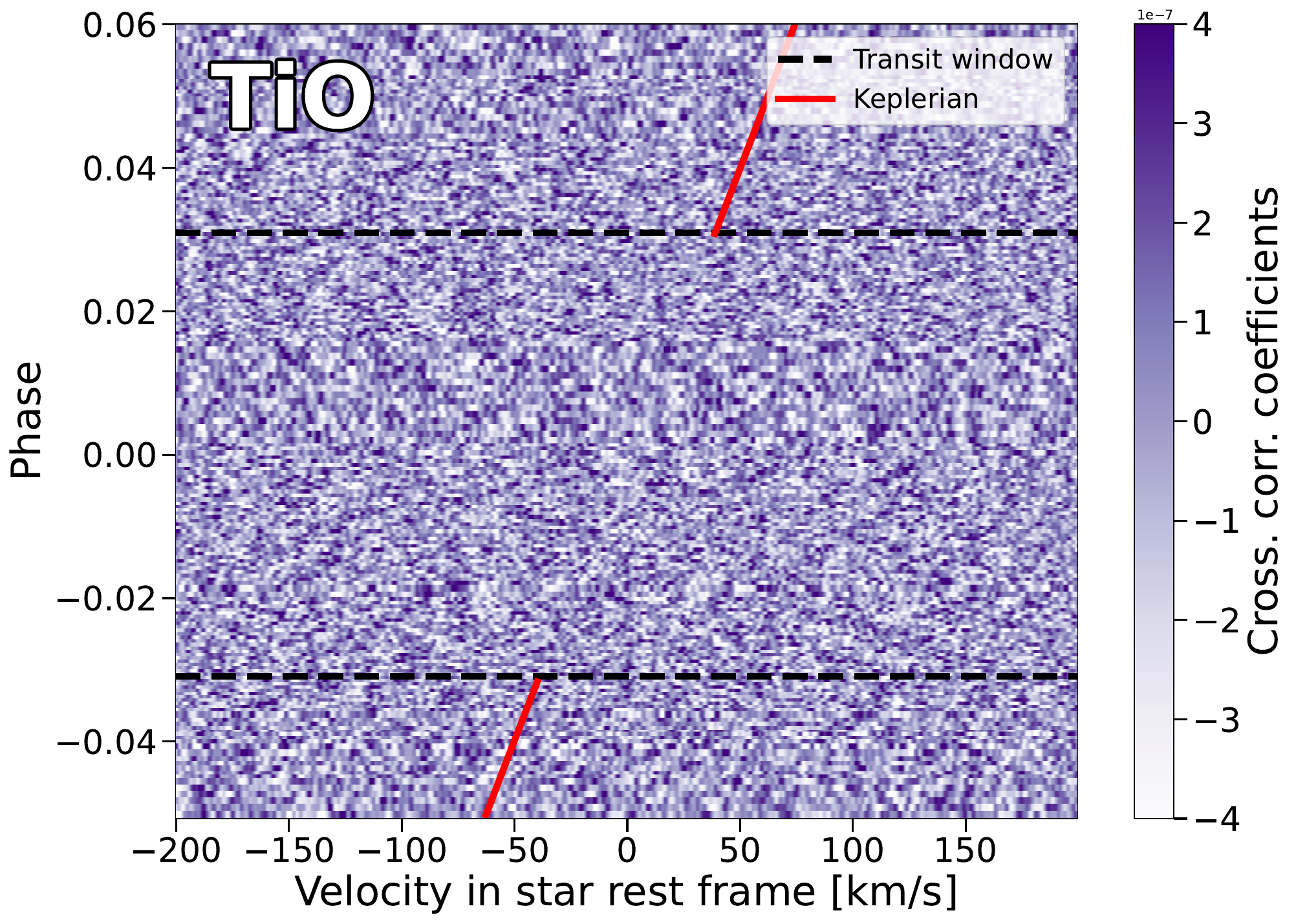}
   \hspace{+0.22cm}
   \includegraphics[width=\columnwidth]{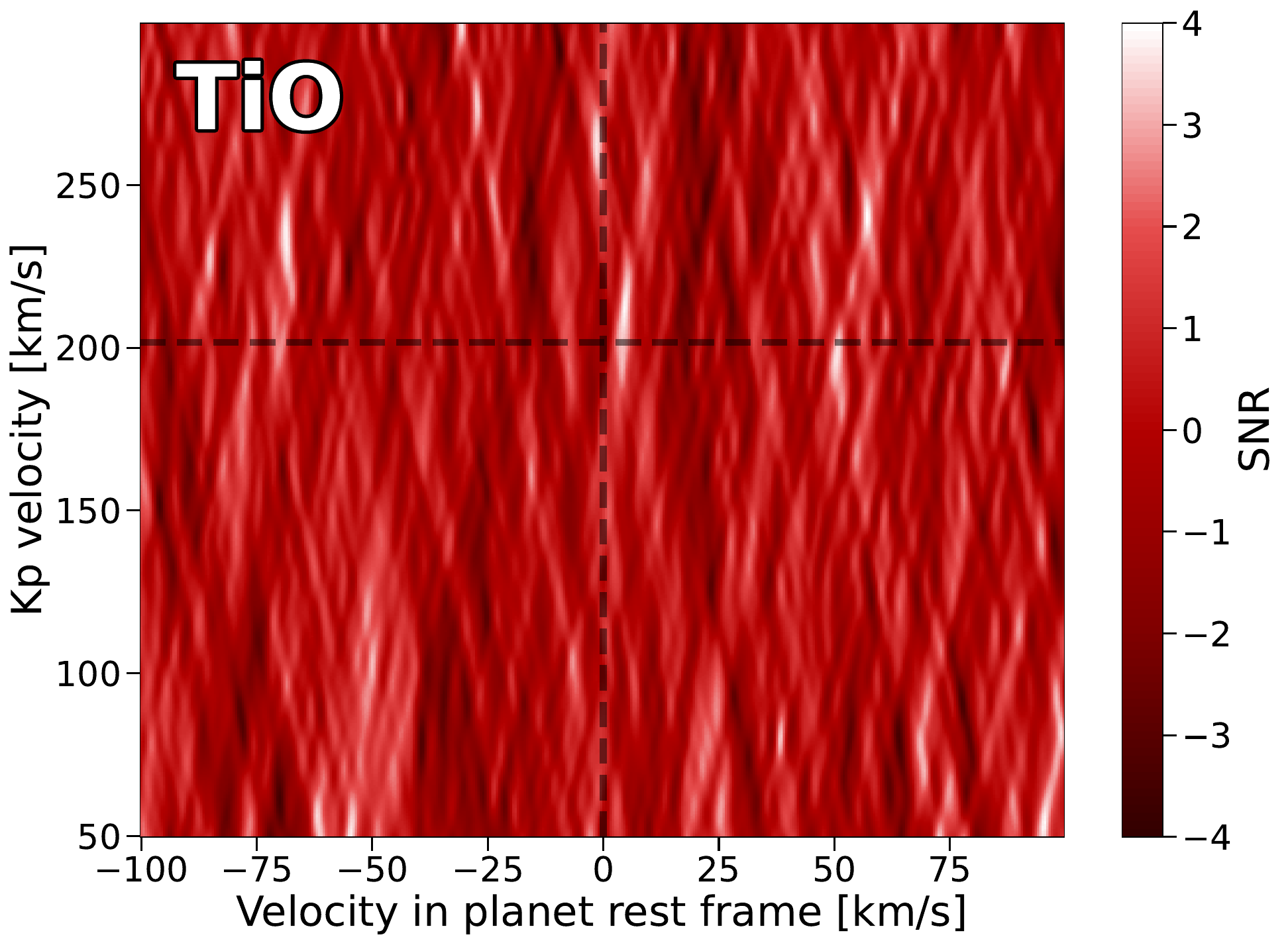} 

   \caption{NIRPS non-detections,~\ref{Figure:NIRPS_non_detections} continued.}
\end{figure*}
\begin{figure*}[!h] \label{Figure:h2o_residuals}
   \includegraphics[width=\columnwidth]{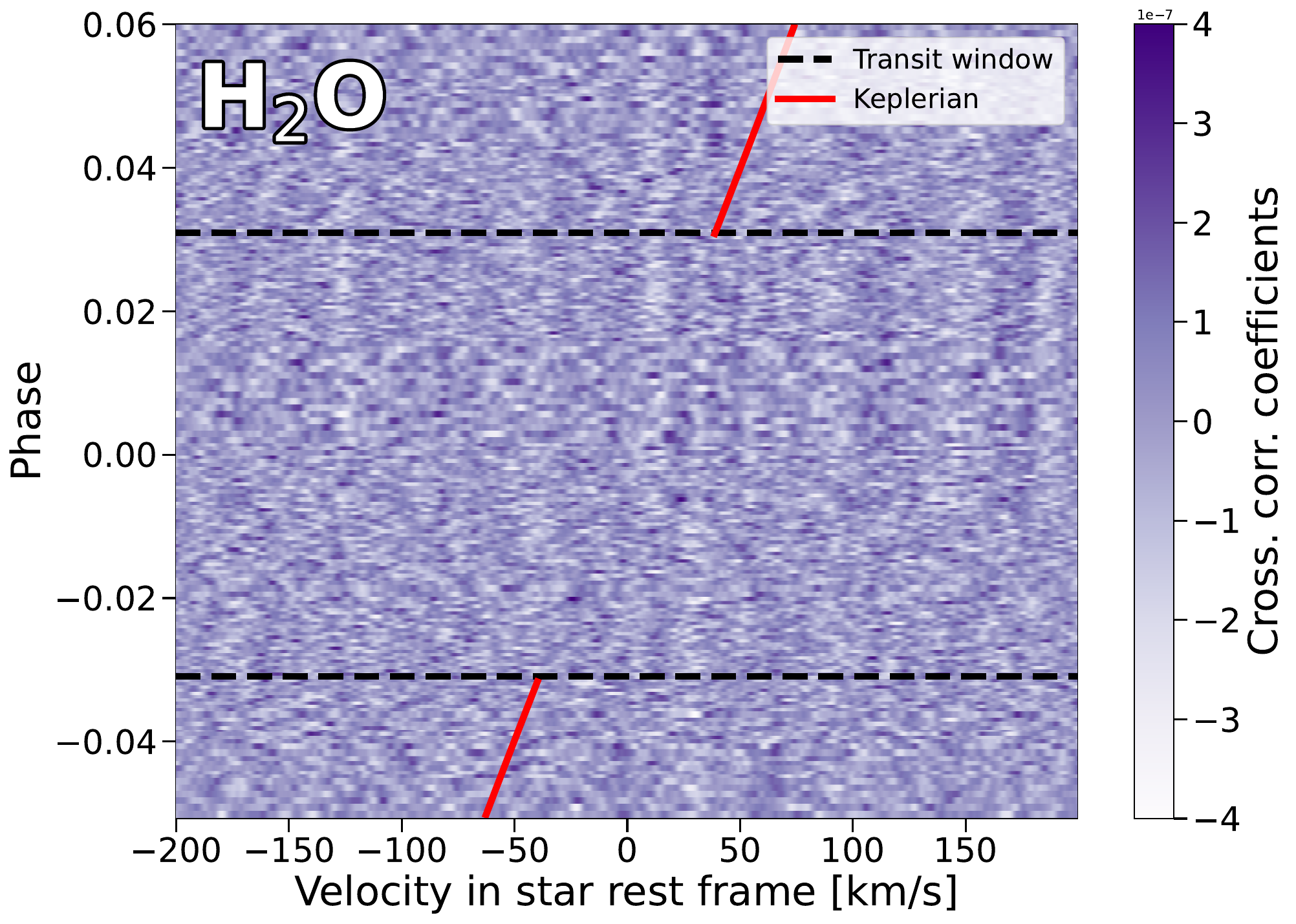}
   \hspace{+0.22cm}
   \includegraphics[width=\columnwidth]{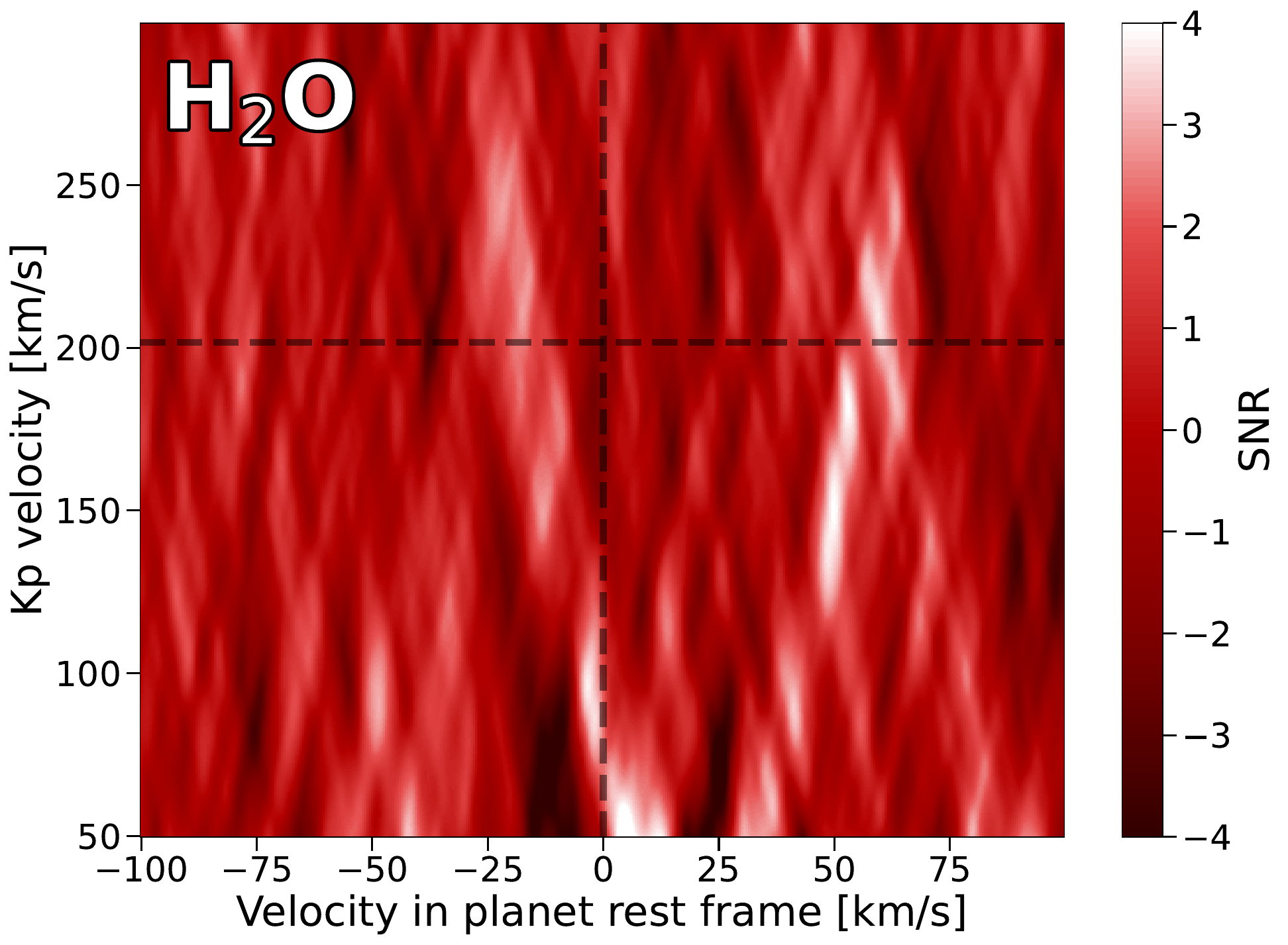} 
   \includegraphics[width=\columnwidth]{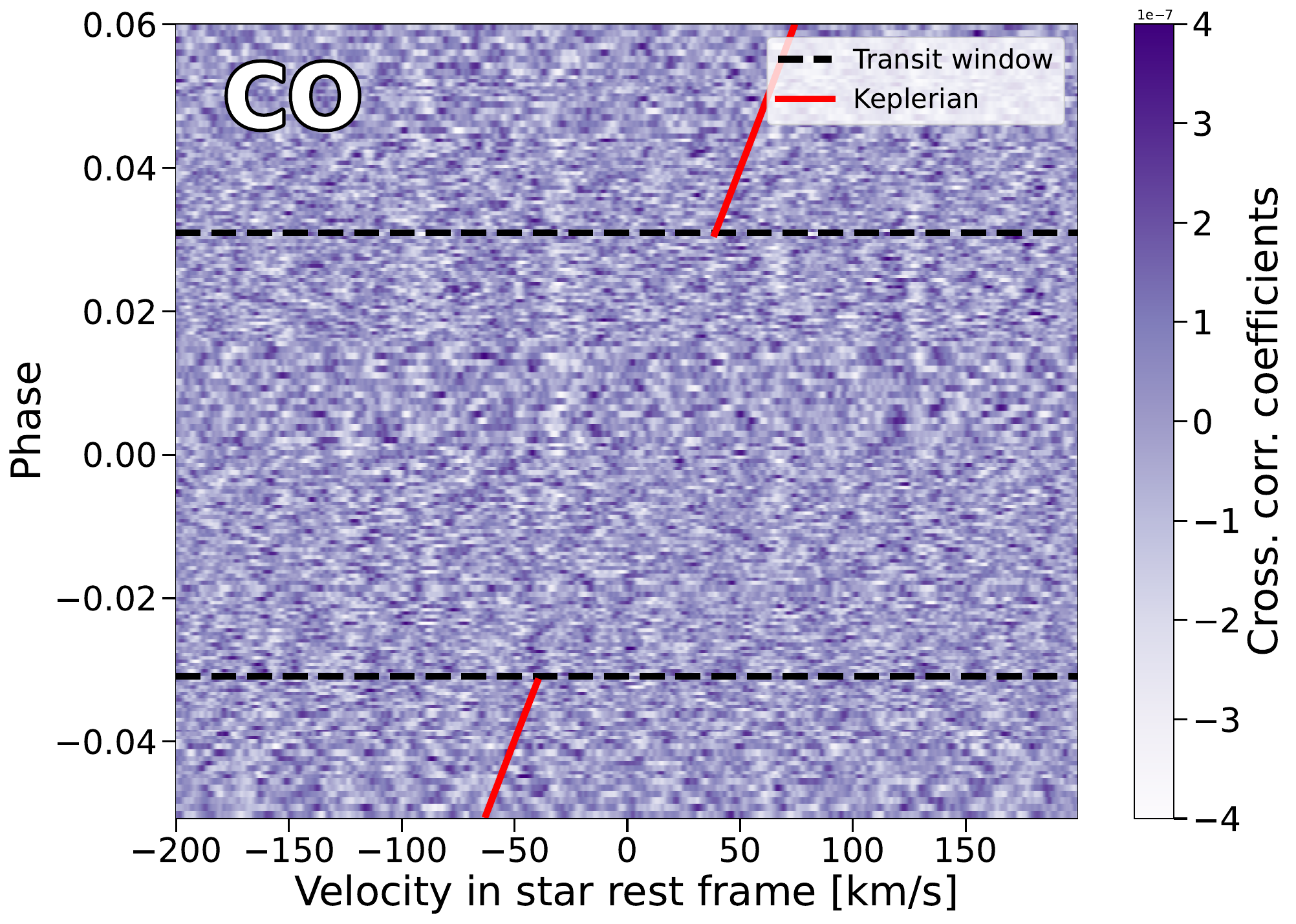}
    \includegraphics[width=\columnwidth]{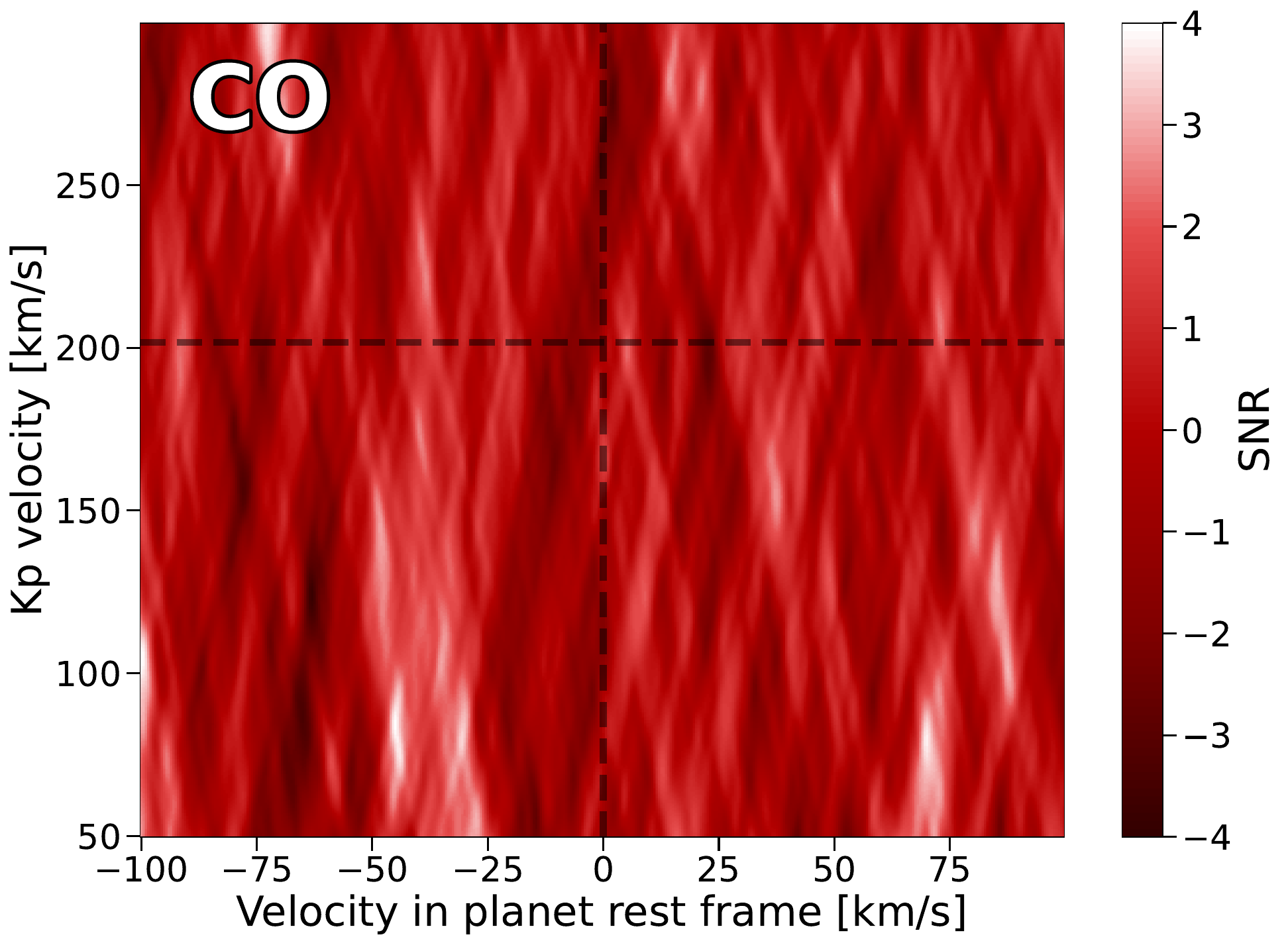}
   \includegraphics[width=\columnwidth]{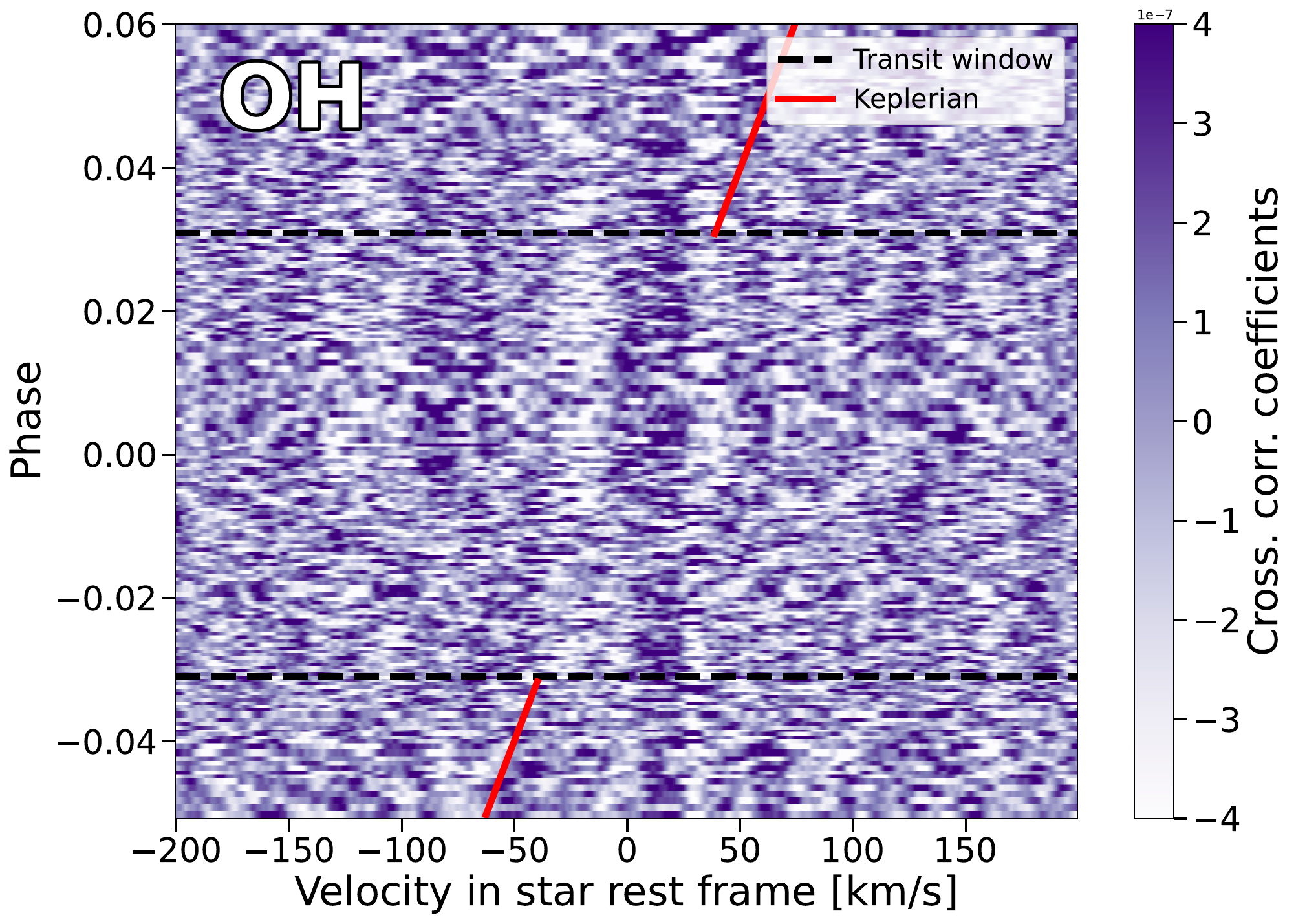}
   \hspace{+0.22cm}
   \includegraphics[width=\columnwidth]{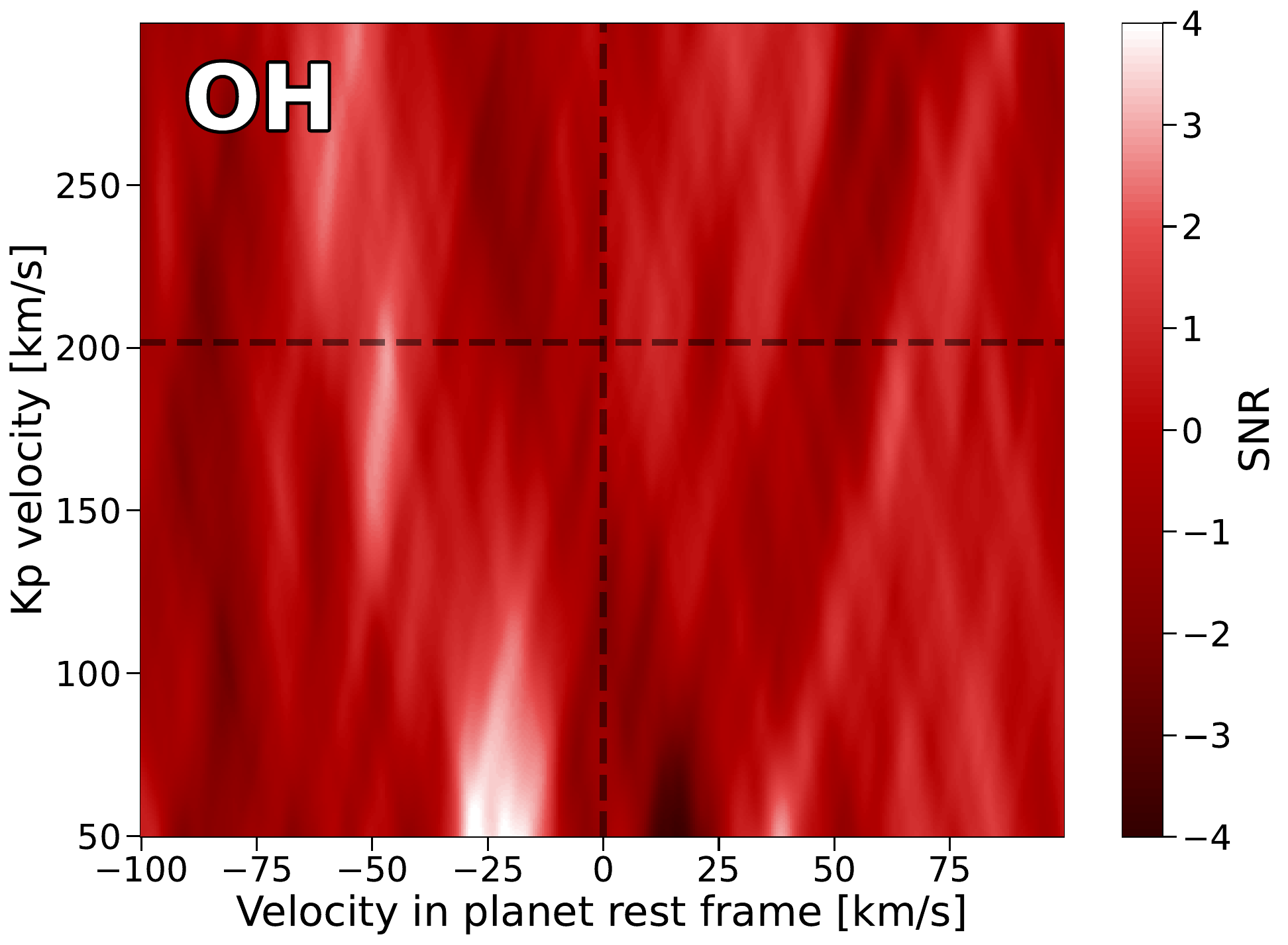} 
    
   \caption{NIRPS non-detections,~\ref{Figure:NIRPS_non_detections} continued. Remnants of the telluric correction (beyond and within in-transit phases) are visible in the water and OH cross-correlation maps.}
\end{figure*}

\clearpage
\section{HARPS non-detections in the transit spectrum of WASP-189b}
\label{Appendix-Sec:HARPS non-detections in the transit spectrum of WASP-189b}

\begin{table*}[!t]
    \begin{minipage}{0.6\textwidth}
        \raggedright
        \caption{Line-contrast upper limits calculated for non-detections of listed chemical species in the HARPS transmission spectrum of WASP-189b.}
        \label{Table:HARPS_upper_limits}
        \begin{tabular}{lc}
            \toprule
            \toprule
            Species & 3-$\sigma$ line-contrast upper limit (HARPS) \\
            \midrule
            Fe & 4.2 $\times$ 10$^{-5}$\\
            Ti & 3.7 $\times$ 10$^{-5}$\\  
            V  & 3.2 $\times$ 10$^{-5}$\\
            Mn & 2.4 $\times$ 10$^{-5}$\\
            Mg & 1.2 $\times$ 10$^{-5}$\\
            Ca & 1.1 $\times$ 10$^{-5}$\\
            Cr & 2.8 $\times$ 10$^{-5}$\\
            Ni & 1.8 $\times$ 10$^{-5}$\\
            Y  & 5.1 $\times$ 10$^{-6}$\\
            Ba & 2.0 $\times$ 10$^{-6}$\\
            Sc & 7.4 $\times$ 10$^{-6}$\\
            Na & 2.5 $\times$ 10$^{-6}$\\
            Fe$^+$ & 2.9 $\times$ 10$^{-7}$\\
            Ti$^+$ & 1.2 $\times$ 10$^{-5}$\\
            TiO & 8.4 $\times$ 10$^{-6}$\\
            H$_2$O & 5.3 $\times$ 10$^{-7}$\\
            OH & 4.2 $\times$ 10$^{-6}$\\
            \bottomrule
            \bottomrule
        \end{tabular}
        \vspace{0.3cm}
        
    \end{minipage}
\end{table*}

\FloatBarrier  

\begin{figure*}[!h]
    \centering
    \includegraphics[width=0.47\linewidth]{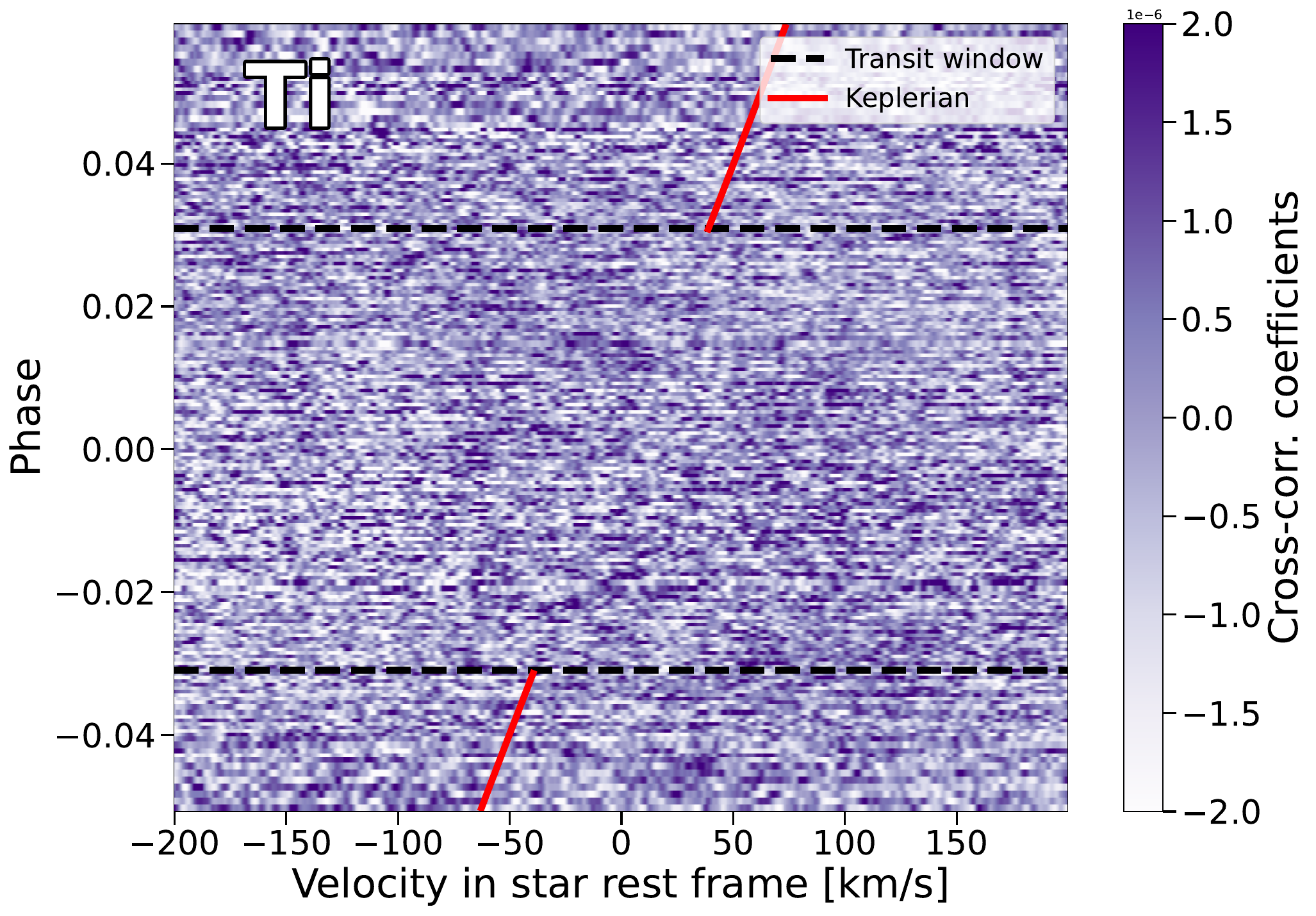}
    \hspace{0.22cm}
    \includegraphics[width=0.47\linewidth]{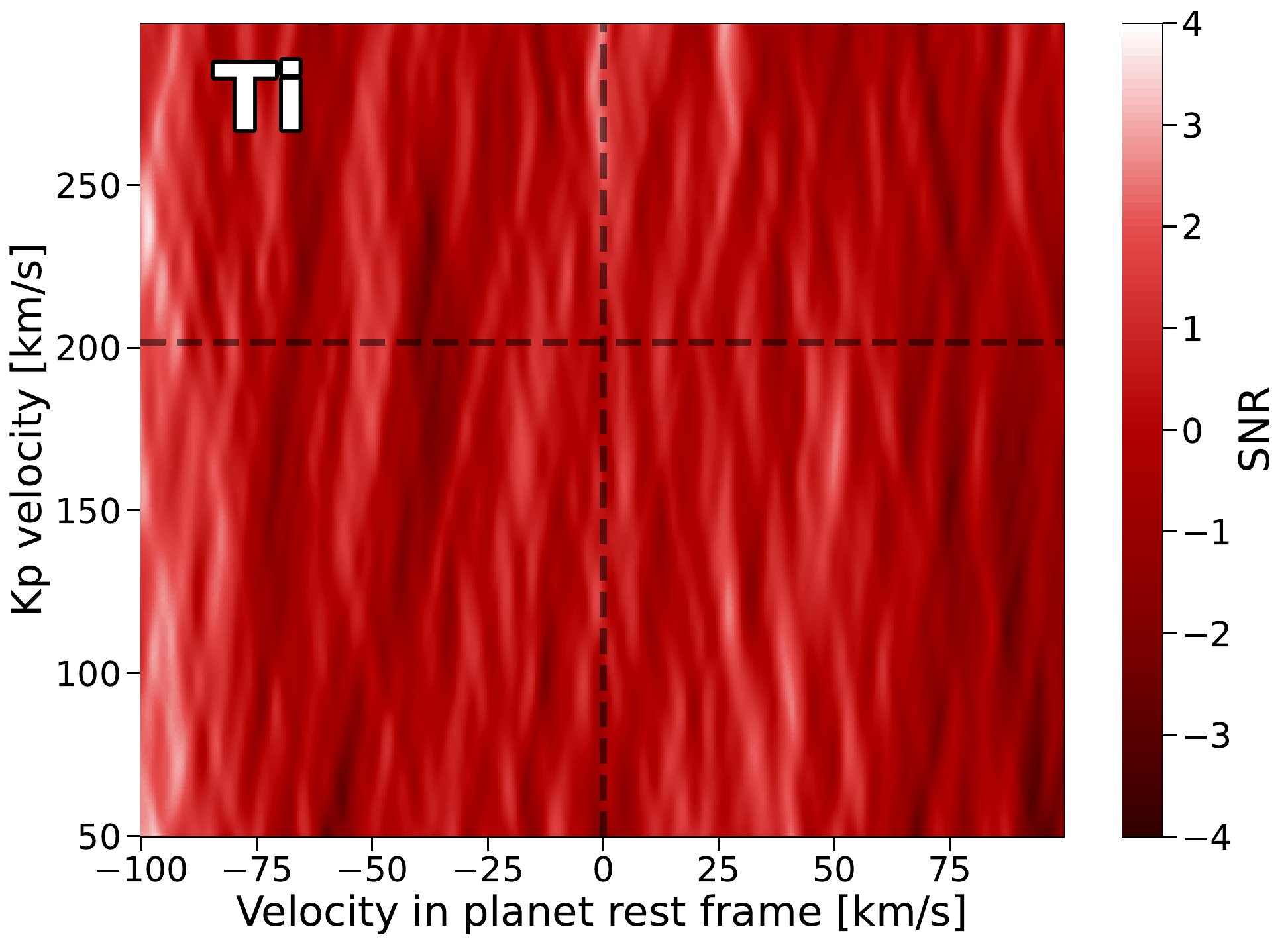}

    \includegraphics[width=0.47\linewidth]{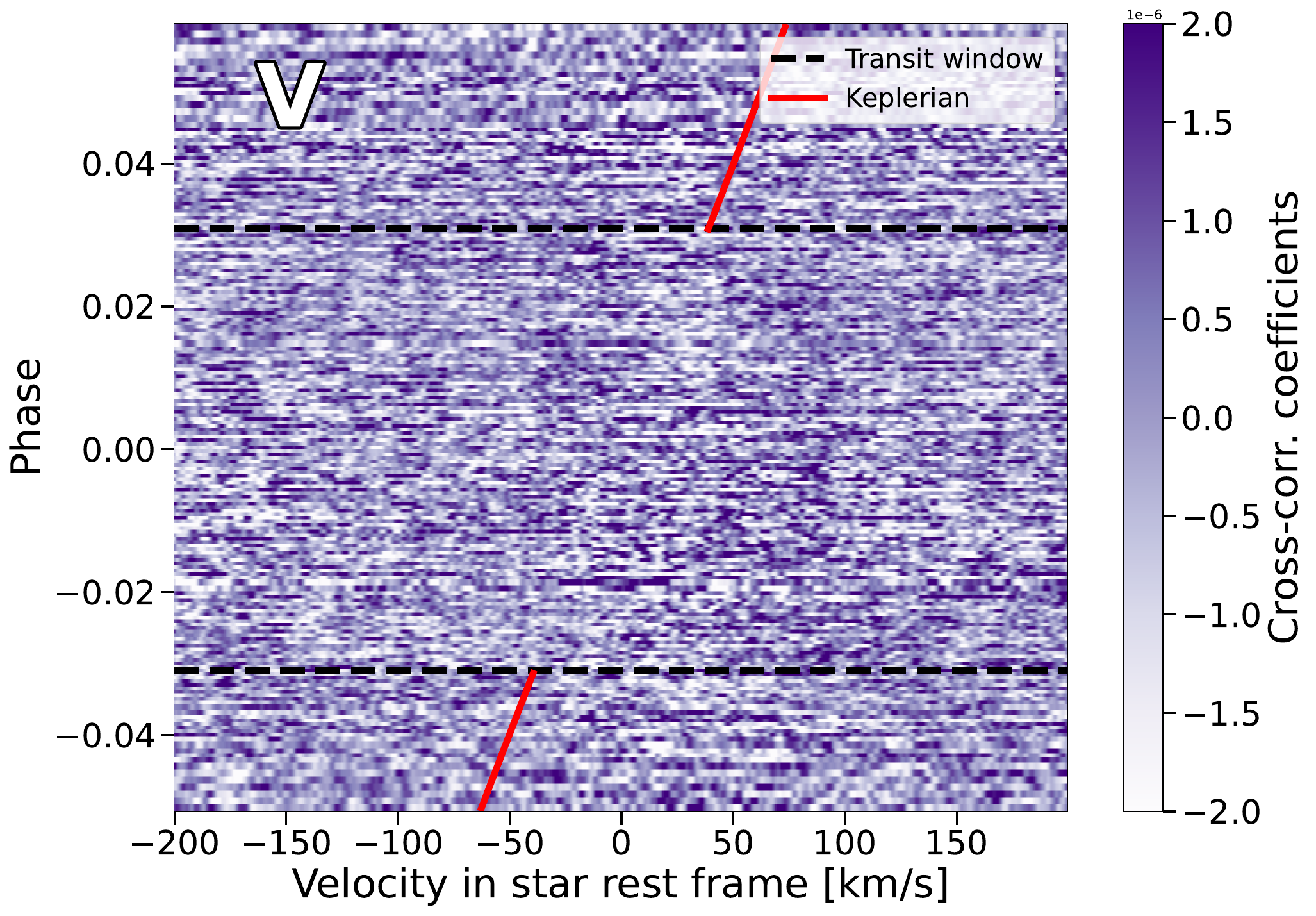}
    \hspace{0.22cm}
    \includegraphics[width=0.47\linewidth]{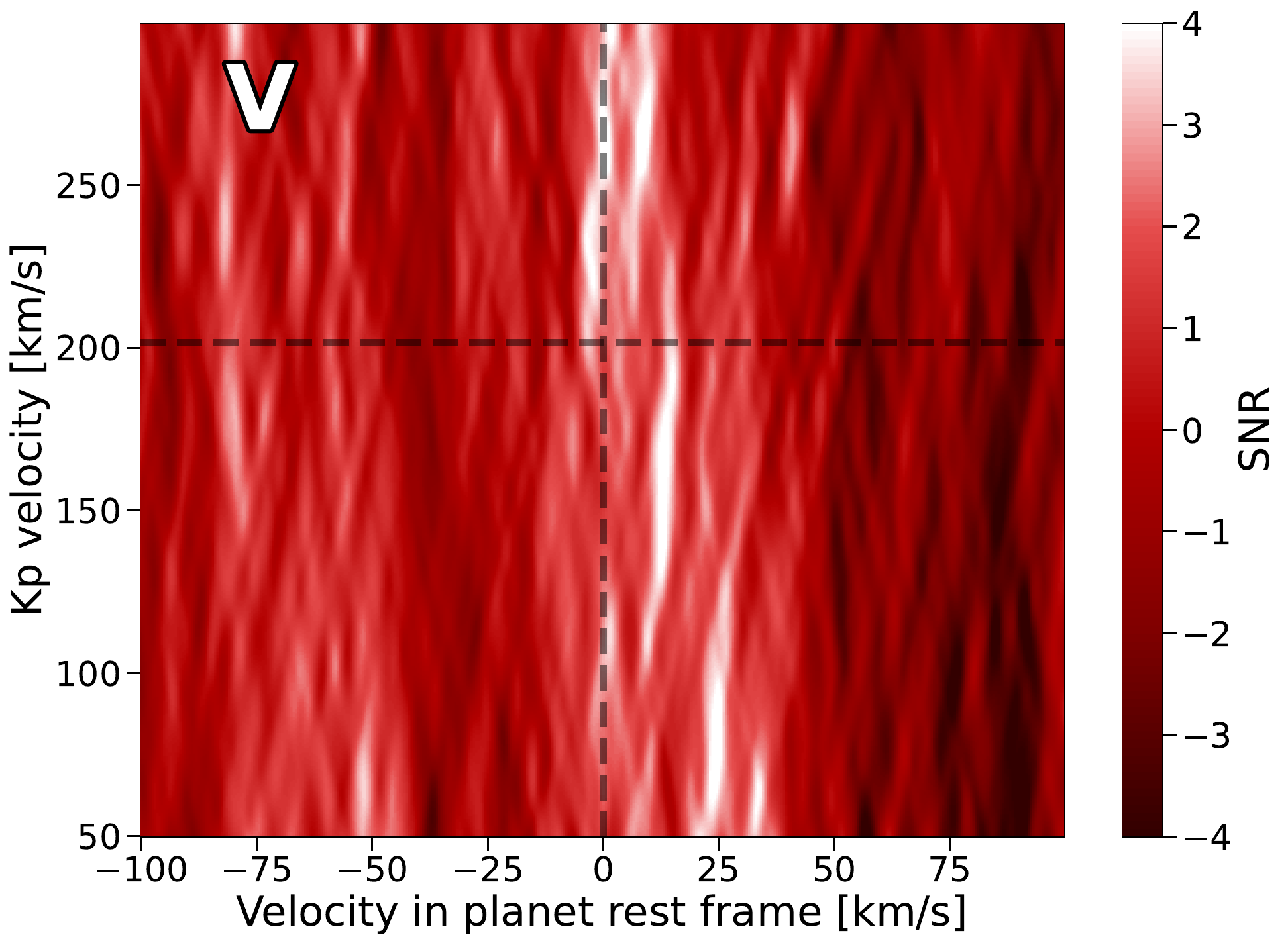}

    \caption{HARPS cross-correlation results showing non-detected chemical species.}
    \label{Figure:HARPS_non_detections}
\end{figure*}

\begin{figure*}[!h] 
   \includegraphics[width=\columnwidth]{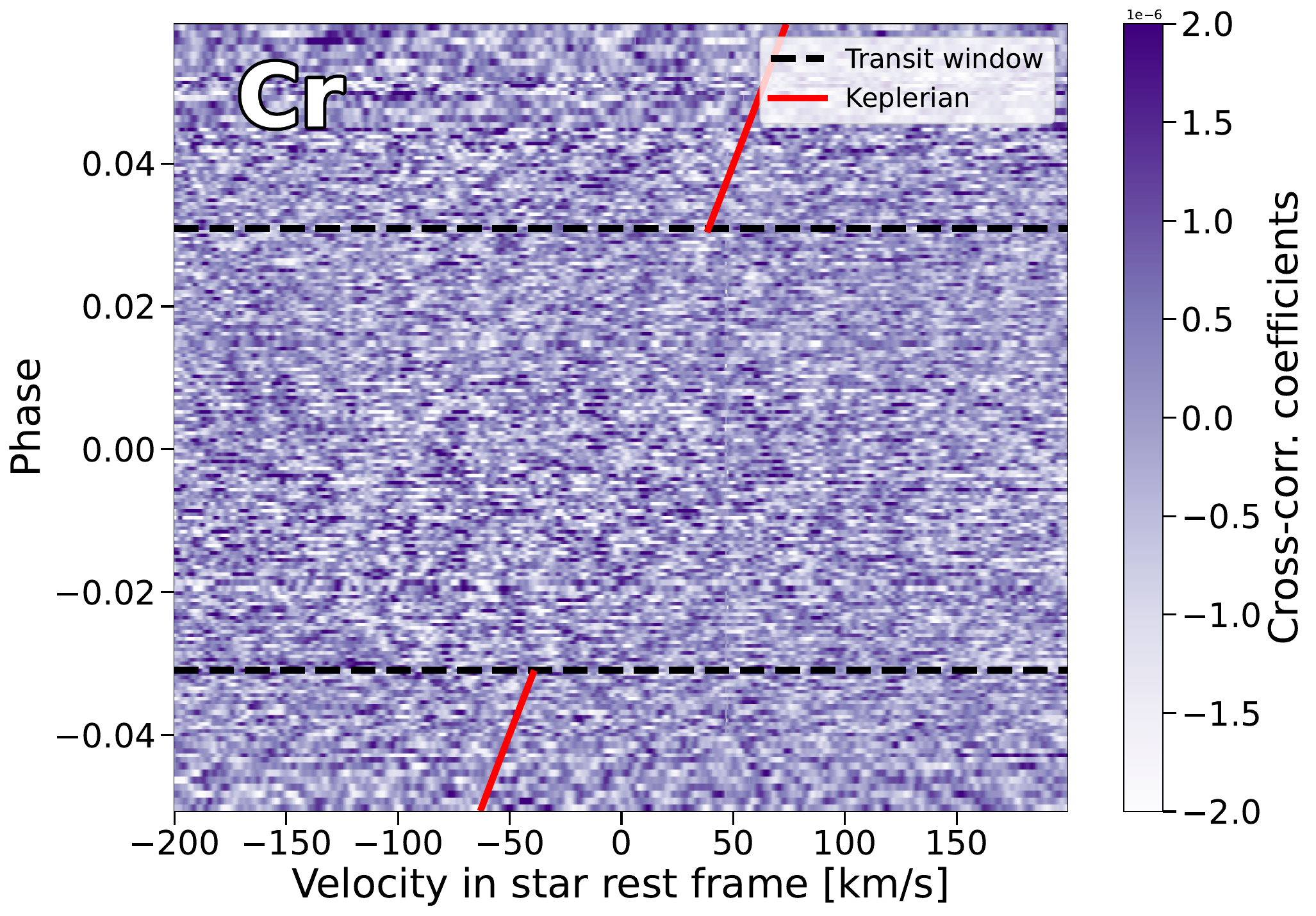}
   \hspace{+0.22cm}
    \includegraphics[width=\columnwidth]{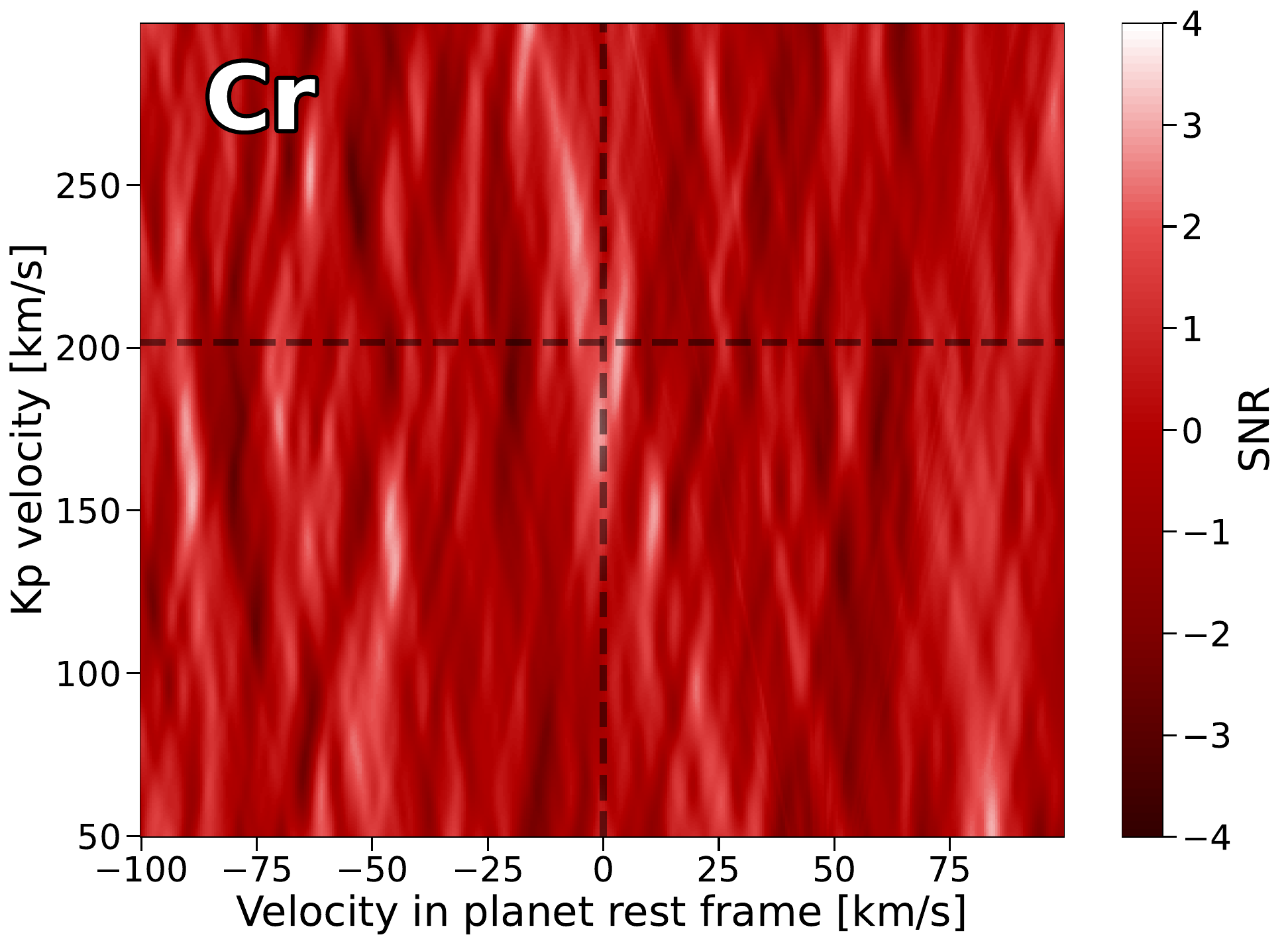}
   \includegraphics[width=\columnwidth]{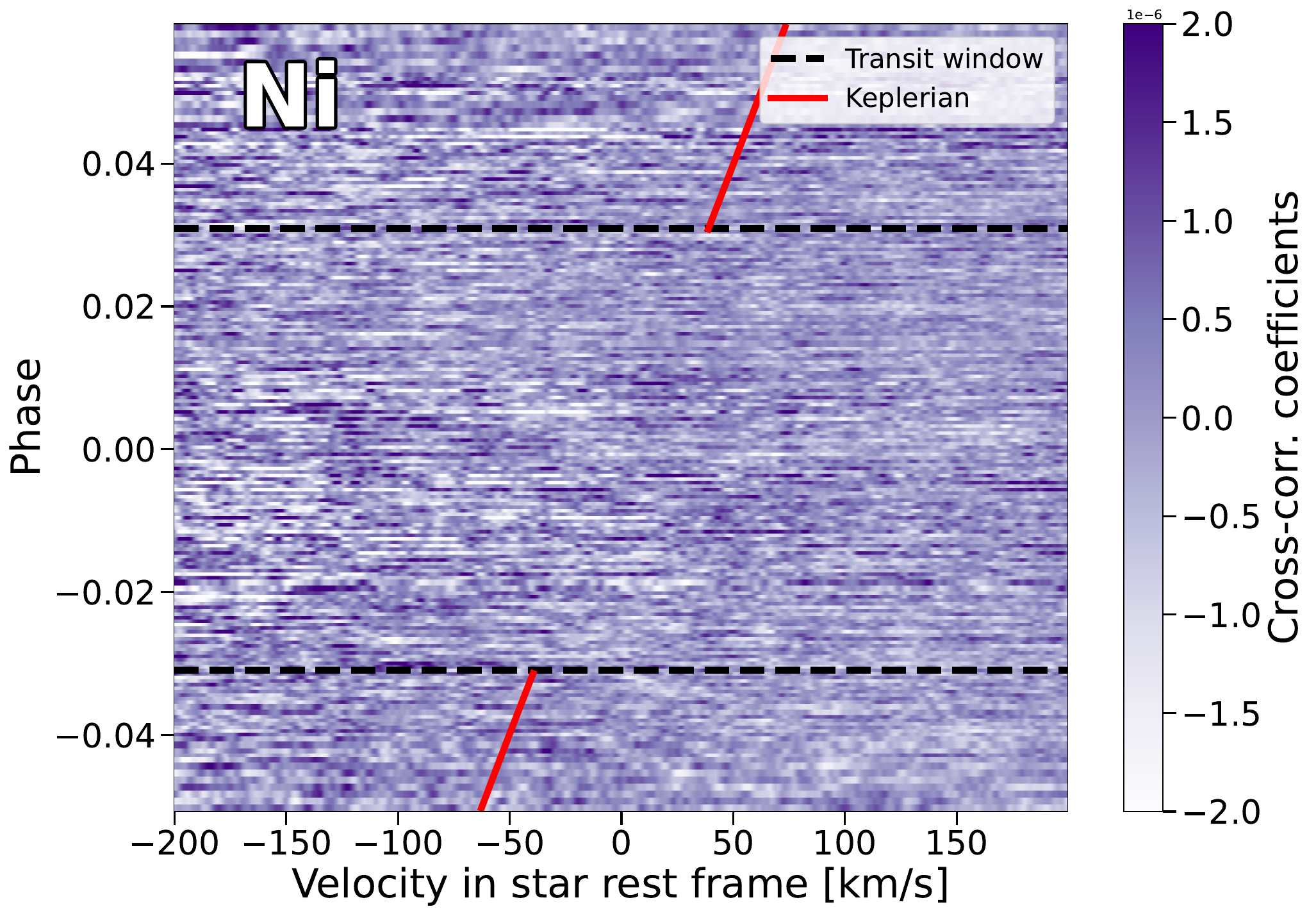}
   \hspace{+0.22cm}
    \includegraphics[width=\columnwidth]{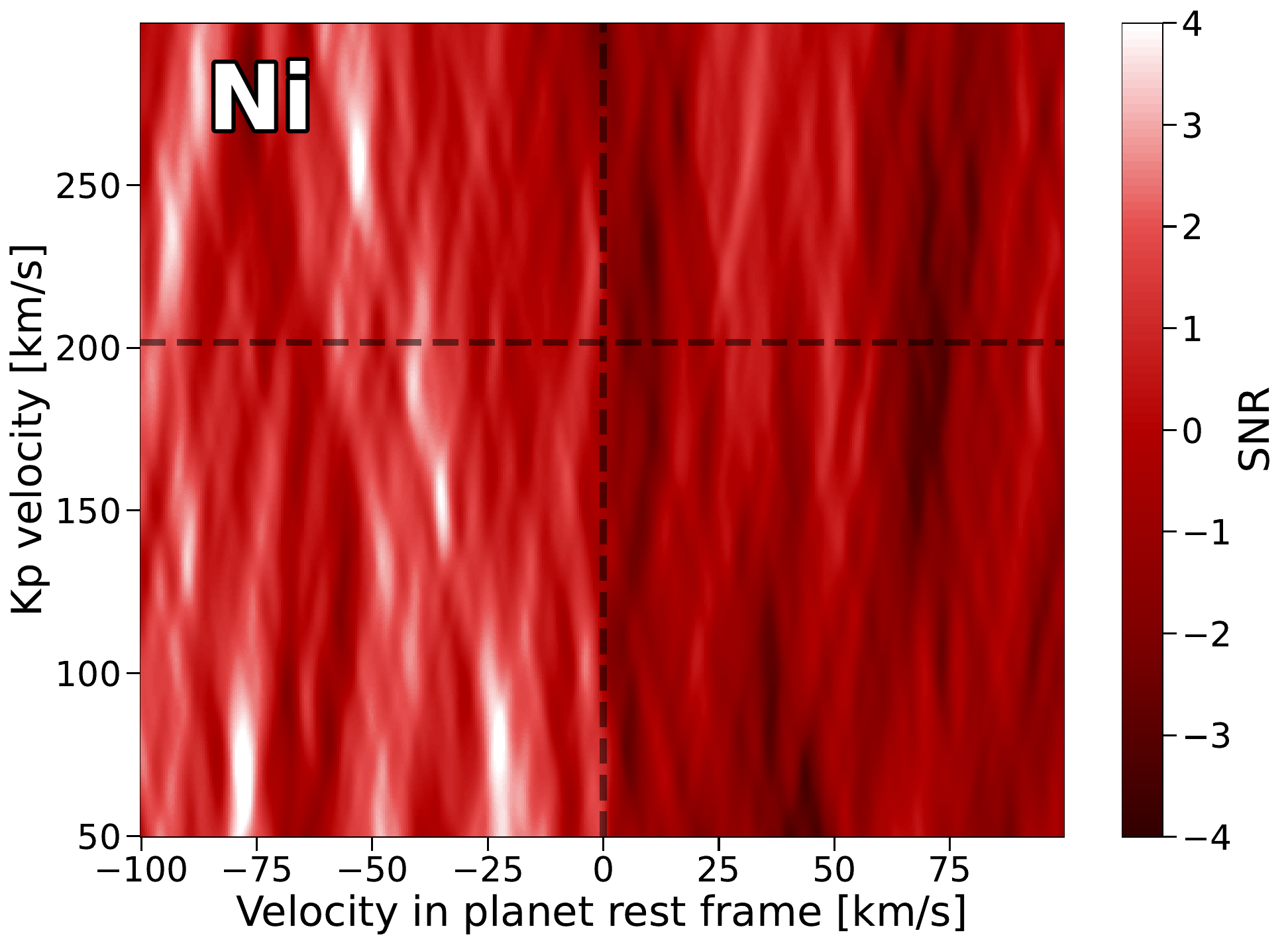}
   \includegraphics[width=\columnwidth]{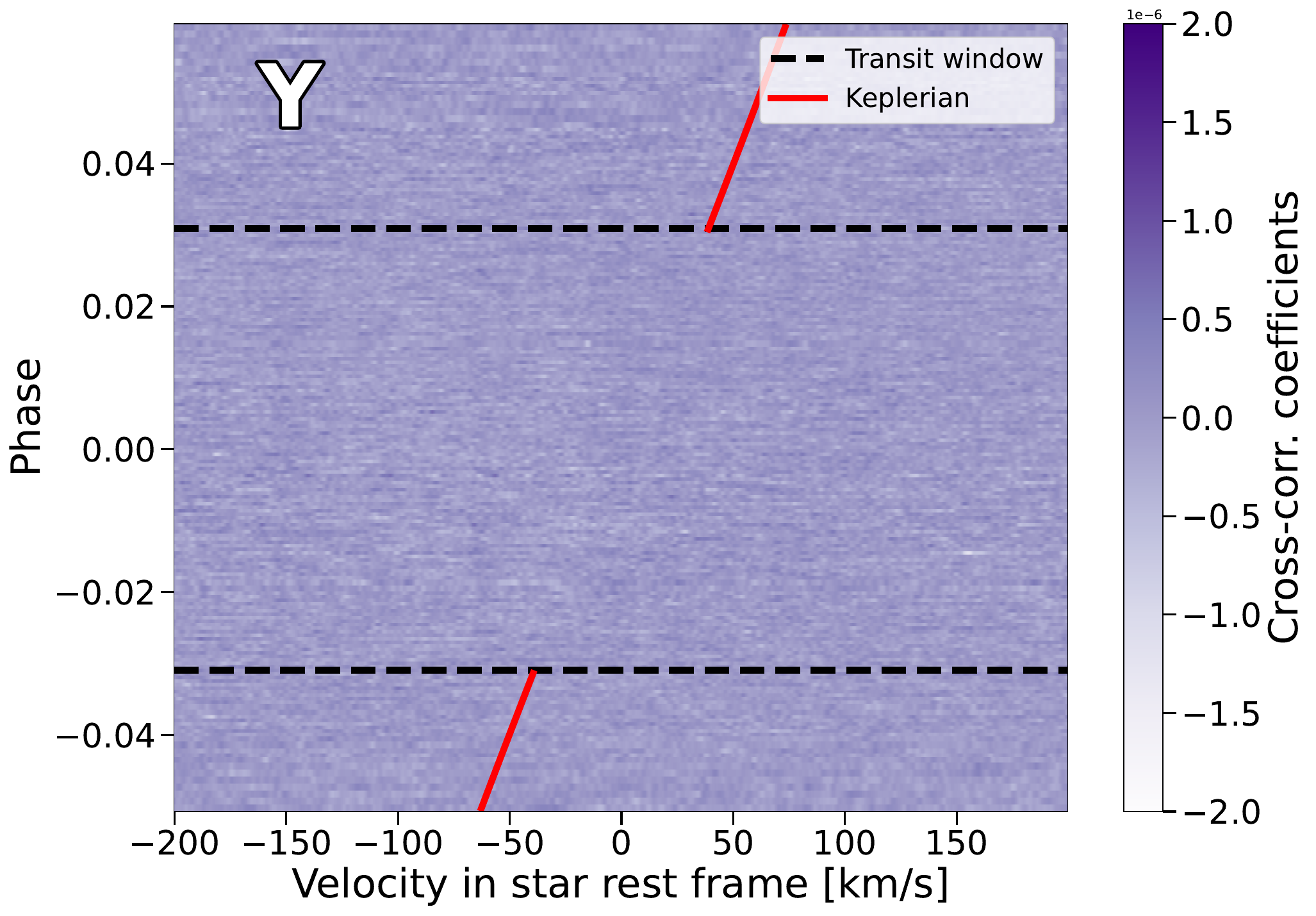}
   \hspace{+0.22cm}
   \includegraphics[width=\columnwidth]{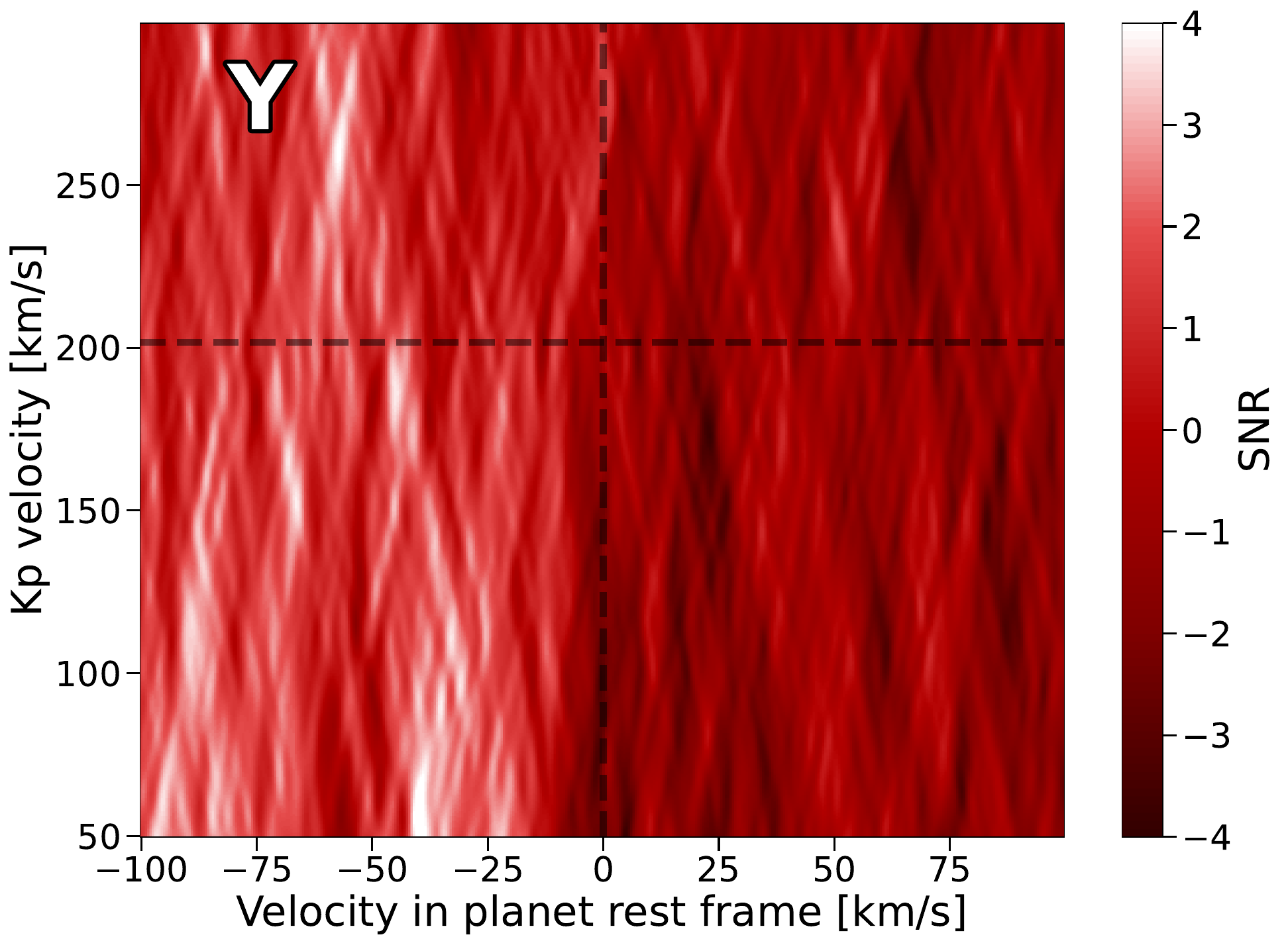}
   \caption{HARPS non-detections,~\ref{Figure:HARPS_non_detections} continued.}
\end{figure*}
\begin{figure*}[!h] 
    \includegraphics[width=\columnwidth]{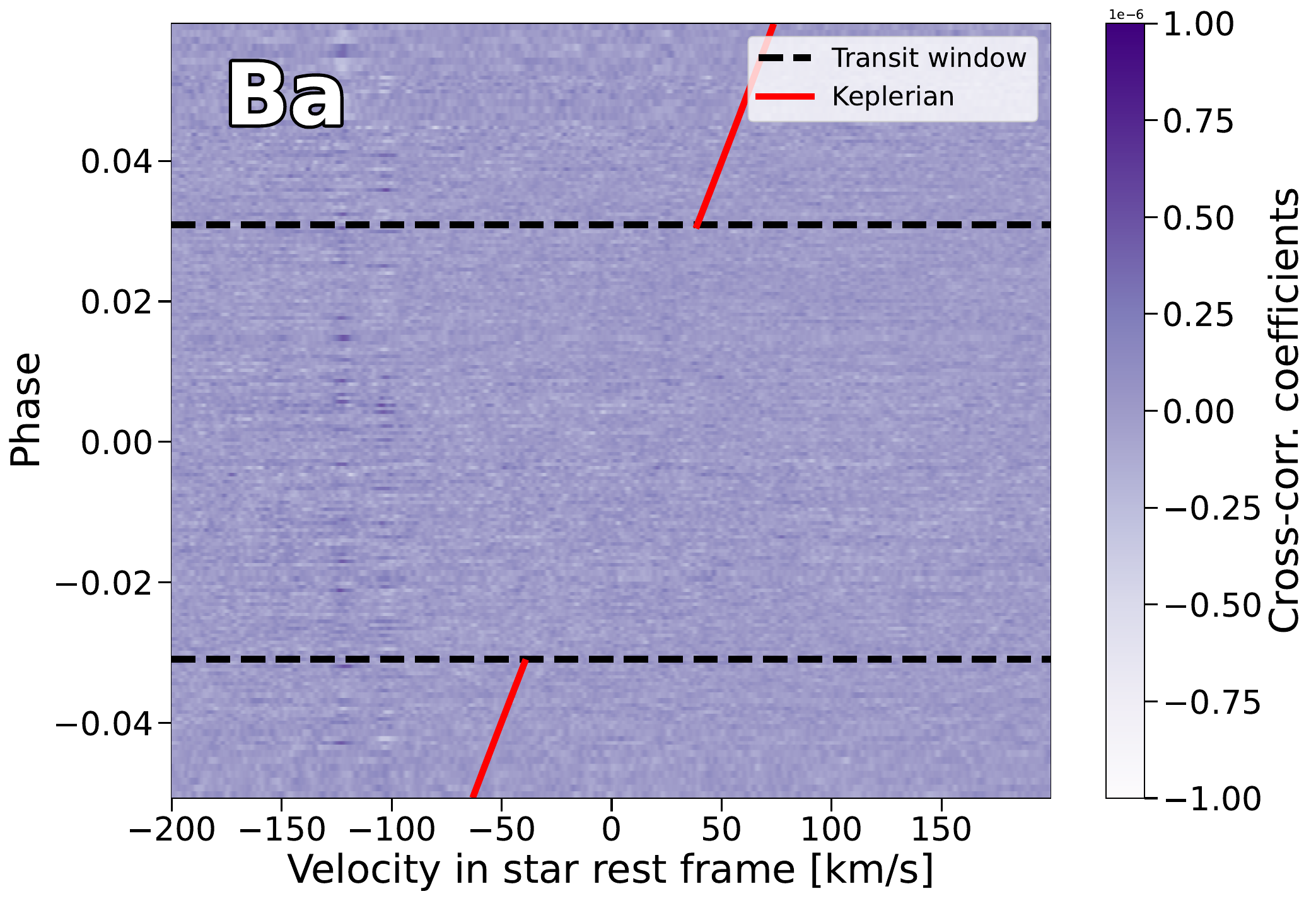}
   \hspace{+0.22cm}
   \includegraphics[width=\columnwidth]{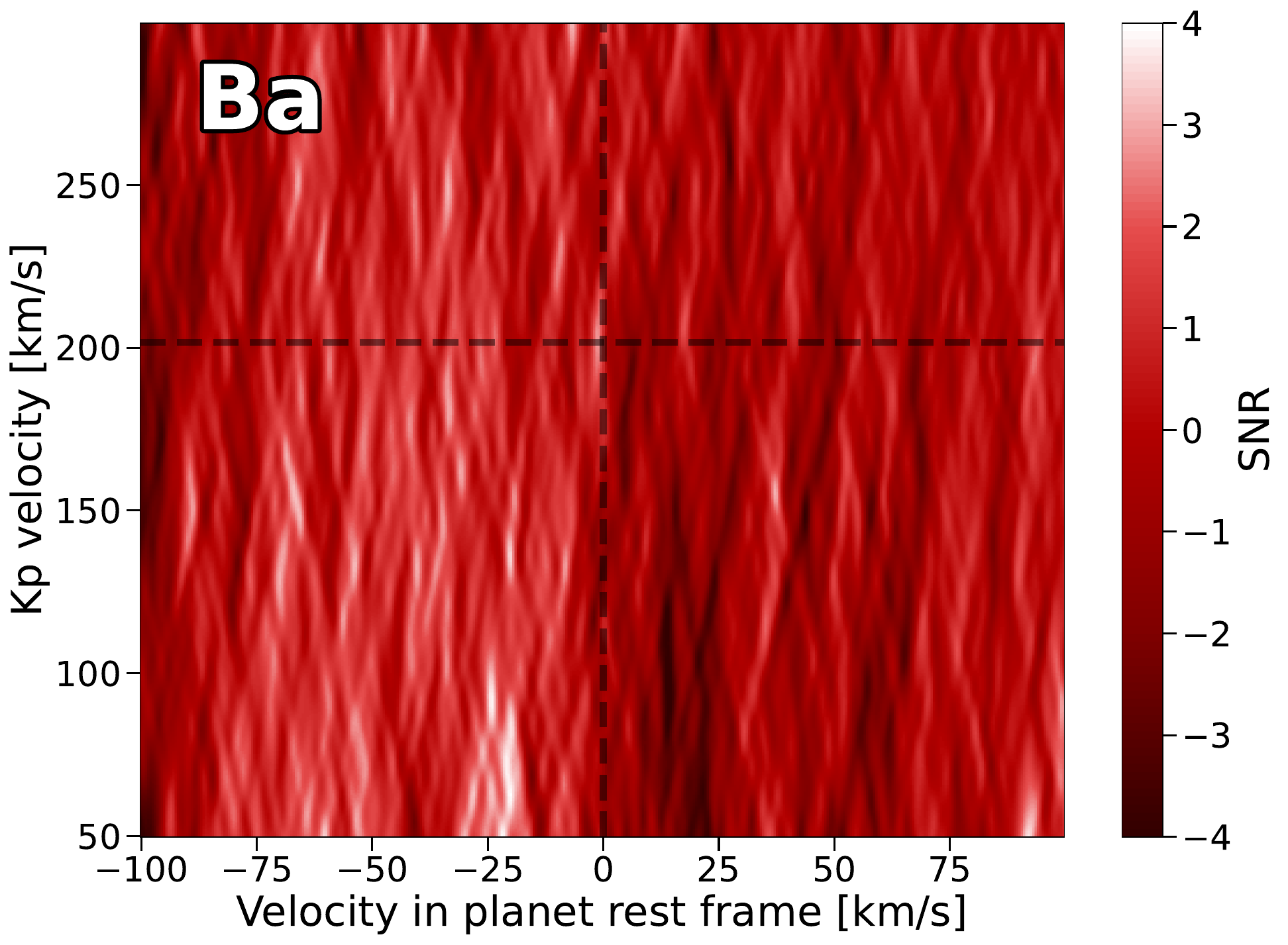} 
   \includegraphics[width=\columnwidth]{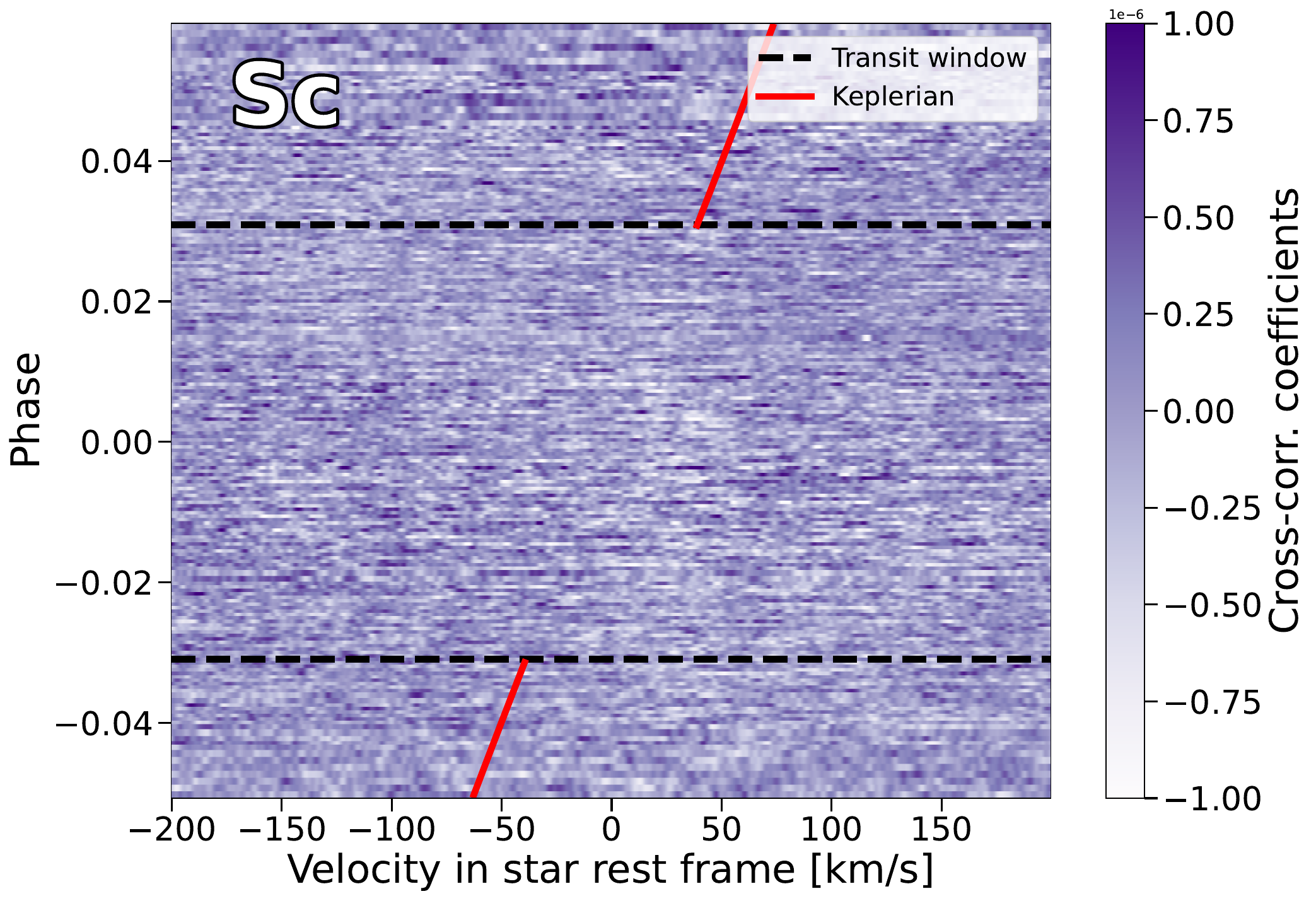}
    \includegraphics[width=\columnwidth]{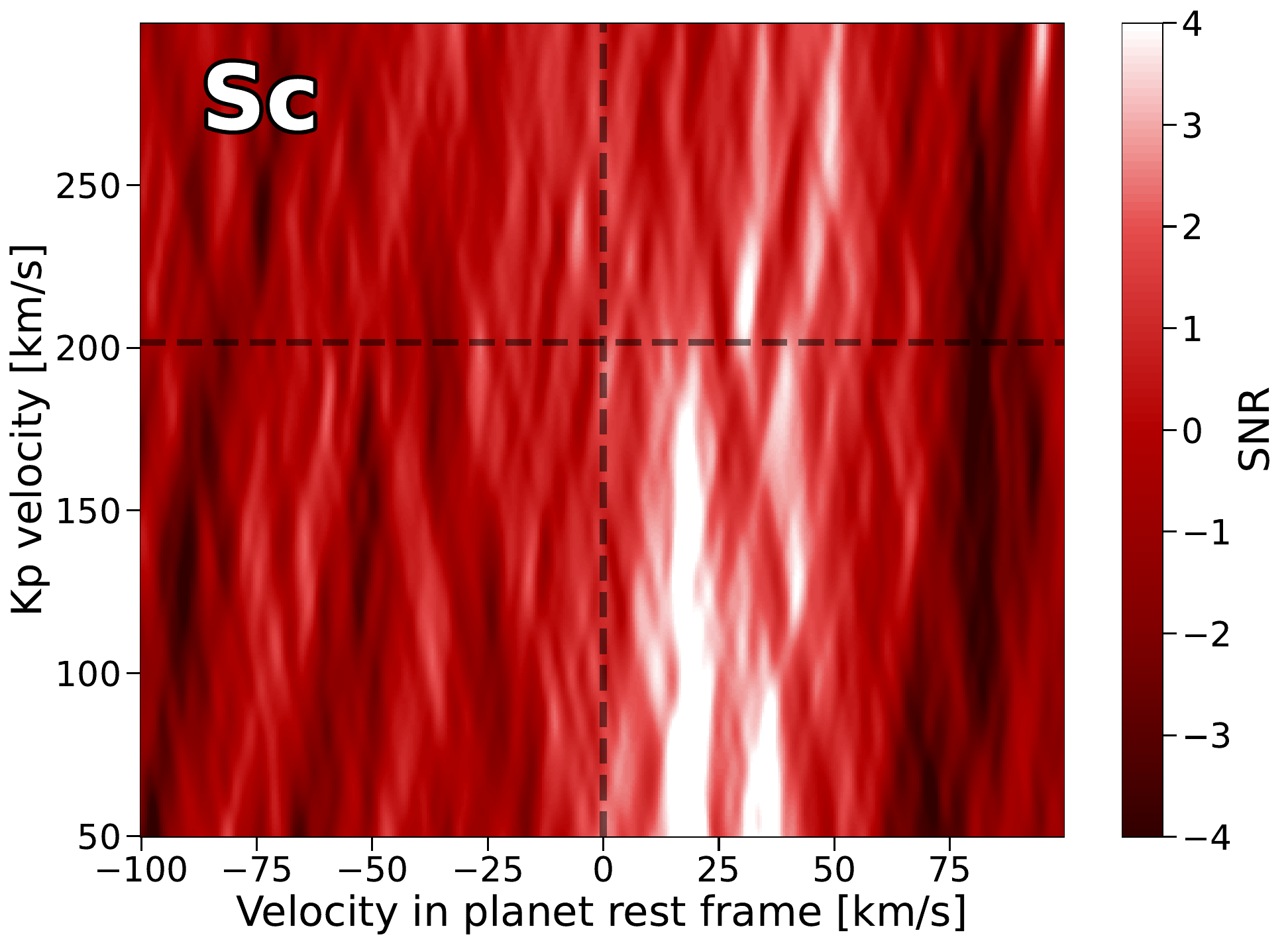}
   \includegraphics[width=\columnwidth]{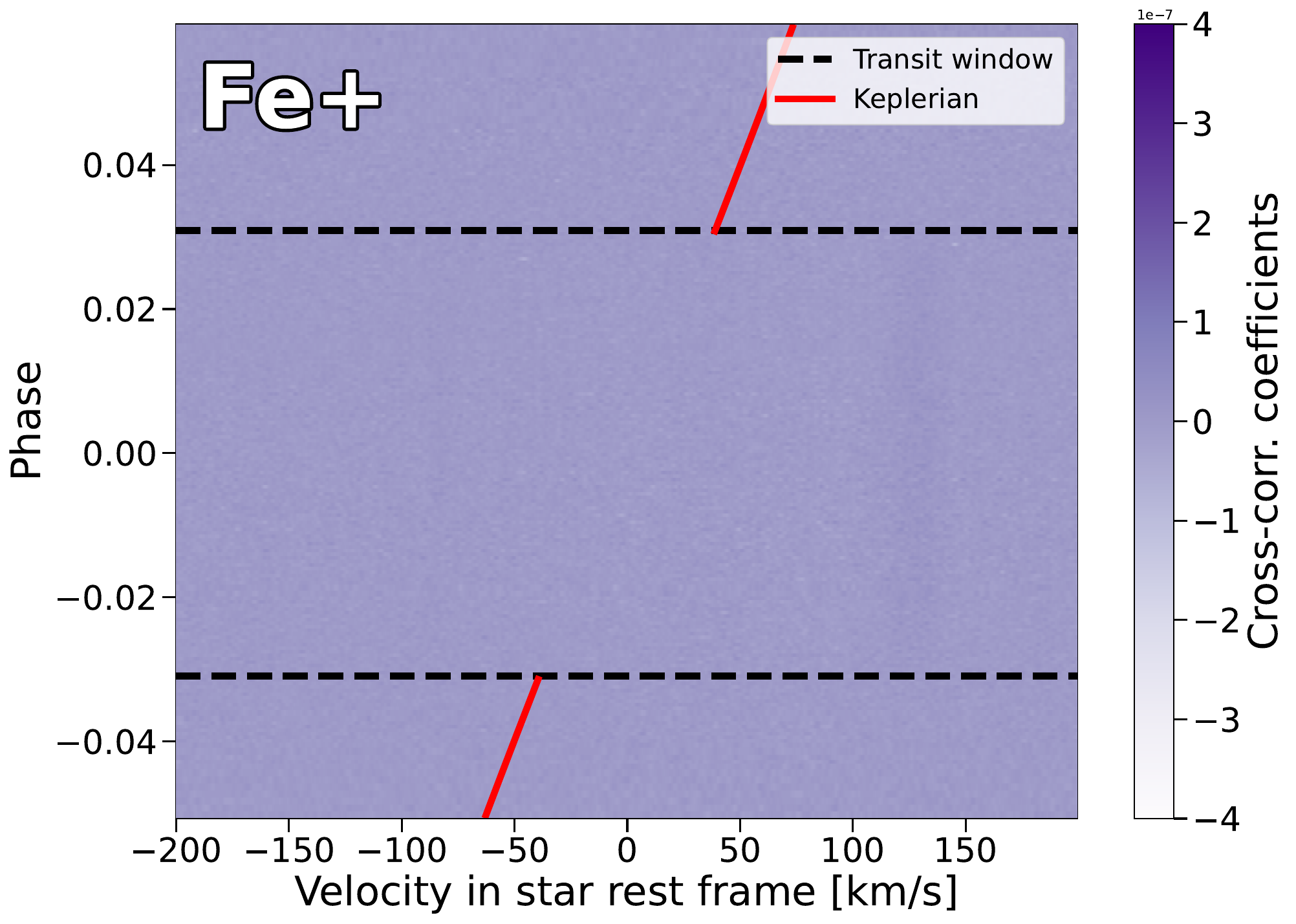}
   \hspace{+0.22cm}
   \includegraphics[width=\columnwidth]{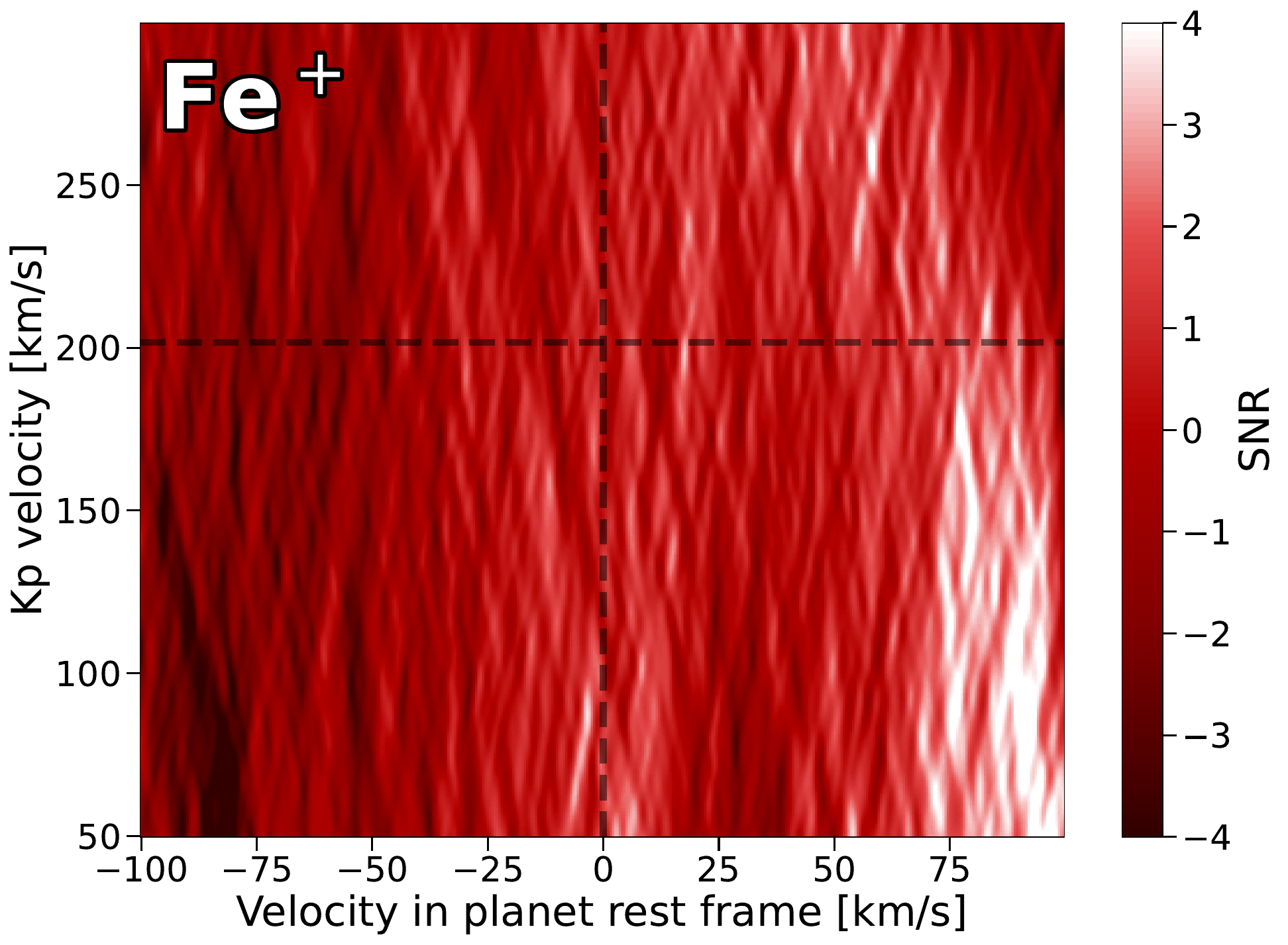}
   \caption{HARPS non-detections,~\ref{Figure:HARPS_non_detections} continued.}
\end{figure*}
\begin{figure*}[!h]    
   \includegraphics[width=\columnwidth]{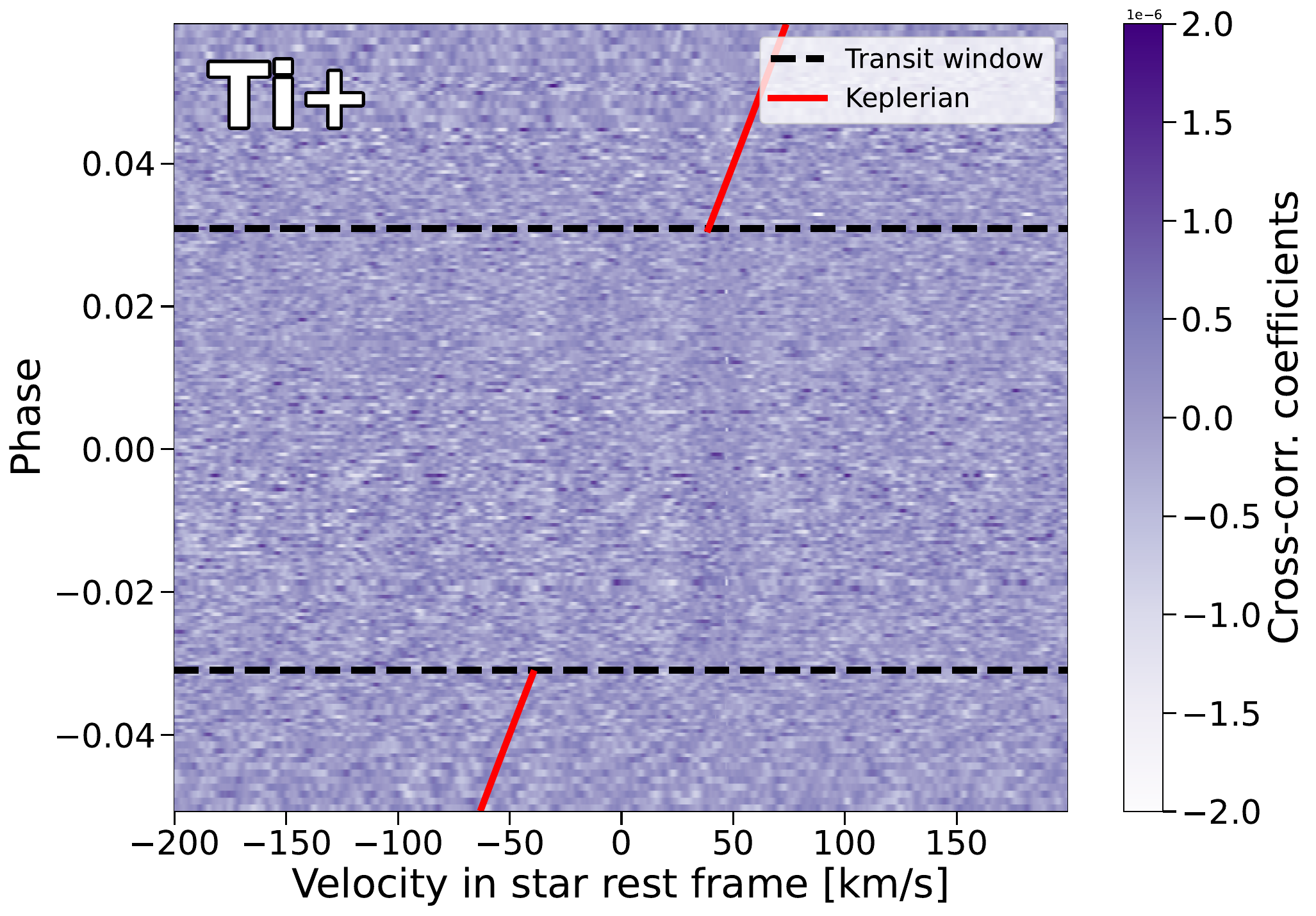}
   \hspace{+0.22cm}
   \includegraphics[width=\columnwidth]{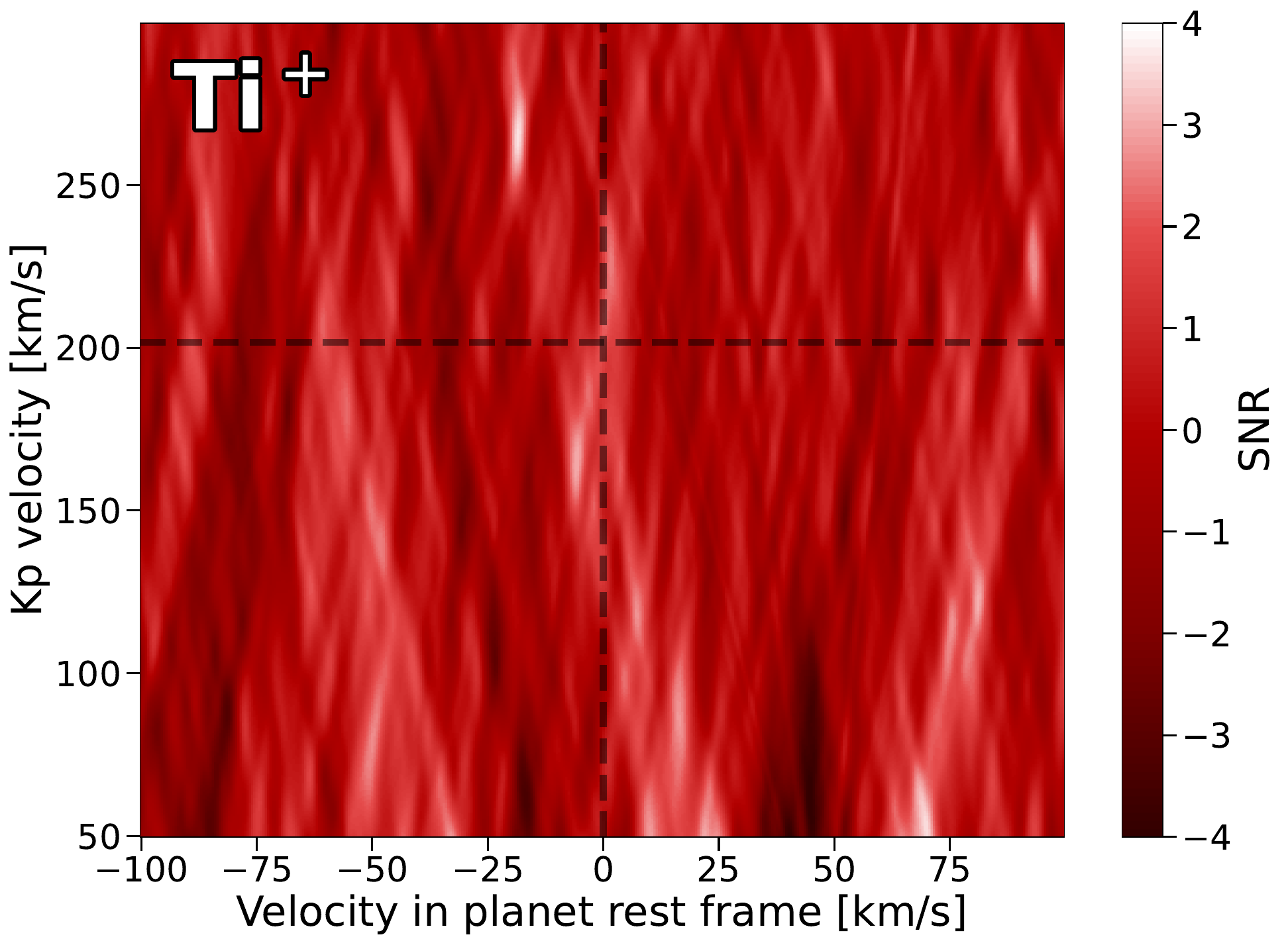}
   \includegraphics[width=\columnwidth]{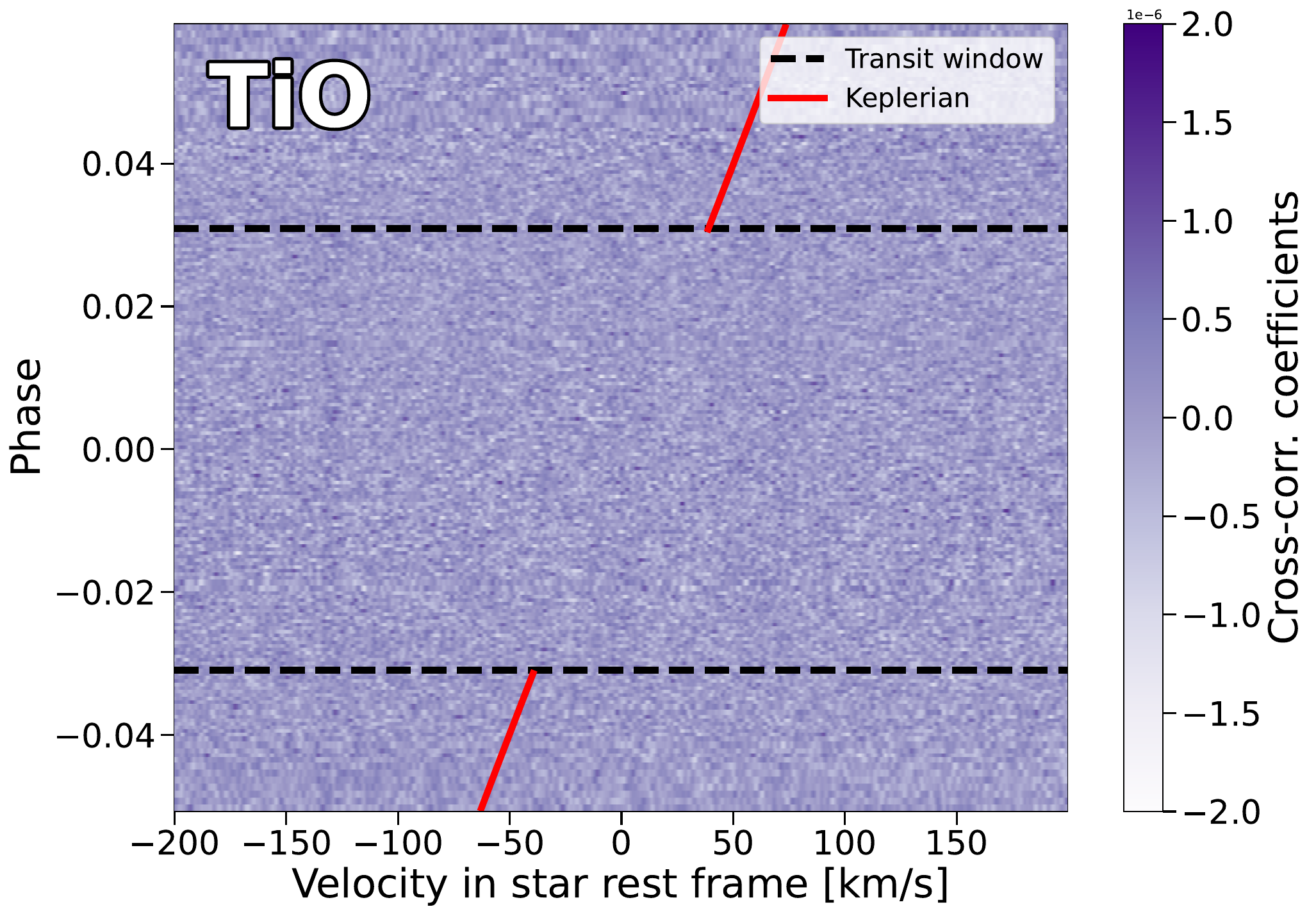}
    \includegraphics[width=\columnwidth]{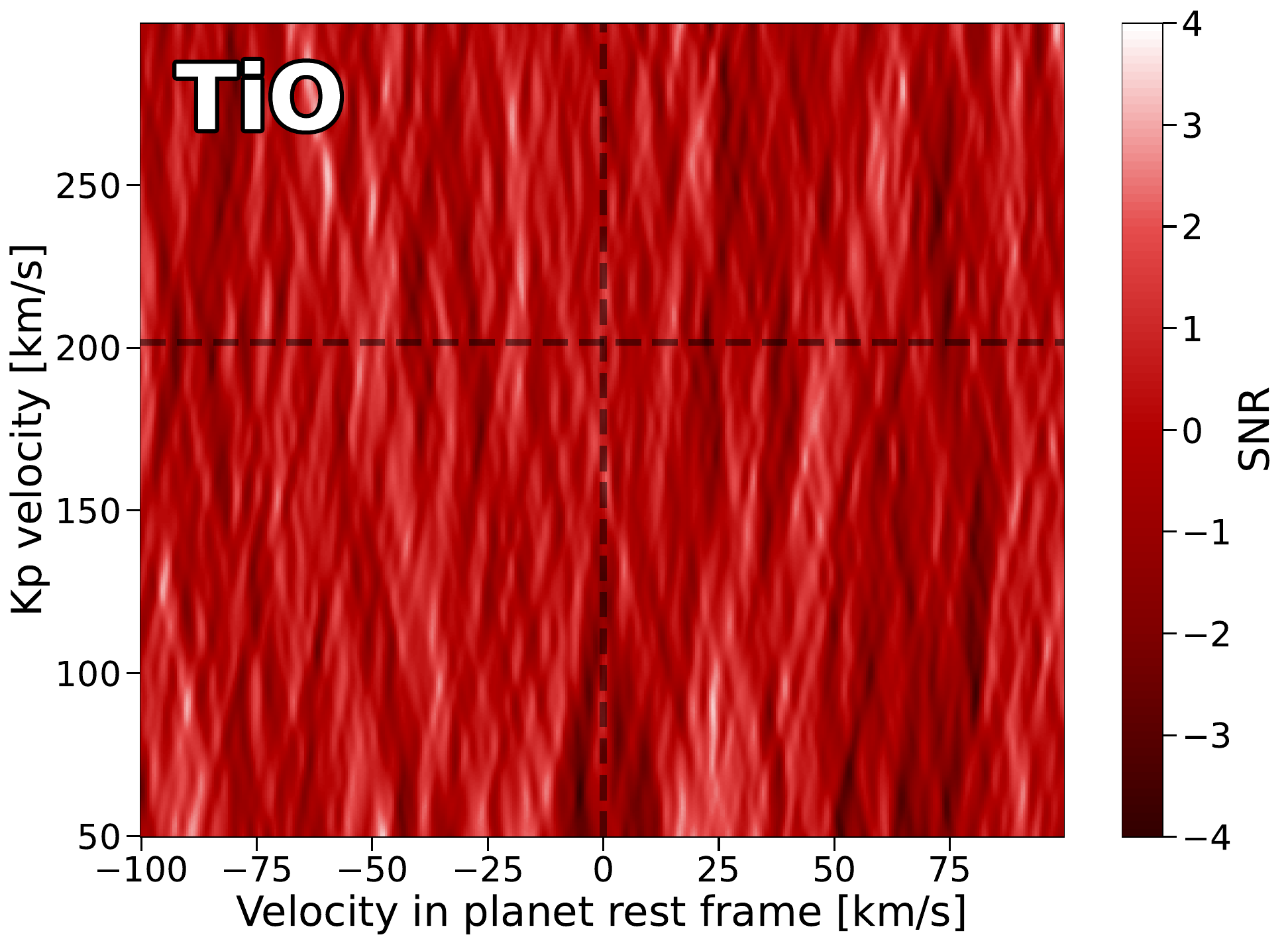}
   \includegraphics[width=\columnwidth]{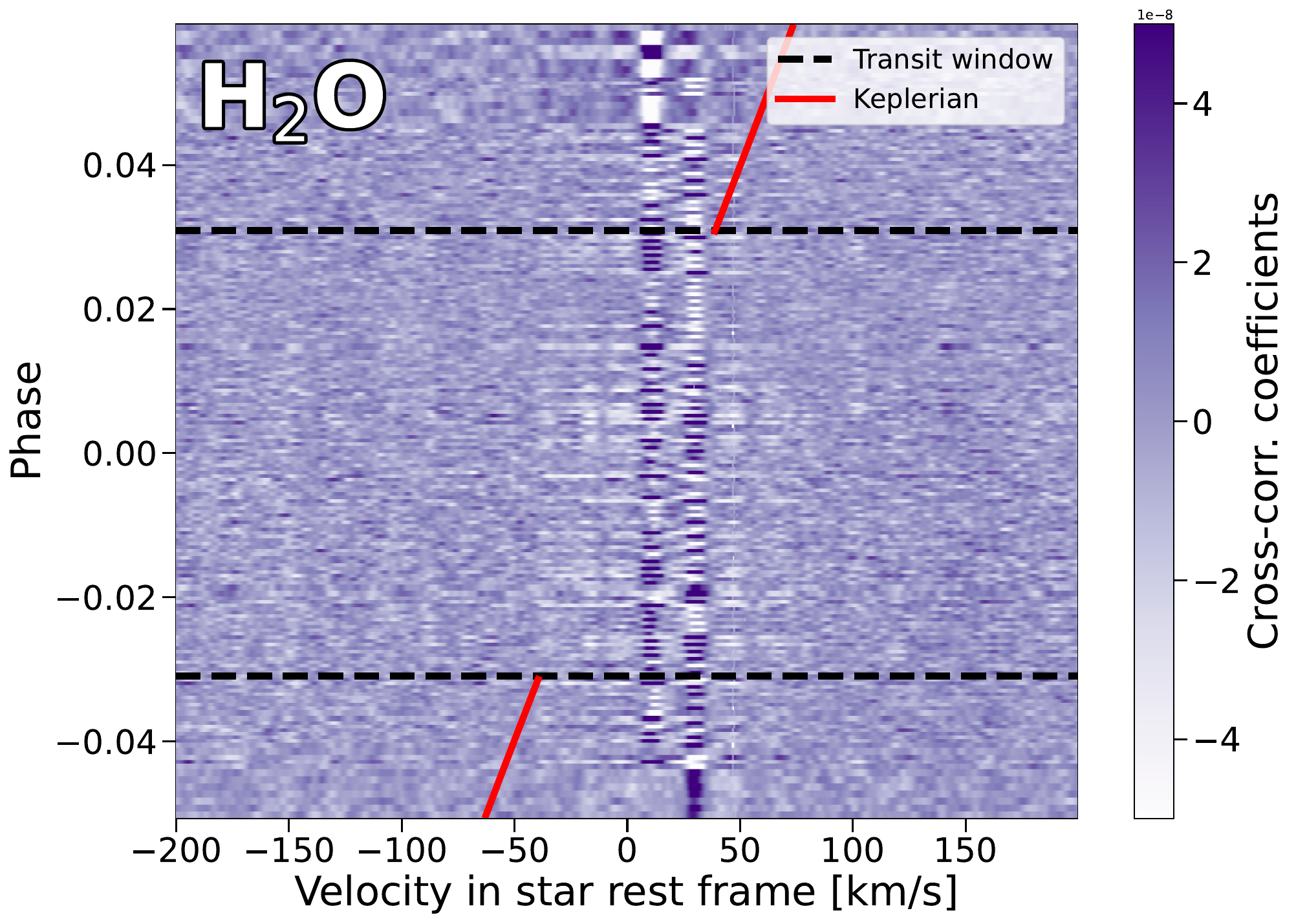}
   \hspace{+0.22cm}
   \includegraphics[width=\columnwidth]{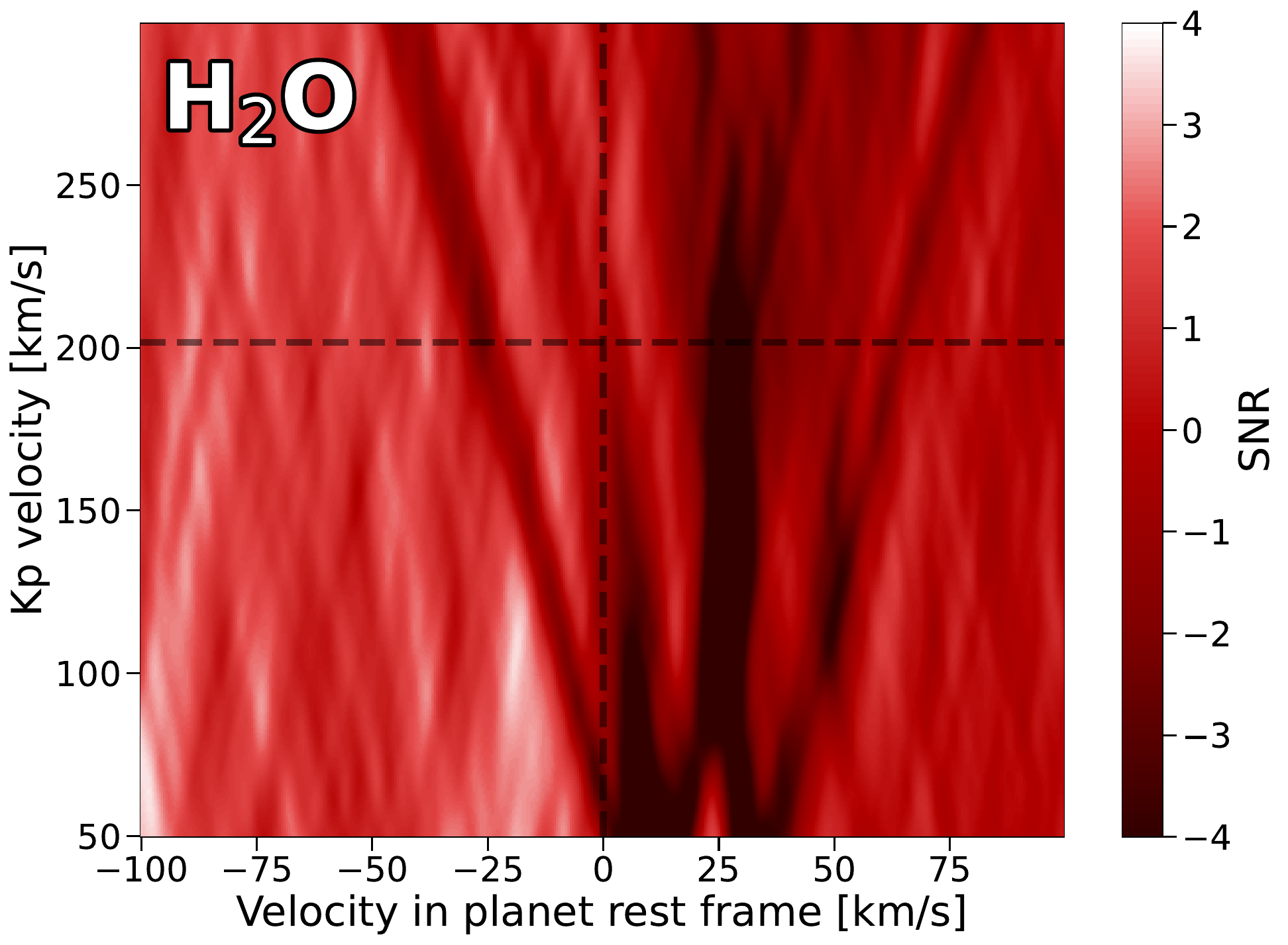} 

   \caption{HARPS non-detections,~\ref{Figure:HARPS_non_detections} continued. The cross-correlation maps for H$_2$O present features around 25 km~s$^{-1}$ due to remnant artefacts of the telluric correction.}
\end{figure*}
\clearpage
\begin{figure*}[!h]
   \includegraphics[width=\columnwidth]{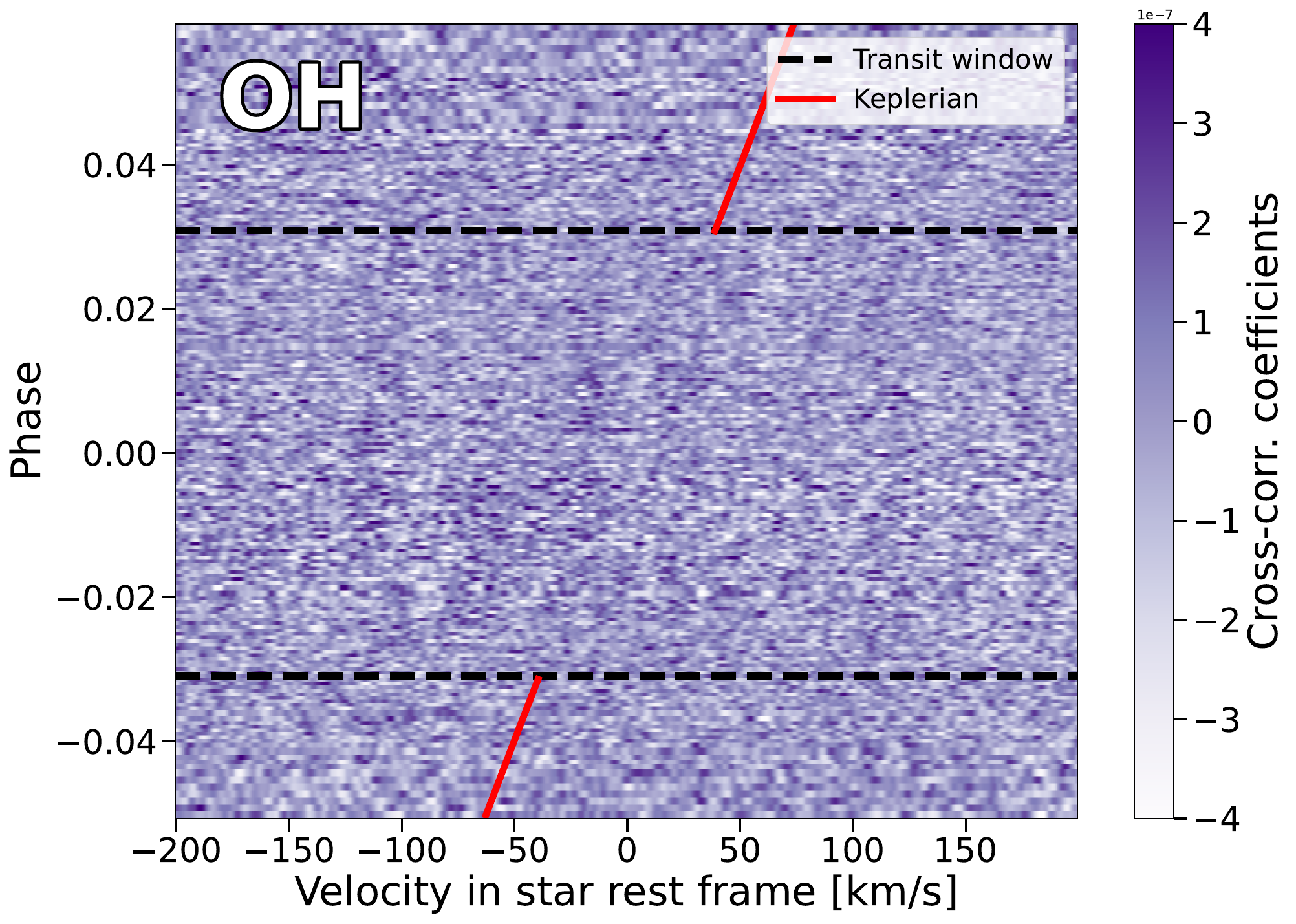}
   \hspace{+0.22cm}
   \includegraphics[width=\columnwidth]{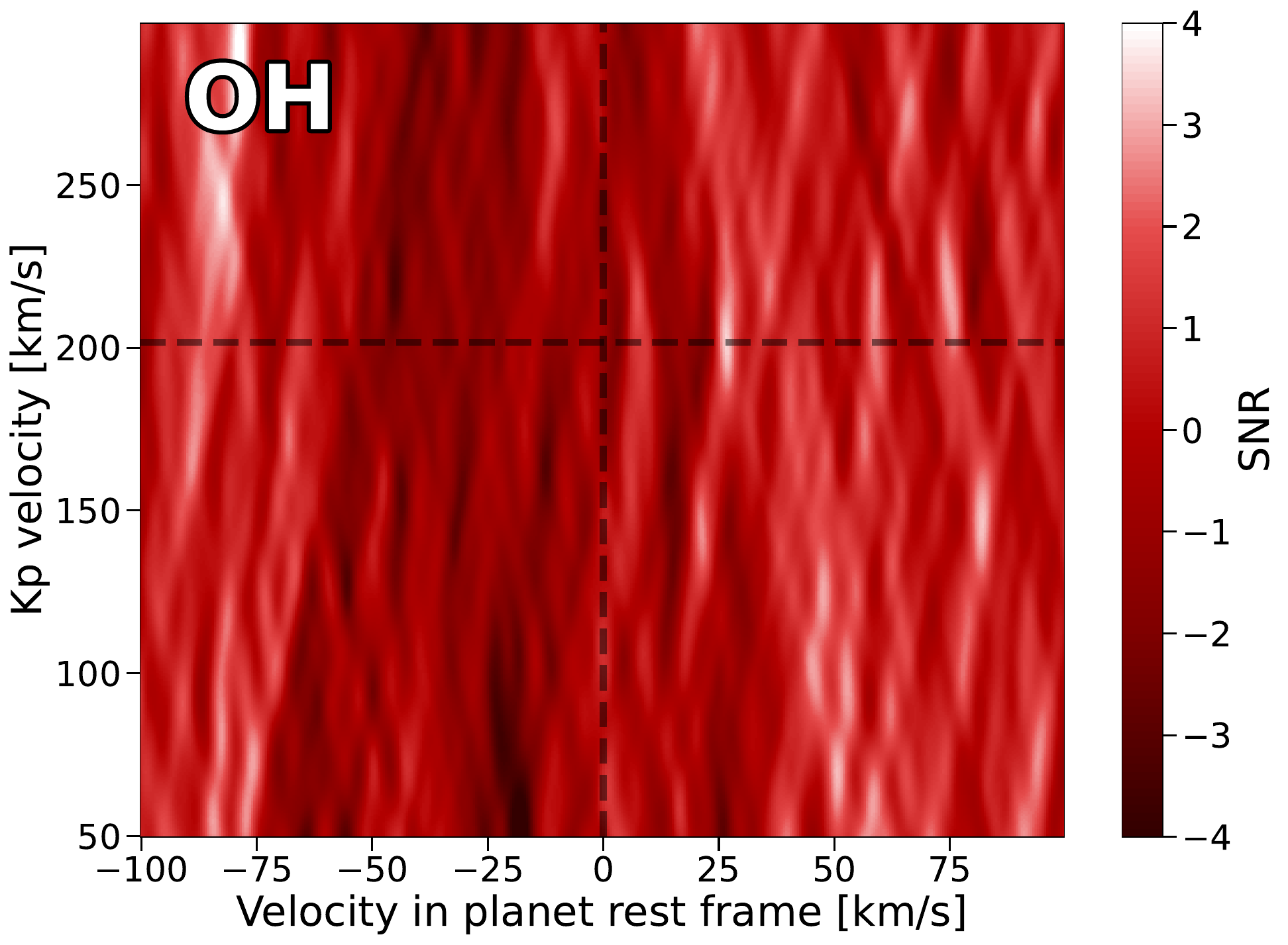} 
    
   \caption{HARPS non-detections,~\ref{Figure:HARPS_non_detections} continued.}
\end{figure*}


\begin{figure*}[!h]
\section{PCA-based CCF analysis}
\label{PCA_CCF}

    \includegraphics[width=\columnwidth]{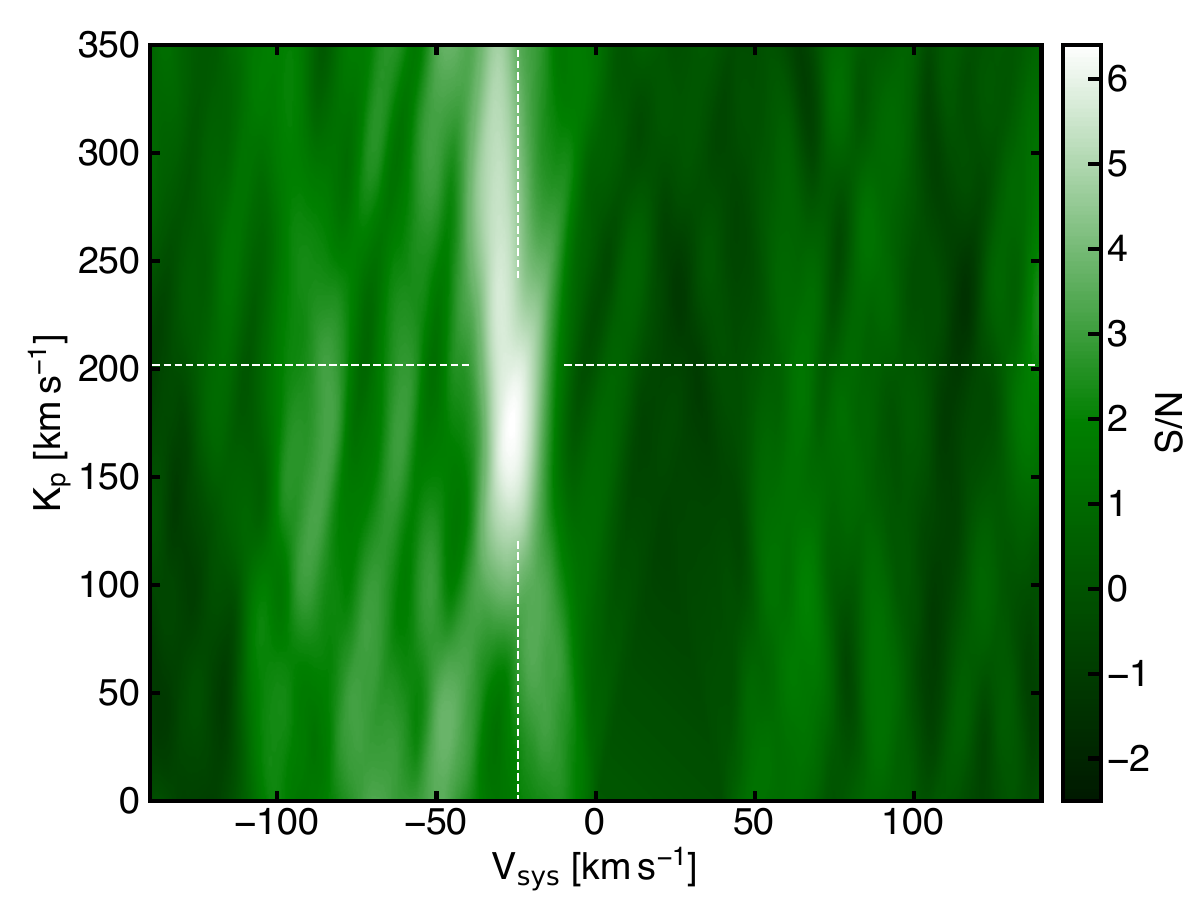}
    \caption{Cross-correlation detection of Fe from the HARPS transits, but now using a blind data-driven PCA-based detrending algorithm instead.  The white dashed lines denote the expected $K_\mathrm{p}$ and $v_\mathrm{sys}$ \citep[$K_\mathrm{p}$=200.7 km~s$^{-1}$ and $v_\mathrm{sys}$=$-$24.45 km~s$^{-1}$;][]{Prinoth+2022, Anderson+2018} of WASP-189b assuming a uniform and static atmosphere. The white blob near the expected location shows the observed signal of WASP-189b. Differently from Figure~\ref{Figure:WASP189b_HARPS_FeDetectionDopplerShadowCorrected},~\ref{Figure:WASP189b_HARPS_FeDetectionDopplerShadowCorrected_Blueshift_DELTAkp},~\ref{Figure:WASP189b_NIRPS_FeNonDetection}, and,~\ref{Figure:HARPS_non_detections},~\ref{Figure:NIRPS_non_detections}, this map has been generated from transmission spectra non-shifted by the systemic velocity.
    The uniformness of the CCF map between 0 and 50 km\,s$^{-1}$ at low $K_\mathrm{p}$ values is a result of the Rossiter-McLaughlin mask excluding values overlapping with the Doppler shadow in radial velocity space.
    The data product that produced this CCF map is the input for the atmospheric retrieval analysis (Section \ref{subsec: Atmospheric retrieval results}).  }
    \label{Figure:PCA_Fe_CCF}
\end{figure*}

\closeappendix

\clearpage

%
%

%

\clearpage

\end{document}